\documentclass[article,a4paper,oneside,10pt]{memoir}

\usepackage{graphics}

\usepackage{amsmath,amsthm}
\usepackage{amssymb}
\usepackage{amstext}
\usepackage{graphicx}
\usepackage{stmaryrd}

\usepackage{hyperref} 

\usepackage{color}
\usepackage[applemac]{inputenc}

\usepackage{pict2e}

\usepackage{xparse}

\makeatletter
\DeclareRobustCommand{\lintprod}{%
  \mathbin{\mathpalette\int@prod{(0,0)(0.8,0)(0.8,0.6)}}%
}
\DeclareRobustCommand{\rintprod}{%
  \mathbin{\mathpalette\int@prod{(0.1,0.6)(0.1,0)(0.9,0)}}}

\newcommand{\int@prod}[2]{%
  \begingroup
  \sbox\z@{$\m@th#1+$}%
  \setlength\unitlength{\wd\z@}%
  \linethickness{0.09\unitlength}%
  \begin{picture}(1,1)
  \roundcap
  \polyline#2
  \end{picture}%
  \endgroup
}
\makeatother

\usepackage[retainorgcmds]{IEEEtrantools}

\usepackage{tikz}

\newcommand{\Ab}{{\mathbf A}}
\newcommand{\Bb}{{\mathbf B}}

\newcommand{\Eb}{{\mathbf E}}
\newcommand{\Fb}{{\mathbf F}}

\newcommand{\Jb}{{\mathbf J}}
\newcommand{\Lb}{{\mathbf L}}

\newcommand{\Sb}{{\mathbf S}}
\newcommand{\Tb}{{\mathbf T}}
\newcommand{\ebf}{{\mathbf e}}

\newcommand{\fb}{{\mathbf f}}

\newcommand{\xb}{{\mathbf x}}

\newcommand{\ub}{{\mathbf u}}
\newcommand{\ubf}{{\mathbf u}}
\newcommand{\vb}{{\mathbf v}}
\newcommand{\wb}{{\mathbf w}}

\DeclareDocumentCommand \spinori { o } {%
  \IfNoValueTF {#1} {%
    \sigma%
  }{%
    \sigma_{#1}%
  }%
}

\DeclareDocumentCommand \indG { o o } {%
  \IfNoValueTF {#1} {%
    \iota%
  }{%
    \IfNoValueTF {#2} {%
      \iota_{#1}%
    }{%
      \iota_{#1}(#2)%
    }
  }%
}

\DeclareDocumentCommand \indGc { o o } {%
  \IfNoValueTF {#1} {%
    \bar\iota%
  }{%
    \IfNoValueTF {#2} {%
      \bar\iota_{#1}%
    }{%
      \bar\iota_{#1}(#2)%
    }
  }%
}

\DeclareDocumentCommand \indGinv { o o } {%
  \IfNoValueTF {#1} {%
    \bar\iota%
  }{%
    \IfNoValueTF {#2} {%
      \iota^{-1}_{#1}%
    }{%
      \iota^{-1}_{#1}(#2)%
    }
  }%
}

\DeclareDocumentCommand \coefG { o o } {%
  \IfNoValueTF {#1} {%
    \gamma%
  }{%
    \IfNoValueTF {#2} {%
      \gamma_{#1}%
    }{%
      \gamma_{#1}(#2)%
    }
  }%
}

\newcommand{\abf}{\mathbf{a}}
\newcommand{\bbf}{\mathbf{b}}

\newcommand{\wbf}{\mathbf{w}}

\renewcommand{\Ab}{\vp}
\renewcommand{\Fb}{\mf}

\newcommand{\emset}{\mathbf{T}_\text{em}}
\newcommand{\emseti}{T^\text{em}}

\DeclareDocumentCommand \act { o } {%
  \IfNoValueTF {#1} {%
    \mathcal S%
  }{%
    \mathcal S_{\text{#1}}%
  }%
}

\DeclareDocumentCommand \ld { o } {%
  \IfNoValueTF {#1} {%
    \mathcal L%
  }{%
    \mathcal L_{\text{#1}}%
  }%
}

\newcommand{\drm}{\,\mathrm{d}}

\newcommand{\deltabf}{{\boldsymbol \partial}}
\newcommand{\xibf}{{\boldsymbol \xi}}
\newcommand{\Omegabf}{{\boldsymbol \Omega}}

\newcommand{\xpert}[1][]{\varepsilon_{#1}}
\newcommand{\xbpert}{\pmb{\varepsilon}}

\newcommand{\mf}{{\mathbf F}}
\newcommand{\mfi}{F}

\newcommand{\sd}{{\mathbf J}}

\newcommand{\vp}{{\mathbf A}}
\newcommand{\vpi}{{A}}

\newcommand{\gf}{{\mathbf G}}

\newcommand{\Gb}{\mathbf{G}}
\makeatletter
\def\trMs{\@ifnextchar[{\@with}{\@without}}
\def\@with[#1]{\Gb_{\xbpert}^{#1}}
\def\@without{\Gb_{\xbpert}^s}
\makeatother

\DeclareMathOperator{\gr}{gr}

\newcommand{\hodgeinv}{{\scriptscriptstyle\mathcal{H}^{-1}}}

\makeatletter
\newcommand*{\boxwedge}{%
  \mathbin{%
    \mathpalette\@boxwedge{}%
  }%
}
\newcommand*{\@boxwedge}[2]{%
  \sbox0{$#1\boxplus\m@th$}%
  \dimen2=.5\dimexpr\wd0-\ht0-\dp0\relax % side bearing
  \dimen@=\dimexpr\ht0+\dp0\relax
  \def\lw{.06}% linw width as factor for height of \boxplus
  \kern\dimen2 % side bearing
  \tikz[
    line width=\lw\dimen@,
    line join=round,
    x=\dimen@,
    y=\dimen@,
  ]
  \draw
    (\lw/2,0) rectangle (1-\lw,1-\lw)
    (\lw,0) -- (.5,1-\lw-\lw/2) -- (1-\lw-\lw/2 ,0)
  ;%
  \kern\dimen2 % side bearing
}
\makeatletter

\makeatletter
\renewcommand*{\owedge}{%
  \mathbin{%
    \mathpalette\@owedge{}%
  }%
}
\newcommand*{\@owedge}[2]{%
  \sbox0{$#1\oplus\m@th$}%
  \dimen2=.5\dimexpr\wd0-\ht0-\dp0\relax % side bearing
  \dimen@=\dimexpr\ht0+\dp0\relax
  \def\lw{.04}% line width as factor for height of \oplus
  \def\radius{.5-\lw/2}%
  \kern\dimen2 % side bearing
  \tikz[
    line width=\lw\dimen@,
    line join=round,
    x=\dimen@,
    y=\dimen@,
    baseline=\dimexpr-.5\dimen@+\dp0\relax,
  ]
  \draw
    (0,0) circle[radius=\radius]
    (225:\radius) -- (0,.5-\lw) -- (-45:\radius)
  ;%
  \kern\dimen2 % side bearing  
}
\makeatother

\newtheoremstyle{question}
  {\topsep}   % ABOVESPACE
  {\topsep}   % BELOWSPACE
  {\upshape}  % BODYFONT
  {0pt}       % INDENT (empty value is the same as 0pt)
  {\itshape}  % HEADFONT
  {.}         % HEADPUNCT
  {5pt plus 1pt minus 1pt} % HEADSPACE
  {\thmname{#1} \thesection.\thmnumber{\itshape#2}\thmnote{(#3)}} % CUSTOM-HEAD-SPEC

\makeatletter
    \def\@endtheorem{\hfill$\P$\endtrivlist\@endpefalse }
\makeatother
\theoremstyle{question}

\makeatletter
\let\save@mathaccent\mathaccent
\newcommand*\if@single[3]{%
  \setbox0\hbox{${\mathaccent"0362{#1}}^H$}%
  \setbox2\hbox{${\mathaccent"0362{\kern0pt#1}}^H$}%
  \ifdim\ht0=\ht2 #3\else #2\fi
  }
\newcommand*\rel@kern[1]{\kern#1\dimexpr\macc@kerna}
\newcommand*\widebar[1]{\@ifnextchar^{{\wide@bar{#1}{0}}}{\wide@bar{#1}{1}}}
\newcommand*\wide@bar[2]{\if@single{#1}{\wide@bar@{#1}{#2}{1}}{\wide@bar@{#1}{#2}{2}}}
\newcommand*\wide@bar@[3]{%
  \begingroup
  \def\mathaccent##1##2{%
    \let\mathaccent\save@mathaccent
    \if#32 \let\macc@nucleus\first@char \fi
    \setbox\z@\hbox{$\macc@style{\macc@nucleus}_{}$}%
    \setbox\tw@\hbox{$\macc@style{\macc@nucleus}{}_{}$}%
    \dimen@\wd\tw@
    \advance\dimen@-\wd\z@
    \divide\dimen@ 3
    \@tempdima\wd\tw@
    \advance\@tempdima-\scriptspace
    \divide\@tempdima 10
    \advance\dimen@-\@tempdima
    \ifdim\dimen@>\z@ \dimen@0pt\fi
    \rel@kern{0.6}\kern-\dimen@
    \if#31
      \overline{\rel@kern{-0.6}\kern\dimen@\macc@nucleus\rel@kern{0.4}\kern\dimen@}%
      \advance\dimen@0.4\dimexpr\macc@kerna
      \let\final@kern#2%
      \ifdim\dimen@<\z@ \let\final@kern1\fi
      \if\final@kern1 \kern-\dimen@\fi
    \else
      \overline{\rel@kern{-0.6}\kern\dimen@#1}%
    \fi
  }%
  \macc@depth\@ne
  \let\math@bgroup\@empty \let\math@egroup\macc@set@skewchar
  \mathsurround\z@ \frozen@everymath{\mathgroup\macc@group\relax}%
  \macc@set@skewchar\relax
  \let\mathaccentV\macc@nested@a
  \if#31
    \macc@nested@a\relax111{#1}%
  \else
    \def\gobble@till@marker##1\endmarker{}%
    \futurelet\first@char\gobble@till@marker#1\endmarker
    \ifcat\noexpand\first@char A\else
      \def\first@char{}%
    \fi
    \macc@nested@a\relax111{\first@char}%
  \fi
  \endgroup
}
\makeatother

\newcounter{aside}
   {\refstepcounter{aside}
   \begin{adjustwidth}{\parindent}{-1.5\parindent}
	\small \textbf{Aside \thechapter.\theaside~#1}
	}%\label{#1}}%
{\par\end{adjustwidth}}

\usepackage{ulem}
\counterwithout{section}{chapter}
\settrimmedsize{297mm}{210mm}{*}
\settypeblocksize{664pt}{498.13pt}{*}
\setulmargins{3cm}{*}{*}
\setlrmargins{*}{*}{0.75}
\setmarginnotes{17pt}{51pt}{\onelineskip}
\setheadfoot{\onelineskip}{2\onelineskip}
\setheaderspaces{*}{2\onelineskip}{*}
\checkandfixthelayout

\providecommand{\keywords}[1]
{
  \small	
  \textbf{\textit{Keywords---}} #1
}

\newcommand{\Icb}{\boldsymbol{\mathcal{I}}}
\newcommand{\Mc}{\mathbf{M}}
\newcommand{\Mcb}{\boldsymbol{\mathbf{M}}}
\newcommand{\Nb}{\mathbf{N}}
\newcommand{\xcr}{\boldsymbol{\alpha}}

\newcommand{\zetabf}{{\boldsymbol \zeta}}

\newcommand{\xisigmai}{\xi_{\bar\ell,\sigma,i}}
\newcommand{\xisigmaj}{\xi_{\bar\ell,\sigma,j}}
\newcommand{\xisigmal}{\xi_{\bar\ell,\sigma,\ell}}

\newcommand{\xisigmabi}{\xi_{\bar\ell,\bar\sigma,i}}
\newcommand{\xisigmabj}{\xi_{\bar\ell,\bar\sigma,j}}
\newcommand{\xisigmabl}{\xi_{\bar\ell,\bar\sigma,\ell}}

\newcommand{\xisigmaoi}{\xi_{\bar\ell,\sigma_1,i}}

\newcommand{\xisigmatj}{\xi_{\bar\ell,\sigma_2,j}}

\newcommand{\xisigmatt}{\xi_{\bar\ell,\sigma_2,t}}
\newcommand{\xibfsigma}{\xibf_{\bar\ell,\sigma}}
\newcommand{\xibfsigmab}{\xibf_{\bar\ell,\bar\sigma}}
\newcommand{\xibfsigmao}{\xibf_{\bar\ell,\sigma_1}}
\newcommand{\xibfsigmat}{\xibf_{\bar\ell,\sigma_2}}
\newcommand{\xibfpm}{\xibf_{\bar\ell,\pm}}
\newcommand{\xibfmp}{\xibf_{\bar\ell,\mp}}

\newcommand{\xiplusj}{\xi_{\bar\ell,+,j}}
\newcommand{\xiplusl}{\xi_{\bar\ell,+,\ell}}

\newcommand{\ximinusj}{\xi_{\bar\ell,-,j}}
\newcommand{\ximinusl}{\xi_{\bar\ell,-,\ell}}
\newcommand{\xibfplus}{\xibf_{\bar\ell,+}}
\newcommand{\xibfminus}{\xibf_{\bar\ell,-}}

\newcommand{\zetabfplus}{\zetabf_{\bar\ell,+}}

\newcommand{\Omegaell}{\boldsymbol{\Omega}_{\xcr}^{\ell}}
\newcommand{\Piell}{\boldsymbol{\Pi}^{\ell}}
\newcommand{\Sbell}{\mathbf{S}^{\ell}}
\newcommand{\Nbell}{\mathbf{N}^{\ell}}

\newcommand{\Lbell}{\mathbf{L}^{\ell}}

\makeatletter
\newlength{\negph@wd}
\DeclareRobustCommand{\negphantom}[1]{%
  \ifmmode
    \mathpalette\negph@math{#1}%
  \else
    \negph@do{#1}%
  \fi
}
\newcommand{\negph@math}[2]{\negph@do{$\m@th#1#2$}}
\newcommand{\negph@do}[1]{%
  \settowidth{\negph@wd}{#1}%
  \hspace*{-\negph@wd}%
}
\makeatother
\makeatletter
\newlength{\negphhf@wd}
\DeclareRobustCommand{\negphantomhalf}[1]{%
  \ifmmode
    \mathpalette\negphhf@math{#1}%
  \else
    \negphhf@do{#1}%
  \fi
}
\newcommand{\negphhf@math}[2]{\negphhf@do{$\m@th#1#2$}}
\newcommand{\negphhf@do}[1]{%
  \settowidth{\negphhf@wd}{#1}%
  \hspace*{-0.5\negphhf@wd}%
}
\makeatother

\RequirePackage{latexsym}
\begin{document}

\title{On the Angular Momentum and Spin of Generalized Electromagnetic Field for $r$-Vectors in $(k,n)$ Space-Time Dimensions
\thanks{This work has been funded in part by the Spanish Ministry of Science, Innovation and Universities under grants TEC2016-78434-C3-1-R, BES-2017-081360 and PID2020-116683GB-C22. Alfonso Martinez is a Serra-H\'{u}nter Fellow. Published in: \textbf{\textit{Eur. Phys. J. Plus} 136, 1047 (2021). DOI:} \href{https://doi.org/10.1140/epjp/s13360-021-02023-5}{10.1140/epjp/s13360-021-02023-5}.}}

\author{\scshape Alfonso Martinez\thanks{alfonso.martinez@ieee.org}, Ivano Colombaro\thanks{ivano.colombaro@upf.edu}, Josep Font-Segura\thanks{josep.font@ieee.org}
\thanks{Authors are with the Department of Information and Communication Technologies, Universitat Pompeu Fabra, Barcelona, Spain.}
}                

\maketitle

\begin{abstract}
This paper studies the relativistic angular momentum for the generalized electromagnetic field, described by $r$-vectors in $(k,n)$ space-time dimensions, with exterior-algebraic methods. First, the angular-momentum tensor is derived from the invariance of the Lagrangian to space-time rotations (Lorentz transformations), avoiding the explicit need of the canonical tensor in Noether's theorem. The derivation proves the conservation law of angular momentum for generic values of $r$, $k$, and $n$. Second, an integral expression for the flux of the tensor across a $(k+n-1)$-dimensional surface of constant $\ell$-th space-time coordinate is provided in terms of the normal modes of the field; this analysis is a natural generalization of the standard analysis of electromagnetism, i.\,e.~a three-dimensional space integral at constant time. Third, a brief discussion on the orbital angular momentum and the spin of the generalized electromagnetic field, including their expression in complex-valued circular polarizations, is provided for generic values of $r$, $k$, and $n$.  
\end{abstract}

\keywords{Angular Momentum, Spin, Electromagnetism. Maxwell Equations, Exterior Algebra, Exterior Calculus, Tensor Calculus }
%\PACS{
%      {03.50.-z}{Classical field theories}   \and
%      {03.50.De}{Classical electromagnetism, Maxwell equations}
%      \and
%      {03.50.Kk}{Other special classical field theories}
%      \and
%      {11.10.Ef}{Lagrangian and Hamiltonian approach}
%      \and
%      {11.10.Kk}{Field theories in dimensions other than four}
%      \and
%      {02.30.Em}{Potential theory}
%     } % end of PACS codes
%end of abstract
%
%\maketitle
%
%\newcommand{\fieldA}{\abf}
%\newcommand{\fieldApert}{\abf_\varepsilon}
%\newcommand{\fieldB}{\bbf}
%\newcommand{\fieldAi}{a}
%\newcommand{\fieldBi}{b}
%\newcommand{\Mat}{\mathbf{A}}
%\newcommand{\Mati}{A}

\section{Introduction: Preliminaries, Notation, and Main Results}

\subsection{Generalized Maxwell Equations}
 
For a given natural number $r$, the generalized Maxwell field $\mf(\xb)$ and source density $\sd(\xb)$ are  characterized by multivector fields of respective grades $r$ and $r-1$ at every point $\xb$ of a flat $(k,n)$-space-time with $k$ temporal and $n$ spatial dimensions~\cite[Sec.~3]{colombaro2020generalizedMaxwellEquations}.
%The Maxwell field $\mf$ and the source density $\sd$ have respectively $\binom{k+n}{r}$ and $\binom{k+n}{r-1}$ components at each space-time point;
%When convenient, we may drop the explicit dependence on $\xb$ and write $\mf$ and $\sd$. 
For any $0 \leq s \leq k+n$, grade-$s$ multivectors belong to a vector space with basis elements $\ebf_I$, where $I$ is an ordered list of $s$ non-repeated space-time indices; we represent space-time indices by Latin letters. We denote by $\mathcal{I}_s$ the set of all such ordered lists of $s$ space-time indices; we let $I_0 = \O$ and we write $\mathcal I$ for $\mathcal I_1$. Let $\Delta_{II} = \ebf_I\cdot\ebf_I$ for $I\in\mathcal{I}_s$ be the space-time metric, where $\cdot$ denotes the dot product~\cite[Eqs.~(12)--(13)]{colombaro2020generalizedMaxwellEquations}.
The temporal (resp.~spatial) basis elements are $\ebf_0$ to $\ebf_{k-1}$ (resp.~$\ebf_k$ to $\ebf_{k+n-1}$) and have metric $-1$ (resp.~$+1$).
The generalized Maxwell equations for arbitrary $r$, $k$, and $n$ are the following pair of coupled differential equations:
\begin{gather} 
\deltabf\lintprod \mf = \Jb, \label{eq:maxwell-Gen1}\\
\deltabf\wedge \mf = 0, \label{eq:maxwell-Gen2}
\end{gather}
in units such that $c = 1$. The interior derivative (or divergence), expressed with the left interior product ($\lintprod$) in~\eqref{eq:maxwell-Gen1}, and the exterior derivative, expressed in terms of the wedge product ($\wedge$) in~\eqref{eq:maxwell-Gen2}, are both defined in~\cite[Sec.~2]{colombaro2020generalizedMaxwellEquations} or~\cite[Sec.~2]{colombaro2019introductionSpaceTimeExteriorCalculus} and the operator $\deltabf$ is given by $\smash{\deltabf = \sum_{i\in\mathcal{I}}\Delta_{ii}\partial_i}$. For $r = 2$, $k = 1$, and $n = 3$, Eqs~\eqref{eq:maxwell-Gen1}--\eqref{eq:maxwell-Gen2} coincide with the standard Maxwell equations, with the identification of $\mf$ as the (antisymmetric) Faraday tensor of the electromagnetic field, in contravariant form and $\deltabf$ the four-gradient \cite[Ch.~4]{landau1982classicalTheoryFields}, \cite[Ch.~11]{jackson}.

The Maxwell equations can be derived by an application of the principle of stationary action \cite[Ch.\ 19]{feynman1977lecturesPhysicsvol2}, \cite[Sec.\ 8]{landau1982classicalTheoryFields}. For a field theory, the action is a quantity given by the integral over a $(k+n)$-dimensional space-time of a scalar Lagrangian density $\ld(\xb)$. For generalized electromagnetism, the basic field in this formulation is taken to be the vector potential $\vp(\xb)$, a multivector field of grade $r-1$, such that %with $\binom{k+n}{r-1}$ components, such that 
\begin{equation}	\label{eq:field-potential}
	\mf = \deltabf\wedge\vp.
\end{equation}
The Lagrangian density $\ld$ is expressed in terms of the multivector dot (scalar) product \cite[Sec.~2]{colombaro2020generalizedMaxwellEquations} as  the sum of two terms: a free-field density, $\ld_\text{em} = \frac{(-1)^{r-1}}{2}\mf\cdot\mf$, and an interaction term, $	\ld_\text{int} = \sd\cdot\vp$, that is 
\begin{align}\label{eq:lagrangian_em}
	\ld &= \frac{(-1)^{r-1}}{2}\mf\cdot\mf + \vp\cdot\sd.
\end{align}
The Euler-Lagrange equations %\cite[Sec.\ 1.3]{zee2003quantumFieldTheoryNutshell}, \cite[Sec.\ 3.1]{maggiore2005modernIntroductionQFT}, \cite[Sec.\ 7.2]{weinberg1995quantumTheoryFieldsvol1}, 
%\cite{}
for the Lagrangian density $\ld$ in \eqref{eq:lagrangian_em} give indeed the Maxwell equation~\eqref{eq:maxwell-Gen1} as vector derivatives of $\ld$ with respect to the potential $\vp$ and its exterior derivative $\deltabf\lintprod\vp$, namely \cite[Sec.~3.2]{colombaro2021eulerLagrange}
\begin{equation}
\partial_\vp \ld = (-1)^{r-1}\deltabf \lintprod \bigl( \partial_{\deltabf \wedge \vp} \ld \bigr). %+ (-1)^{r} \deltabf \wedge \bigl( \partial_{\deltabf \lintprod \vp} \ld \bigr). \label{eq:eleqs_exterior_intro}
\end{equation}

If we replace the potential $\vp$ by a new field $\vp' = \vp + \bar\vp + \deltabf\wedge\gf$, where $\bar\vp$ is a constant $(r-1)$-vector and $\gf$ is an $(r-2)$-vector gauge field, the homogenous Maxwell equation~\eqref{eq:maxwell-Gen2} is unchanged~\cite[Sec.~3]{colombaro2020generalizedMaxwellEquations}. For a given Maxwell field, there is therefore some unavoidable (gauge) ambiguity on the value of the vector potential if $r \geq 2$. Of special interest for this work are the Coulomb-$\ell$ gauge and the Lorenz gauge. For a space-time index $\ell$, let us define the differential operator $\smash{\deltabf_{\bar\ell} = \sum_{i\in\mathcal{I}}\Delta_{ii}\partial_i}$. In the Coulomb-$\ell$-gauge, the following two conditions are imposed:
\begin{gather}\label{eq:coul-l-gauge-1}
	\ebf_\ell\lintprod\vp = 0, \\
	\deltabf_{\bar\ell}\lintprod\vp = 0. \label{eq:coul-l-gauge-2}
\end{gather}
In classical electromagnetism, setting $\ell = 0$ recovers the Coulomb or radiation gauge. In the Coulomb-$\ell$ gauge, it also holds that $\deltabf\lintprod\vp = 0$. In the less restrictive Lorenz gauge, it simply holds that
\begin{equation}\label{eq:lorenz_gauge}
	\deltabf\lintprod\vp = 0.
\end{equation}
The multivectorial equation in~\eqref{eq:lorenz_gauge} has $\smash{\binom{k+n}{r-2}}$ components, i.\,e.~a scalar equation for $r = 2$. 

\subsection{Energy-Momentum Tensor and Lorentz Force}

Energy-momentum can be transferred from the field to the source through a process modelled as a force acting on the source.
The generalized Lorentz force density $\fb$ is a grade-1 vector with $k+n$ components given by~\cite[Sec.~4]{colombaro2020generalizedMaxwellEquations} 
\begin{equation}\label{eq:lorentz}
	\fb = \Jb\lintprod \Fb = (\deltabf\lintprod \mf)\lintprod \mf.
\end{equation}
The volume integral of the Lorentz force density $\fb$ over a $(k+n)$-dimensional hypervolume $\mathcal{V}^{k+n}$ quantifies the transfer of energy-momentum to the source in that volume. The conservation law relating the Lorentz force~\eqref{eq:lorentz} and the stress-energy-momentum  tensor $\emset$ of the free Maxwell field $\mf$ is given by~\cite[Sec.~4]{colombaro2020generalizedMaxwellEquations}, \cite[Sec.~4.3]{martinez2021exteriorAlgebraicStressEnergyMomentumTensor},
\begin{equation}\label{eq:conservation-EM}
	\fb + \deltabf\lintprod \emset = 0,
\end{equation}	
where $\emset$ is a symmetric rank-2 tensor for all values of $r$, $k$, and $n$. In analogy to the (antisymmetric) multivector basis elements $\ebf_{I}$, we denote the rank-$s$ symmetric-tensor basis elements by $\ubf_I$, where $I\in\mathcal{J}_s$ is an ordered list of $s$, possibly repeated, space-time indices and $\mathcal{J}_s$ denotes the set of all such lists. The interior derivative (divergence) $\deltabf\lintprod \emset$ is computed according to the interior product \cite[Eq.~(25)]{martinez2021exteriorAlgebraicStressEnergyMomentumTensor}, and indeed satisfies~\eqref{eq:conservation-EM}, cf.~\cite[Eq.~(40)]{martinez2021exteriorAlgebraicStressEnergyMomentumTensor}. 

The tensor $\emset$ is expressed in terms of the $\odot$ and $\owedge$ tensor products \cite[Sec.~2.4]{martinez2021exteriorAlgebraicStressEnergyMomentumTensor}. Given two multivectors $\abf$ and $\bbf$ of the same grade $s$, the  $\abf\odot\bbf$ and $\abf\owedge\bbf$ are two rank-2 tensors \cite[Sec.~2.4]{martinez2021exteriorAlgebraicStressEnergyMomentumTensor} with basis elements $\wbf_{ij} = \ebf_i\otimes\ebf_j$ and respective $(i,j)$-th components given by
\begin{gather}
 \abf\odot\bbf\bigr|_{ij} = (\Delta_{ii}\ebf_{i}\lintprod \abf)\cdot (\bbf\rintprod \Delta_{jj}\ebf_{j}), \label{eq:odot-ij} \\
 \abf\owedge\bbf\bigr|_{ij} = (\Delta_{ii}\ebf_{i}\wedge\abf)\cdot(\bbf \wedge \Delta_{jj}\ebf_{j}), \label{eq:owedge-ij}
\end{gather}
where $\Delta_{ii}$ and $\Delta_{jj}$ are the space-time metric defined previously. In general, neither $\abf\odot\bbf$ nor $\abf\owedge\bbf$ are symmetric; however, the sum $\abf\odot\bbf + \abf\owedge\bbf$ is symmetric in its components \cite[Sec.~2.4]{martinez2021exteriorAlgebraicStressEnergyMomentumTensor}. For all values of $r$, $k$, and $n$, the tensor $\emset$ is expressed in terms of the $\odot$ and $\owedge$ tensor products \cite[Sec.~4.2]{colombaro2020generalizedMaxwellEquations}, \cite[Sec.~4.3]{martinez2021exteriorAlgebraicStressEnergyMomentumTensor}, as
\begin{equation}\label{eq:totalT}
\emset = -\frac{1}{2}(\mf\odot\mf+\mf\owedge\mf).
\end{equation}
%The $(i,j)$-th component of the tensor $\emset$ is thus given by
%\begin{align}
% \emseti_{ij} &= -\frac{1}{2}\bigl((\Delta_{ii}\ebf_{i}\lintprod \mf)\cdot (\mf\rintprod \Delta_{jj}\ebf_{j}) + (\Delta_{ii}\ebf_{i}\wedge\mf)\cdot(\mf \wedge \Delta_{jj}\ebf_{j})\bigr).
%\end{align}
The diagonal, $\emseti_{ii}$, and off-diagonal, $\emseti_{ij}$ with $i < j$, components of $\emset$ are explicitly given by~\cite[Eqs~(38)--(39)]{martinez2021exteriorAlgebraicStressEnergyMomentumTensor}
\begin{gather}
	\emseti_{ii} = \frac{(-1)^{r-1}}{2}\Delta_{ii}\Biggl(\sum_{L\in{\mathcal{I}_{r}}:i\notin L}\negphantom{\scriptscriptstyle L}\Delta_{LL}\mfi_{L}^2-\negphantom{\scriptscriptstyle L}\sum_{L\in{\mathcal{I}_{r}}:i\in L}\negphantom{\scriptscriptstyle L}\Delta_{LL}\mfi_{L}^2\Biggr), \label{eq:emset_ii} \\
	\emseti_{ij} = -\negphantom{\scriptscriptstyle L\in}\sum_{L\in{\mathcal{I}_{r-1}}:i,j\notin L}\negphantom{\scriptscriptstyle \notin L}\Delta_{LL}\sigma(L,i)\sigma(j,L)\mfi_{\varepsilon(i,L)}\mfi_{\varepsilon(j,L)}, \label{eq:emset_ij}
\end{gather}
where for two disjoint lists $I$ and $J$ of non-repeated space-time indices, $\sigma(I,J)$ is the signature of the permutation that sorts the concatenated list $(I,J)$, and $\varepsilon(I,J)$ is the sorted concatenated list $(I,J)$. If the lists $I$ and $J$ are not disjoint, we adopt the convention that $\sigma(I,J)  = 0$.

For later use, let us define the product $\boxwedge$ between basis elements $\ebf_i$ and $\ubf_{I}$, $I = (i_1,i_2)\in\mathcal{J}_2$ as
\begin{equation}\label{eq:boxwedge}
 \ebf_i \boxwedge \ubf_{I} = \sum_{I^\pi \in I!} \sigma(i,i_2^\pi)\wbf_{i_1^\pi,\varepsilon(i,i_2^\pi)}.
\end{equation}
Here $I!$ denotes the set of all permutations (not necessarily ordered) of $I$, and $I^\pi = (i_1^\pi,i_2^\pi)$ denotes one such permuted list. The condition $i_2^\pi\neq i$ is implicitly enforced by the permutation signature $\sigma(i,i_2^\pi)$.

Both the conservation law \eqref{eq:conservation-EM} and the formula for the symmetric tensor $\emset$ \eqref{eq:totalT} can be derived by exterior-algebraic methods from the invariance of the free-field action with density $\ld_\text{em}$ to infinitesimal space-time translations \cite{martinez2021exteriorAlgebraicStressEnergyMomentumTensor}. This exterior-algebraic derivation directly gives a symmetric tensor, without recurring to the Belinfante-Rosenfeld procedure to symmetrize the canonical tensor that appears in a standard application of Noether's theorem to the invariance of the action \cite{Forger_2004,Blaschke_2016}, \cite[Sec.\ 3.2]{maggiore2005modernIntroductionQFT}, \cite[Sec.\ 2.5]{diFrancesco1997conformalFieldTheory}. In Sec.~\ref{sec:angular-momentum-law} of this paper, we show how a formula for the relativistic angular-momentum tensor can be derived by exterior-algebraic methods from the invariance of the action for the free field with density $\ld_\text{em}$ to infinitesimal space-time rotations. %This direct derivation again avoids the non-symmetric canonical tensor and bypasses the explicit need of Noether's theorem.

Generalizing the usual  electromagnetic analysis of flux as a three-dimensional space integral at constant time, the energy-momentum flux $\smash{\Piell}$ across the $(k+n)$-dimensional half space-time $\mathcal{V}_\ell^{k+n}$ of  fixed $\ell$-th space-time coordinate $x_\ell$, for $\ell \in \{0,\dotsc,k+n-1\}$, can be expressed in terms of the transverse normal modes of the field \cite[Eq.~(86)]{colombaro2020generalizedMaxwellEquations} as a multidimensional integral over $\mathbf\Xi_\ell$, the set of values  of $\xibf_{\bar\ell}$ for which $\Delta_{\ell\ell}  \xibf_{\bar\ell}\cdot\xibf_{\bar\ell} \leq 0$,  where $\xibf_{\bar\ell} = \xibf - \xi_\ell\ebf_\ell$, namely
\begin{equation}\label{eq:Pi_ell} 
	\Piell = 4\pi^2(-1)^r\sigma(\ell,\ell^c)\int_{\mathbf\Xi_\ell}\negphantom{\scriptscriptstyle \ell}
\frac{\drm\xi_{\ell^c}}{2\chi_\ell}  \, 
\xibfplus \lvert\hat{\Ab}(\xibfplus)\rvert^2,
\end{equation}
where $\drm\xi_{\ell^c}$ is an infinitesimal element \cite[Sec.~3.1]{colombaro2019introductionSpaceTimeExteriorCalculus} along all coordinates except the $\ell$-th, the frequency $\chi_\ell$ is given by $\chi_\ell = +\sqrt{-\Delta_{\ell\ell}  \xibf_{\bar\ell}\cdot\xibf_{\bar\ell}}$, and $\xibfplus = \xibf_{\bar\ell} + \chi_\ell\ebf_\ell$; the complex-valued normal field components are denoted by $\smash{\hat{\Ab}(\xibfplus)}$.
In Sec.~\ref{sec:flux-angular-momentum-tensor} of this paper, we provide an analogous formula for the angular-momentum flux and its split into center-of-motion, orbital angular momentum, and spin components, as described in the next section.

\subsection{Relativistic Angular Momentum: Background and Summary of Main Results}

In classical mechanics, the angular momentum $\Lb$ is an axial vector (or pseudovector) with three spatial components. The relativistic angular momentum $\Omegabf$ is an antisymmetric tensor of rank 2, or a bivector, that combines the angular momentum $\Lb$ and the polar vector $\Nb$ for the velocity of the center-of-mass (also known as moment of energy). In fact, the way $\Omegabf$ is constructed is the same as the way the electromagnetic field bivector $\mf$ is constructed from the axial magnetic field and the polar electric field, that is $\Omegabf = \ebf_0\wedge \Nb + \Lb^{\mathcal{H}}$ \cite[Sec.~3.1]{colombaro2020generalizedMaxwellEquations}, where $\Lb^{\mathcal{H}}$ is the spatial Hodge dual of $\Lb$ \cite[Eq.~(18)]{colombaro2020generalizedMaxwellEquations}, i.\,e.~the bivector corresponding to the axial vector. In $(k,n)$-space-time, relativistic angular momentum $\Omegabf$ is a grade-2 multivector with $\smash{\binom{k+n}{2}}$ components. 

In analogy to energy-momentum, a conservation law relates the transfer of angular momentum over a $(k+n)$-dimensional hypervolume $\mathcal{V}^{k+n}$ to the divergence of an angular-momentum tensor $\Mcb_{\xcr}$ with rotation center $\xcr$. In contrast to $\emset$, the basis elements of $\Mcb_{\xcr}$ are of the form $\wbf_{i,I} = \ebf_{i}\otimes\ebf_{I}$, where $i \in \mathcal{I}$ and $I \in \mathcal{I}_2$. For classical electromagnetism, with $r = 2$, $k = 1$, and $n = 3$, this tensor is given in contravariant form as \cite[Sec.~12.10.B]{jackson}
\begin{equation}
	\Mc_{\xcr}^{\alpha\beta\gamma} = T^{\alpha\beta}(x^{\gamma}-\alpha^{\gamma}) -	T^{\alpha\gamma}(x^{\beta}-\alpha^{\beta}),
\end{equation}
where $T^{\alpha\beta}$ are the components of the symmetric stress-energy-momentum tensor. In our notation, $T^{\alpha\beta} = \emseti_{\varepsilon(\alpha,\beta)}$.
The vectors $\Lb$ and $\Nb$ are given by volume integrals of some appropriate functions of $\Mcb_{\xcr}$. For instance, for $\xcr = 0$, the spatial angular momentum  vector $\Lb$ of the electromagnetic field is given \cite[Prob.~7.27]{jackson} in terms of the standard cross product of the spatial position vector $\xb$ and electric and magnetic fields $\Eb$ and $\Bb$ by:
\begin{align}\label{eq:Lb}
\Lb = \int_{\mathbf R^{3}}\negphantom{\scriptscriptstyle 3}\drm x_{123}\, \bigl(\xb \times (\Eb \times \Bb)\bigr).
\end{align}	
Since the spatial relativistic angular momentum bivector is the space-Hodge-dual $\Lb^{\mathcal{H}}$, using~\cite[Eq.~(36)]{colombaro2020generalizedMaxwellEquations} we have
\begin{align}
\Lb^{\mathcal{H}} = \int_{\mathbf R^{3}}\negphantom{\scriptscriptstyle 3}\drm x_{123}\, \bigl(\xb \wedge (\Eb \times \Bb)\bigr).
\end{align}	
Moreover, the $j$-th component of the Poynting vector $\Eb\times\Bb$ coincides with $\emseti_{ij}$ in~\eqref{eq:emset_ij}, with $i = 0$,
\begin{align}
	\emseti_{0j} %&= \sum_{m\in{\mathcal{I}}:0,j\neq m}\sigma(j,m)\mfi_{\varepsilon(0,m)}\mfi_{\varepsilon(j,m)} \\
	&= \negphantom{\scriptscriptstyle m}\sum_{m\in{\mathcal{I}}:m\neq 0,j}\negphantom{\scriptscriptstyle ,j}\sigma(j,m)\mfi_{\varepsilon(0,m)}\mfi_{\varepsilon(j,m)} \\
	&= (\Eb \times \Bb)\bigl|_{j},
\end{align}
where we have used that $r = 2$ to rewrite $L$ as $m \in \mathcal{I}$, that $\Delta_{mm} = 1$ for the spatial indices, and that $\sigma(m,0) = -1$ for any spatial $m$, as well as the definition of the cross-product $\Eb \times \Bb$. The $(i,j)$-th component of $\Lb^{\mathcal{H}}$ in~\eqref{eq:Lb} is thus given by the volume integral of the quantity
\begin{equation}
	x_i\emseti_{0j}\sigma(i,j) + x_j\emseti_{0i}\sigma(j,i),
\end{equation}
which in turn can be identified with the component in $\wbf_{0,ij}$ of the product $\xb\boxwedge\emset$ defined in~\eqref{eq:boxwedge}. In Sec.~\ref{sec:angular-momentum-law}, we prove that this is no coincidence, and that in general it holds that 
\begin{equation}
	\Mcb_{\xcr} = (\xb-\xcr)\boxwedge\emset.
\end{equation}	
The proof is built on the principle	of invariance of the action to infinitesimal space-time rotations around $\xcr$. 

In Sec.~\ref{sec:flux-angular-momentum-tensor}, we provide a formula for the relativistic angular momentum $\Omegaell$ of the generalized electromagnetic field, including $\Lb$ and the center-of-mass velocity $\Nb$, for any values of $k$, $n$, and $r$, as the flux of the tensor $\Mcb_{\xcr}$ across a $(k+n-1)$-dimensional surface of constant $\ell$-th space-time coordinate (Eqs~\eqref{eq:flux-T} and~\eqref{eq:flux_Til}), for any $\ell$,
\begin{align}\label{eq:flux-T-int}
	\Omegaell &= \int_{\partial\mathcal{V}^{k+n}}\negphantom{\scriptscriptstyle n}\drm^{k+n-1}\xb^{\hodgeinv}\times \Mcb_{\xcr} \\
	&= \sigma(\ell,\ell^c)\sum_{i,j\in\mathcal{I}}\sigma(i,j)\ebf_{\varepsilon(i,j)}\int_{\mathbf R^{k+n-1}}\negphantom{\scriptscriptstyle -1}\drm x_{\ell^c} (x_i-\alpha_i)T_{\varepsilon(\ell,j)},  \label{eq:flux_Til-int}  
\end{align}	
where the flux integral is carried out with respect to the inverse Hodge of the infinitesimal element $\drm\xb$~\cite[Eq.~(19)]{colombaro2019introductionSpaceTimeExteriorCalculus}.
The total angular momentum $\Omegaell$ can be decomposed as $\Omegaell = \Nbell + \Lbell + \Sbell -\xcr\wedge\Piell$, i.\,e.~the center-of-mass component $\Nbell$, the orbital angular momentum $\Lbell$, and the spin $\Sbell$. 
In terms of the transverse normal modes of the field, evaluated in the Coulomb-$\ell$ gauge, these three terms are, respectively, expressed (cf.~Eqs~\eqref{eq:Nbell}--\eqref{eq:Sbell}), as
\begin{gather} \label{eq:Nbell-intro}
	\Nbell = x_\ell\wedge\Piell + j\pi(-1)^{r}\sigma(\ell,\ell^c) 
	\int_{\mathbf\Xi_\ell}\negphantom{\scriptscriptstyle \ell} \frac{\drm\xi_{\ell^c}}{2\chi_\ell}  \, 
	\chi_\ell\ebf_\ell\wedge \Bigl( \bigl(\deltabf_{\xibf_{\bar\ell}}\otimes\hat\vp^*(\xibfplus)\bigr)\times \hat{\Ab}(\xibfplus)  - \text{cc}  \Bigr), \\
	\Lbell = j\pi(-1)^{r}\sigma(\ell,\ell^c) \label{eq:Lbell-intro}
	\int_{\mathbf\Xi_\ell}\negphantom{\scriptscriptstyle \ell} \frac{\drm\xi_{\ell^c}}{2\chi_\ell}  \, 
	\xibf_{\bar\ell}\wedge \Bigl( \bigl(\deltabf_{\xibf_{\bar\ell}}\otimes\hat\vp^*(\xibfplus)\bigr)\times \hat{\Ab}(\xibfplus)  - \text{cc}  \Bigr), \\
	\Sbell = -j 2\pi\sigma(\ell,\ell^c)
	\int_{\mathbf\Xi_\ell}\negphantom{\scriptscriptstyle \ell} \frac{\drm\xi_{\ell^c}}{2\chi_\ell}  \, 
	\Bigl(\hat{\Ab}^*(\xibfplus) \odot \hat{\Ab}(\xibfplus)  - \text{cc}  \Bigr), \label{eq:Sbell-intro}
\end{gather}
where cc stands for the complex conjugate.
Expressions for the bivector components of $\Lbell$ and $\Sbell$ are given in~\eqref{eq:L_I_ell} and~\eqref{eq:S_I_ell}. Of special interest are the circular-polarization-basis formulas for the orbital angular momentum and the spin, respectively given in~\eqref{eq:L_I_ell_jones} and~\eqref{eq:S_I_ell_jones}. 
For the standard electromagnetic field, the spatial components of the orbital angular momentum and spin in~\eqref{eq:Lbell-intro}--\eqref{eq:Sbell-intro}, computed for $\ell = 0$, $r = 2$, $k = 1$, and $n = 3$, coincide with the well-known values \cite[Eq.~(16) in B$_\text{I}$.2]{cohen-tannoudji1997photonsAtoms}, respectively given in vector notation, rather than as a bivector, by
\begin{gather} 
	\Lb = -j\pi \label{eq:Lbell-em}
	\int_{\mathbf{R}^3}\negphantom{\scriptscriptstyle 3} \frac{\drm\xi_{123}}{2\chi_0}  \, 
	\sum_{m=1}^n\xibf_{\bar\ell}\times \Bigl( \bigl(\deltabf_{\xibf_{\bar\ell}}\hat\vpi_m(\xibfplus)\bigr) \hat{\vpi}_m^*(\xibfplus)  - \text{cc}  \Bigr), \\
	\Sb = -j 2\pi 
	\int_{\mathbf{R}^3}\negphantom{\scriptscriptstyle 3} \frac{\drm\xi_{123}}{2\chi_0}  \, 
	\Bigl(\hat{\Ab}^*(\xibfplus) \times \hat{\Ab}(\xibfplus)  - \text{cc}  \Bigr). \label{eq:Sbell-em}
\end{gather}
By construction, the components of the angular momentum and spin bivectors that include the index $\ell$ are zero. 

The feasibility of the separation of angular momentum into orbital and spin parts in a gauge-invariant manner, as well as its possible operational meaning, have been subject to some discussion, particularly in a quantum context \cite{darwin-1932,vanEnk1994spinOrbital,Leader-2018}. Since the consideration of quantum aspects is beyond the scope of this work, and it seems unlikely that statements about the generalized electromagnetic field can be supported by experimental observations to settle the issue, we do not dwell on this matter in this paper, apart from noting that we carry out our analysis in the Coulomb-$\ell$ gauge (or equivalently for the transverse normal modes of the field \cite[Sec.~B$_\text{I}$]{cohen-tannoudji1997photonsAtoms}), the condition that has been found to be in best empirical agreement with observations for the standard electromagnetic field \cite{Leader-2018}.

\section{Angular-Momentum Conservation Law for the Free Generalized Electromagnetic Field}
\label{sec:angular-momentum-law}

In this section, we exploit the invariance of the action with Lagrangian density $\ld_\text{em}$ to infinitesimal space-time rotations, e.\,g.~Lorentz transformations, to 
derive a conservation law and an expression for the relativistic angular-momentum tensor by direct exterior-algebraic methods, avoiding the non-symmetric canonical tensor and the related currents in Noether's theorem. For the sake of notational compactness, we remove the subscript $\text{em}$ in the tensor.

\subsection{Conservation Law for Angular Momentum}
Let us shift the origin of coordinates by an infinitesimal perturbation $\xbpert$.
For a translation, each of the $k+n$ components is an independent function of space-time $\xbpert_\text{t}$. For a space-time rotation (Lorentz transformation) around a center point $\xcr$, and given an infinitesimal bivector $\xbpert_\text{r}$ with $\binom{k+n}{2}$ components, it holds that 
\begin{equation}\label{eq:rotations-st}
	\xbpert = \xbpert_\text{r} \rintprod (\xb - \xcr).
\end{equation} 
%[Does it hold both for boosts (space-time index pair) as well as pure rotations (space-time index pair)?]
Let $\smash{\{\ebf'\}}$ denote the rotated (perturbed) basis elements, expressed in the original basis $\smash{\{\ebf\}}$. %In general, and with some abuse of notation, we denote the components of a multivector $\fieldA$ by $\fieldA$ and $\fieldA'$ in the original and new coordinates respectively
Along the $i$-th coordinate, the basis element $\ebf_i$ is perturbed to first order by an infinitesimal amount
\begin{equation}
	\ebf_i' = \ebf_i\times(\mathbf{1}+\deltabf\otimes\xbpert), \label{eq:trans_cov}
\end{equation}
where $\mathbf{1} = \sum_{i\in\mathcal{I}}\Delta_{ii}\wbf_{ii}$ is the identity matrix and
 the Jacobian partial-derivative matrix $\deltabf\otimes\xbpert$ is given by
\begin{align}
	\deltabf\otimes\xbpert = \sum_{i,j\in\mathcal I}\Delta_{ii}\partial_i\xpert[j]\wbf_{ij}.
\end{align}
The $j$-th column of the Jacobian matrix contains the exterior derivative, i.\,e.~gradient, of the $j$-th component of the perturbation in the coordinates, $\xpert[j]$. 
As proved in \cite[Sec.~3.3]{martinez2021exteriorAlgebraicStressEnergyMomentumTensor}, a similar general expression holds for the transformation of multivector basis elements of grade $s$,
\begin{equation}
	\ebf_I' = \ebf_I\times(\mathbf{1}_s+\trMs), \label{eq:trans_cov_s}
\end{equation}
where $\mathbf{1}_s = \sum_{I\in\mathcal{I}_s}\Delta_{II}\wbf_{I,I}$ is the grade-$s$ identity matrix and 
the matrix $\trMs$ is given by \cite[Eq.~(70)]{martinez2021exteriorAlgebraicStressEnergyMomentumTensor}
\begin{align}
	\trMs &= (-1)^{s-1}\sum_{I\in\mathcal{I}_s}\sum_{i\in I} \sum_{j\in\mathcal{I}\setminus\{I\setminus i\}}\negphantom{\scriptscriptstyle \}}\Delta_{II} \sigma(I\setminus i,i)\sigma(j,I\setminus i)\,\partial_i\xpert[j]\, \wbf_{I,\varepsilon(j,I\setminus i)}.
\end{align}

Writing the action functional over a closed region $\mathcal{R}$ in the new perturbed coordinates involves changing the integrand and the differentials according to \eqref{eq:trans_cov} and \eqref{eq:trans_cov_s}. For the Lagrangian density $\ld_\text{em}$, given by a scalar product of two multivectors, the full details are given in \cite[Sec.~3.4--3.5]{martinez2021exteriorAlgebraicStressEnergyMomentumTensor}. Let us assume that the fields vanish at infinity sufficiently fast, e.\,g.~the integral of $\xbpert\lintprod\Tb = (\xbpert_\text{r}\rintprod(\xb - \xcr))\lintprod\Tb$ at infinity (the boundary of the volume in the action) vanishes. Then, the change of action $\delta \act_{\ld_\text{em}}$ is expressed in terms of the rank-2 manifestly symmetric tensor $\Tb$, the stress-energy-momentum tensor~\eqref{eq:totalT} of the free generalized electromagnetic field, as
\begin{align}\label{eq:change_action_a}
	\delta \act_{\ld_\text{em}} &= \int_\mathcal{R}\! \drm^{k+n}\xb\,(\deltabf\otimes\xbpert )\cdot \Tb \\
	&= -\int_\mathcal{R}\! \drm^{k+n}\xb\,(\deltabf\lintprod\Tb)\cdot\xbpert,  \label{eq:change_action}
\end{align}
having assumed that the integration region $\mathcal{R}$ is large enough to make the physical system closed, and that the fields decay fast enough over $\mathcal{R}$ so that the flux of the fields over the boundary of $\mathcal{R}$ is negligible.
This formula for the change of action~\eqref{eq:change_action} holds for arbitrary grades of the generalized electromagnetic field $\mf$.

The integrand in \eqref{eq:change_action} can be rewritten using \eqref{eq:rotations-st} and \cite[Eq.~(27)]{colombaro2020generalizedMaxwellEquations} as
\begin{align}\label{eq:19}
	(\deltabf\lintprod\Tb)\cdot\bigl(\xbpert_\text{r} \rintprod (\xb - \xcr)\bigr) = \bigl((\xb - \xcr)\wedge(\deltabf\lintprod\Tb)\bigr)\cdot \xbpert_\text{r}.  
\end{align}
Assuming that infinitesimal space-time rotations are a symmetry of the system and that the fields decay sufficiently fast, the fact that the variation of the action $\delta \act_{\ld_\text{em}}$ must be zero for all infinitesimal perturbations $\xbpert_\text{r}$ implies that 
\begin{equation}\label{eq:xwedgedivT}
	(\xb - \xcr)\wedge(\deltabf\lintprod\Tb) = 0.
\end{equation}
This expression characterizes the conservation law related to angular momentum, in the absence of external currents.
Differently from the condition $\deltabf\lintprod\Tb = 0$ that appears in the context of invariance to translations and gives a the conservation law for the energy-momentum, invariance to infinitesimal rotations requires the interior derivative (divergence) of the stress-energy-tensor to be radial, or equivalently parallel to the relative-position vector $\xb - \xcr$. 

In the following section, we provide an expression for a rank-3 angular-momentum tensor, valid for any number of space-time dimensions and grade of the electromagnetic field.

\subsection{Relativistic Angular-Momentum Tensor}

In this section, we prove that~\eqref{eq:xwedgedivT} can be expressed as the matrix derivative (divergence) of a rank-3 tensor, which we will identify with the relativistic angular-momentum tensor of the generalized electromagnetic field. 

To start, we expand the bivector equation~\eqref{eq:xwedgedivT} in components as
\begin{align}
	(\xb - \xcr)\wedge(\deltabf\lintprod\Tb) &= \sum_{i\in\mathcal{I}}(x_i-\alpha_i)\ebf_i\wedge\Biggl(\sum_{j,\ell\in\mathcal{I}} \partial_j T_{\varepsilon(j,\ell)}\ebf_\ell\Biggr) \\
	&= \sum_{i,j,\ell\in\mathcal{I}} \sigma(i,\ell)(x_i-\alpha_i)\partial_j T_{\varepsilon(j,\ell)}\ebf_{\varepsilon(i,\ell)}. \label{eq:22}
\end{align}
Consider now a bivector of a similar form, where $(x_i-\alpha_i)$ and $T_{\varepsilon(j,\ell)}$ are swapped, i.\,e.~$(x_i-\alpha_i)\partial_j T_{\varepsilon(j,\ell)}$ is replaced by $T_{\varepsilon(j,\ell)}\partial_j (x_i-\alpha_i)$. Since $\partial_j (x_i-\alpha_i) = \delta_{ji}$, this bivector can be evaluated as the zero bivector,
\begin{align}
		\sum_{i,j,\ell\in\mathcal{I}} \sigma(i,\ell)T_{\varepsilon(j,\ell)}\partial_j (x_i-\alpha_i)\ebf_{\varepsilon(i,\ell)} &= \sum_{i,j,\ell\in\mathcal{I}} \sigma(i,\ell)T_{\varepsilon(j,\ell)}\delta_{ji}\ebf_{\varepsilon(i,\ell)} \\
		&= \sum_{i,\ell\in\mathcal{I}} \sigma(i,\ell)T_{\varepsilon(i,\ell)}\ebf_{\varepsilon(i,\ell)} \\
		&= \negphantom{\scriptscriptstyle i}\sum_{i,\ell\in\mathcal{I}:i<\ell}\negphantom{\scriptscriptstyle \ell} \bigl(\sigma(i,\ell)+\sigma(\ell,i)\bigr)T_{\varepsilon(i,\ell)}\ebf_{\varepsilon(i,\ell)} \label{eq:25},% \\
%		&= 0, \label{eq:26}
\end{align}
where we have used that $\sigma(i,i) = 0$ to keep only the terms with $i\neq \ell$ and then split the summation into the disjoint cases $i < \ell$ and $\ell < i$ and interchanged the roles of $i$ and $\ell$ in the latter case. Since $\sigma(i,\ell) = -\sigma(\ell,i)$, we verify that Eq.~\eqref{eq:25} is zero. Adding this zero bivector to \eqref{eq:22} and applying the Leibniz rule for the derivative gives
\begin{align}
	(\xb-\xcr)\wedge(\deltabf\lintprod\Tb) &= \sum_{i,j,\ell\in\mathcal{I}} \sigma(i,\ell)\bigl((x_i-\alpha_i)\partial_j T_{\varepsilon(j,\ell)}+T_{\varepsilon(j,\ell)}\partial_j (x_i-\alpha_i)\bigr)\ebf_{\varepsilon(i,\ell)} \\
	 &= \sum_{i,j,\ell\in\mathcal{I}} \sigma(i,\ell)\partial_j \bigl((x_i-\alpha_i)T_{\varepsilon(j,\ell)}\bigr)\ebf_{\varepsilon(i,\ell)}. \label{eq:27}
\end{align}

It remains to prove that \eqref{eq:27} is the divergence of a suitably defined tensor field. Let $\Mcb_{\xcr} = (\xb - \xcr) \boxwedge \Tb$ be the angular-momentum tensor field, where the product $\boxwedge$ is defined in~\eqref{eq:boxwedge}. The tensor field $\Mcb_{\xcr}$ is antisymmetric in the second and third components, as its basis elements are given by $\wbf_{i,I} = \ebf_{i} \otimes \ebf_{I}$.
Expanding the product $(\xb - \xcr) \boxwedge \Tb$ with the definition in~\eqref{eq:boxwedge}, the tensor field $\Mcb_{\xcr}$ is given by
\begin{align}
	\Mcb_{\xcr} &= \sum_{i\in\mathcal{I}}\sum_{I\in\mathcal{J}_2} (x_i-\alpha_i)T_{I}\ebf_i\boxwedge \ubf_I \\
	&= \sum_{i\in\mathcal{I}}\sum_{I\in\mathcal{J}_2} (x_i-\alpha_i)T_{I}\Biggl(\sum_{I^\pi \in I!} \sigma(i,i_2^\pi)\wbf_{i_1^\pi,\varepsilon(i,i_2^\pi)}\Biggr) \\
	&= \sum_{i,j\in\mathcal{I}} (x_i-\alpha_i)T_{jj} \sigma(i,j)\wbf_{j,\varepsilon(i,j)} + \negphantom{\scriptscriptstyle i,}\sum_{i,j,\ell\in\mathcal{I}:j<\ell}\negphantom{\scriptscriptstyle <\ell} (x_i-\alpha_i)T_{j\ell}\bigl( \sigma(i,\ell)\wbf_{j,\varepsilon(i,\ell)} + \sigma(i,j)\wbf_{\ell,\varepsilon(i,j)}\bigr),
\end{align}
where we have split the summation over lists $I\in\mathcal{J}_2$ into two, the first one for the lists $I$ of the form $(j,j)$ and the second one for the lists of the form $(j,\ell)$, with $j < \ell$. Splitting further the second summation into two, and renaming $j$ and $\ell$ as $\ell$ and $j$, respectively, we obtain
\begin{align}
	\Mcb_{\xcr} &= \sum_{i,j\in\mathcal{I}} (x_i-\alpha_i)T_{jj} \sigma(i,j)\wbf_{j,\varepsilon(i,j)} + \negphantom{\scriptscriptstyle i,}\sum_{i,j,\ell\in\mathcal{I}:j<\ell}\negphantom{\scriptscriptstyle <\ell} (x_i-\alpha_i)T_{j\ell} \sigma(i,\ell)\wbf_{j,\varepsilon(i,\ell)} + \negphantom{\scriptscriptstyle i,}\sum_{i,j,\ell\in\mathcal{I}:j>\ell}\negphantom{\scriptscriptstyle >\ell} (x_i-\alpha_i)T_{\ell j}\sigma(i,\ell)\wbf_{j,\varepsilon(i,\ell)}\\
	&= \sum_{i,j\in\mathcal{I}} (x_i-\alpha_i)T_{jj} \sigma(i,j)\wbf_{j,\varepsilon(i,j)} + \negphantom{\scriptscriptstyle i,}\sum_{i,j,\ell\in\mathcal{I}:j\neq\ell}\negphantom{\scriptscriptstyle \neq\ell} (x_i-\alpha_i)T_{\varepsilon(j,\ell)} \sigma(i,\ell)\wbf_{j,\varepsilon(i,\ell)} \label{eq:33} \\
	&= \sum_{i,j,\ell\in\mathcal{I}} (x_i-\alpha_i)T_{\varepsilon(j,\ell)} \sigma(i,\ell)\wbf_{j,\varepsilon(i,\ell)}, \label{eq:34}
\end{align}
where we have combined in \eqref{eq:33} the separate summations over $j < \ell$ and $j > \ell$ into one single summation over $j \neq \ell$, and then in \eqref{eq:34} combined this result with the first summand, expressed as a double summation over $j$ and $\ell$ such that $j = \ell$, into a triple summation over indices $i$, $j$, and $\ell$.

Computing the matrix derivative \cite[Eq.~(34)]{martinez2021exteriorAlgebraicStressEnergyMomentumTensor} of $\Mcb_{\xcr}$, denoted by $\deltabf \times \Mcb_{\xcr}$, we recover \eqref{eq:27}, that is
\begin{align}
	\deltabf \times \Mcb_{\xcr} &= \sum_{i,j,\ell\in\mathcal{I}} \partial_j ((x_i-\alpha_i)T_{\varepsilon(j,\ell)}) \sigma(i,\ell)\ebf_{\varepsilon(i,\ell)} \\
	&= (\xb - \xcr)\wedge(\deltabf\lintprod\Tb).
\end{align}
Substituting this expression in \eqref{eq:19} and the result back in \eqref{eq:change_action}, we find that the change of action is given by
\begin{align}\label{eq:change_action_ang}
	\delta \act_{\ld_\text{em}} &= -\int_\mathcal{R}\! \drm^{k+n}\xb\,\bigl(\deltabf \times \Mcb_{\xcr}\bigr)\cdot \xbpert_\text{r}.  
\end{align}
The invariance of the action to rotations, $\smash{\delta \act_{\ld_\text{em}} = 0}$, implies~\eqref{eq:xwedgedivT} and equivalently that $\deltabf \times \Mcb_{\xcr} = 0$. In the presence of sources, the divergence $\deltabf \times \Mcb_{\xcr}$ can be seen as an angular-momentum density, and the volume integral of $\deltabf \times \Mcb_{\xcr}$ across an $(k+n)$-dimensional hypervolume $\mathcal{V}^{k+n}$ gives the transfer of relativistic angular momentum from the field to the sources in the volume. In the next section, we characterize this transfer of angular momentum in terms of the flux of $\Mcb_{\xcr}$, and provide an expression for the flux in terms of the normal modes of the field.

\section{Flux of the Angular-Momentum Tensor: Spin and Orbital Angular Momentum of the Generalized Electromagnetic Field}
\label{sec:flux-angular-momentum-tensor}

\subsection{Integral Form of the Conservation Law and Angular-Momentum Flux}
The angular-momentum conservation law admits an integral form, which we derive next. First, the volume integral of the divergence $\deltabf \times \Mcb_{\xcr}$ over an $(k+n)$-dimensional hypervolume $\mathcal{V}^{k+n}$ gives the transfer of angular momentum from the field to the sources. This volume integral is the flux of the divergence over $\mathcal{V}^{k+n}$ \cite[Eq.~(40)]{colombaro2020generalizedMaxwellEquations}, 
\begin{equation}
	\int_{\mathcal{V}^{k+n}}\negphantom{\scriptscriptstyle n}\drm x_{0,\dotsm,{k+n-1}}\,(\deltabf \times \Mcb_{\xcr}) = \int_{\mathcal{V}^{k+n}}\negphantom{\scriptscriptstyle n}\drm^{k+n}\xb^{\hodgeinv}\lintprod(\deltabf \times \Mcb_{\xcr}),
\end{equation}
where the flux integral is carried out with respect to the inverse Hodge~\cite[Eq.~(10)]{colombaro2019introductionSpaceTimeExteriorCalculus} of the infinitesimal element $\drm^{k+n}\xb$, i.\,e.$\drm x_{0,\dotsm,{k+n-1}}$.
A short adaptation, included in Appendix~\ref{app:stokes_theorem}, of the analysis in~\cite[Sec.~3.5]{colombaro2019introductionSpaceTimeExteriorCalculus}proves a Stokes theorem for the angular-momentum tensor: 
the flux of $\Mcb_{\xcr}$ across the boundary $\partial\mathcal{V}^{m}$ of an $m$-dimensional hypersurface $\mathcal{V}^{m}$ is equal to the flux of the divergence of $\Mcb_{\xcr}$ across $\mathcal{V}^{m}$ for any $m \leq k+n$. For $m = k+n$, this Stokes theorem gives
\begin{equation}\label{eq:flux-T}
	\int_{\mathcal{V}^{k+n}}\negphantom{\scriptscriptstyle n}\drm^{k+n}\xb^{\hodgeinv}\lintprod(\deltabf \times \Mcb_{\xcr}) = \int_{\partial\mathcal{V}^{k+n}}\negphantom{\scriptscriptstyle n}\drm^{k+n-1}\xb^{\hodgeinv}\times \Mcb_{\xcr}.
\end{equation}

As an example, and for some fixed $x_\ell$ and $\ell \in\mathcal{I}$, consider the $(k+n)$-dimensional half space-time region
\begin{equation} \label{eq:integr-region}
	\mathcal{V}_\ell^{k+n} = (-\infty,\infty)\times(-\infty,\infty)\dotsm \times (-\infty,x_\ell)\times\dotsm(-\infty,\infty).
\end{equation}
The boundary of this region is a surface of constant space-time coordinate $\ell$ of value $x_\ell$, given by
\begin{equation} \label{eq:integr-region-boundary}
	\partial\mathcal{V}_\ell^{k+n} = (-\infty,\infty)\times(-\infty,\infty)\dotsm \times \{x_\ell\}\times\dotsm(-\infty,\infty).
\end{equation}
Let $\Omegaell$ denote the flux of the tensor field $\Mcb_{\xcr} = (\xb-\xcr)\boxwedge\Tb$ across the boundary $\partial \mathcal{V}_\ell^{k+n}$.
In this case, the Hodge-dual infinitesimal vector element in the r.\,h.\,s.~of~\eqref{eq:flux-T} is given by \cite[Eq.~(83)]{colombaro2020generalizedMaxwellEquations}
\begin{equation}\label{eq:61}
	\drm^{k+n-1}\xb^\hodgeinv = \drm x_{\ell^c}\sigma(\ell,\ell^c)\Delta_{\ell\ell}\ebf_{\ell},
\end{equation}	
where the factor $\sigma(\ell,\ell^c)$ arises from the orientation such that the normal vector $\ebf_\ell$ points outside the integration region. Using~\eqref{eq:34} in~\eqref{eq:flux-T} and using~\eqref{eq:61}, carrying out the matrix product, and rearranging the expression, yields
\begin{align}
	\Omegaell %&= \int_{\partial\mathcal{V}_\ell^{k+n}}\drm^{k+n-1}\xb^\hodgeinv \times \Mcb_{\xcr} \\ 
	&= \int_{\mathbf R^{k+n-1}}\negphantom{\scriptscriptstyle -1} \drm x_{\ell^c}\sigma(\ell,\ell^c)\Delta_{\ell\ell}\ebf_{\ell} \times \Biggl(\sum_{i,m,j\in\mathcal{I}} (x_i-\alpha_i)T_{\varepsilon(m,j)} \sigma(i,j)\wbf_{m,\varepsilon(i,j)}\Biggr) \label{eq:flux_T} \\
	&= \sigma(\ell,\ell^c)\sum_{i,j\in\mathcal{I}}\sigma(i,j)\ebf_{\varepsilon(i,j)}\int_{\mathbf R^{k+n-1}}\negphantom{\scriptscriptstyle -1}\drm x_{\ell^c} (x_i-\alpha_i)T_{\varepsilon(\ell,j)}.  \label{eq:flux_Til}  
\end{align}

An alternative, slightly more explicit, expression for \eqref{eq:flux_Til} is the following
\begin{align}
	\Omegaell 
	&= \sigma(\ell,\ell^c)\sum_{(i,j)\in\mathcal{I}_2}\int_{\mathbf R^{k+n-1}}\negphantom{\scriptscriptstyle -1}\drm x_{\ell^c} \bigl( \ebf_{ij}(x_{i}-\alpha_{i})T_{\varepsilon(\ell,j)} + \ebf_{ji}(x_{j}-\alpha_{j})T_{\varepsilon(\ell,i)} \bigr).  \label{eq:flux_Til_alt}  
\end{align}

\subsection{Normal Modes of the Field}

Substituting in~\eqref{eq:flux_T} the stress-energy-momentum tensor $\Tb$ by its expression in \eqref{eq:totalT}, the flux $\Omegaell$ of the angular-momentum tensor a surface of constant space-time coordinate $\ell$ of value $x_\ell$ is given by the integral
\begin{align}
\Omegaell &= -\frac 12\Delta_{\ell\ell}\sigma (\ell , \ell^c)\int_{\mathbf R^{k+n-1}}\negphantom{\scriptscriptstyle -1}\drm x_{\ell^c} \ebf_\ell \times\bigl((\xb-\xcr) \boxwedge (\mf\odot\mf+\mf\owedge\mf)\bigr).
\label{eq:proofT3}
\end{align}
The r.\,h.\,s.~of \eqref{eq:proofT3} is computed w.\,r\,.t.~$x_{\ell^c}$, being $\ell^c$ the set of indices excluding $\ell$. We let $\xb_{\bar\ell} = \xb -x_\ell\ebf_\ell$ and similarly $\xibf_{\bar\ell} = \xibf - \xi_\ell\ebf_\ell$ for the frequency vector defined below. We also let $\kappa_\ell = -\frac 12\Delta_{\ell\ell}\sigma (\ell , \ell^c)$. 

In the absence of charges, the free field $\mf$ satisfies the homogeneous wave equation and can be expressed as a linear superposition of complex exponentials $e^{j2\pi\xibf \cdot \xb}$ such that $\xibf\cdot\xibf=0$. Note that here $j = \sqrt{-1}$; the context will make it clear whether $j$ refers to a coordinate label or to the imaginary number. Denoting the coefficient of each complex exponential by $\hat{\mf}$, the Fourier transform of $\mf$, and with the definition $\drm^{k+n} = \drm \xi_0\dotsm \drm\xi_{k+n-1}$, we have
\begin{equation}
\mf (\xb) = \int_{\mathbf R^{k+n}}%\bigintssss_{\mathbf R^{k+n}}
\negphantom{\scriptscriptstyle n}\drm^{k+n}\xibf \, \delta( \xibf\cdot\xibf ) \,
e^{j2\pi\xibf \cdot \xb} \, \hat{\mf} (\xibf).
\label{eq:proofT4}
\end{equation} 
We resolve the Dirac delta by rewriting the condition  $\xibf \cdot\xibf = 0$ in terms of $\xibf_{\bar\ell}$ as $\Delta_{\ell\ell}\xi_\ell^2 + \xibf_{\bar\ell}\cdot\xibf_{\bar\ell} = 0$. 
This equation has real solutions for $\xi_\ell$ only if $\Delta_{\ell\ell}  \xibf_{\bar\ell}\cdot\xibf_{\bar\ell} \leq 0$, namely the two possible values $\xi_\ell  = \pm\chi_\ell$, where $\chi_\ell$ is given by 
\begin{equation}
\chi_\ell = +\sqrt{-\Delta_{\ell\ell}  \xibf_{\bar\ell}\cdot\xibf_{\bar\ell}}.
\label{eq:T_Tij1}
\end{equation}
Let $\mathbf\Xi_\ell$ be the set of values  of $\xibf_{\bar\ell}$ for which $\Delta_{\ell\ell}  \xibf_{\bar\ell}\cdot\xibf_{\bar\ell} \leq 0$.
We define the pair of frequency vectors $\smash{\xibfsigma}$ as 
\begin{gather}
	\xibfsigma = \xibf_{\bar{\ell}} + \sigma\chi_\ell\ebf_\ell. \label{eq:xiplus_minus}
\end{gather}
for $\sigma \in \mathcal{S} = \{+1,-1\}$, respectively, shortened to $+$ and $-$.
Using \cite[p.~184]{Gelfand1966}, we can  write the inverse Fourier transform \eqref{eq:proofT4} w.\,r.\,t.~the integration variables $\xi_{\ell^c}$, now with the appropriate constraints on the integration range so that $\chi_\ell$ exists, in various equivalent forms as
\begin{align}
	\mf (\xb) &= \int_{\mathbf\Xi_\ell}\negphantom{\scriptscriptstyle \ell}
		 \frac{\drm\xi_{\ell^c}}{2\chi_\ell} 
		 \Biggl(\sum_{\sigma \in \mathcal{S}} e^{j2\pi\xibfsigma\cdot \xb} \, \hat{\mf} (\xibfsigma)\Biggr) \\
%		\bigl(e^{j2\pi\xibf_{+}\cdot \xb} \, \hat{\mf} (\xibf_+) + e^{j2\pi\xibf_{-}\cdot \xb} \, \hat{\mf} (\xibf_-)\bigr) \\
%	&= \int_{\Delta_{\ell\ell} \xibf_{\bar\ell} \cdot \xibf_{\bar\ell}\le 0}
%		\drm\xi_{\ell^c} \frac{1}{2\chi_\ell} 
%		\bigl(e^{j2\pi\xibf_{+}\cdot \xb} \, \hat{\mf} (\xibf_+) + e^{-j2\pi\xibf_{+}\cdot \xb} \, \hat{\mf} (-\xibf_+)\bigr) \\
%	&= \int_{\Delta_{\ell\ell} \xibf_{\bar\ell} \cdot \xibf_{\bar\ell}\le 0}
%		\drm\xi_{\ell^c} \frac{1}{2\chi_\ell} 
%		\bigl(e^{j2\pi\xibf_{+}\cdot \xb} \, \hat{\mf} (\xibf_+) + \text{cc} \bigr).
%	\label{eq:proofT4b1}
%\end{align}
%Alternatively, we can  write the Fourier transform \eqref{eq:proofT4} as
%\begin{align}
%	\mf (\xb) 
%	&= \int_{\mathbf\Xi_\ell}\negphantom{\scriptscriptstyle \ell}
%		\frac{\drm\xi_{\ell^c}}{2\chi_\ell} e^{j2\pi\xibf_{\bar\ell}\cdot \xb_{\bar\ell}}
%		\Biggl(\sum_{\sigma \in \mathcal{S}} e^{j2\pi\Delta_{\ell\ell}\sigma\chi_\ell x_\ell} \, \hat{\mf} (\xibfsigma)\Biggr) \\
	&= \int_{\mathbf\Xi_\ell}\negphantom{\scriptscriptstyle \ell}
		\frac{\drm\xi_{\ell^c}}{2\chi_\ell}  e^{j2\pi\xibf_{\bar\ell}\cdot \xb_{\bar\ell}}
		\hat{\mf}^\ell(\xibf_{\bar\ell}),
	\label{eq:proofT4b2}
\end{align}
where we have factored out a common factor $e^{j2\pi\xibf_{\bar\ell}\cdot \xb_{\bar\ell}}$ and defined the function $\hat{\mf}^\ell(\xibf_{\bar\ell})$ as
\begin{equation}
\hat{\mf}^\ell(\xibf_{\bar\ell}) = e^{j2\pi\Delta_{\ell\ell}\chi_\ell x_\ell} \, \hat{\mf} (\xibfplus) + e^{-j2\pi\Delta_{\ell\ell}\chi_\ell x_\ell} \, \hat{\mf} (\xibfminus).
\label{eq:proofT4c}
\end{equation}

We may rewrite the flux~$\Omegaell$ in terms of $\hat{\mf}^\ell$ by substituting~\eqref{eq:proofT4b2} in~\eqref{eq:proofT3}  as
\begin{align}
\Omegaell &= \kappa_\ell \int_{\mathbf R^{k+n-1}} \negphantom{\scriptscriptstyle -1} \drm x_{\ell^c} 
\iint_{\mathbf\Xi_\ell \times \mathbf\Xi_\ell}\negphantom{\scriptscriptstyle \ell}
\frac{\drm\xi_{\ell^c}}{2\chi_\ell} \frac{\drm\xi'_{\ell^c}}{2\chi'_\ell} \, 
e^{j2\pi(\xibf_{\bar\ell}+\xibf_{\bar\ell}')\cdot \xb_{\bar\ell}}
\ebf_\ell \times\Bigl((\xb-\xcr) \boxwedge \bigl(\hat{\mf}^\ell(\xibf_{\bar\ell})\odot\hat{\mf}^\ell(\xibf'_{\bar\ell})+\hat{\mf}^\ell(\xibf_{\bar\ell})\owedge\hat{\mf}^\ell(\xibf'_{\bar\ell})\bigr)\Bigr) \label{eq:proofT5}. % \\
%&= \kappa_\ell \sum_{t\in\mathcal{I}}\underbrace{\int_{\mathbf R^{k+n-1}} \negphantom{\scriptscriptstyle -1} \drm x_{\ell^c} 
%\iint_{\mathbf\Xi_\ell \times \mathbf\Xi_\ell}\negphantom{\scriptscriptstyle \ell}
%\frac{\drm\xi_{\ell^c}}{2\chi_\ell} \frac{\drm\xi'_{\ell^c}}{2\chi'_\ell} \, 
%e^{j2\pi(\xibf_{\bar\ell}+\xibf_{\bar\ell}')\cdot \xb_{\bar\ell}}
%\ebf_\ell \times\Bigl((x_t-\alpha_t) \ebf_t \boxwedge \bigl(\hat{\mf}^\ell(\xibf_{\bar\ell})\odot\hat{\mf}^\ell(\xibf'_{\bar\ell})+\hat{\mf}^\ell(\xibf_{\bar\ell})\owedge\hat{\mf}^\ell(\xibf'_{\bar\ell})\bigr)\Bigr)}_{I_t}, \label{eq:proofT5a}
\end{align}

\subsection{Spin and Angular Momentum of the Generalized Electromagnetic Field}

In Appendix~\ref{app:computationOmegaell} we carry out the rather tedious evaluation of this integral in terms of the transverse normal modes in the Coulomb-$\ell$ gauge. Under the assumption that the various field components commute, we obtain the following formula for the angular momentum as a sum of four components, cf.~Eq.~\eqref{app:B_ellj_t-17}, 
\begin{equation}
	\Omegaell = \Nbell + \Lbell + \Sbell - \xcr\wedge\Piell,
\end{equation}
namely the center-of-mass velocity $\Nbell$, the orbital angular momentum $\Lbell$, and the spin $\Sbell$, 
respectively, given by
\begin{gather} \label{eq:Nbell}
	\Nbell = x_\ell\wedge\Piell + j\pi(-1)^{r}\sigma(\ell,\ell^c) 
	\int_{\mathbf\Xi_\ell}\negphantom{\scriptscriptstyle \ell} \frac{\drm\xi_{\ell^c}}{2\chi_\ell}  \, 
	\chi_\ell\ebf_\ell\wedge \Bigl( \bigl(\deltabf_{\xibf_{\bar\ell}}\otimes\hat\vp^*(\xibfplus)\bigr)\times \hat{\Ab}(\xibfplus)  - \text{cc}  \Bigr), \\
	\Lbell = j\pi(-1)^{r}\sigma(\ell,\ell^c) \label{eq:Lbell}
	\int_{\mathbf\Xi_\ell}\negphantom{\scriptscriptstyle \ell} \frac{\drm\xi_{\ell^c}}{2\chi_\ell}  \, 
	\xibf_{\bar\ell}\wedge \Bigl( \bigl(\deltabf_{\xibf_{\bar\ell}}\otimes\hat\vp^*(\xibfplus)\bigr)\times \hat{\Ab}(\xibfplus)  - \text{cc}  \Bigr), \\
	\Sbell = -j 2\pi\sigma(\ell,\ell^c)
	\int_{\mathbf\Xi_\ell}\negphantom{\scriptscriptstyle \ell} \frac{\drm\xi_{\ell^c}}{2\chi_\ell}  \, 
	\Bigl(\hat{\Ab}^*(\xibfplus) \odot \hat{\Ab}(\xibfplus)  - \text{cc}  \Bigr), \label{eq:Sbell}
\end{gather}
where $\smash{\Piell}$ is the energy-momentum flux across the region in~\eqref{eq:Pi_ell} and contributes to the angular momentum with a term dependent of the origin of coordinates $\xcr$. 
The product $\odot$ could be replaced by $\owedge$ in~\eqref{eq:Sbell} with an overall change of sign, since the off-diagonal transposed components of both products coincide \cite[Eq.~(22)]{colombaro2020generalizedMaxwellEquations}, and the diagonal components vanish in the Coulomb-$\ell$ gauge defined in~\eqref{eq:coul-l-gauge-1}--\eqref{eq:coul-l-gauge-2}.

Using the various product definitions, the $I$-th component, where $I = (i,j) \in\mathcal{I}_2$ and $\ell\notin I$, of the orbital angular momentum and spin are, respectively, given by
\begin{align}\label{eq:L_I_ell}
	L_I^\ell &= j\pi(-1)^{r}\sigma(\ell,\ell^c)
	\int_{\mathbf\Xi_\ell}\negphantom{\scriptscriptstyle \ell} \frac{\drm\xi_{\ell^c}}{2\chi_\ell}  \, 
	\Bigl(\Delta_{jj}\xi_i \bigl(\partial_{\xi_j}^*\hat\vp^*(\xibfplus)\bigr)\cdot \hat{\Ab}(\xibfplus)  - \Delta_{ii}\xi_j \bigl(\partial_{\xi_i}^*\hat\vp^*(\xibfplus)\bigr)\cdot \hat{\Ab}(\xibfplus)  - \text{cc}  \Bigr) \\
	&= j\pi(-1)^{r}\sigma(\ell,\ell^c) \label{eq:Lbell_I}
	\int_{\mathbf\Xi_\ell}\negphantom{\scriptscriptstyle \ell} \frac{\drm\xi_{\ell^c}}{2\chi_\ell}  \, 
	\negphantom{\scriptscriptstyle K}\sum_{K\in\mathcal{I}_{r-1}}\negphantom{\scriptscriptstyle 1}\biggl(\Delta_{KK} \Bigl( \Delta_{jj}\xi_i \bigl(\partial_{\xi_j}\hat\vpi_K^*(\xibfplus)\bigr)\hat{\vpi}_K(\xibfplus) - \Delta_{ii}\xi_j \bigl(\partial_{\xi_i}\hat\vpi_K^*(\xibfplus)\bigr)\hat{\vpi}_K(\xibfplus)   \Bigr) - \text{cc}  \biggr),
\end{align}
and	
\begin{gather}
	S_I^\ell = -j 2\pi\sigma(\ell,\ell^c)
	\int_{\mathbf\Xi_\ell}\negphantom{\scriptscriptstyle \ell} \frac{\drm\xi_{\ell^c}}{2\chi_\ell}
	\Biggl( \sum_{L\in{\mathcal{I}_{r-2}}:i,j\notin L}\negphantom{\scriptscriptstyle L}\Delta_{LL}\sigma(L,i)\sigma(j,L)\hat{\vpi}^*_{\varepsilon(i,L)}(\xibfplus) \hat{\vpi}_{\varepsilon(j,L)}(\xibfplus)  - \text{cc}  \Biggr).\label{eq:Sbell_I}
\end{gather}

By construction, the subspace of vector potential components in~\eqref{eq:Sbell_I} is restricted to those lists disjoint from $I$, with components different from $\ell$ (from the Coulomb-$\ell$ gauge condition in~\eqref{eq:coul-l-gauge-1}), and orthogonal to $\xibfplus$ (from~\eqref{eq:coul-l-gauge-2}). This leaves a total of $k+n-4$ space-time indices, to be distributed in lists of $r-2$ different elements. The dimension of this subspace is thus $\smash{\binom{k+n-4}{r-2}}$. This dimension might be related to the classification of distinct pairs of spin-1 particles linked to the direction $\xibfplus$, a possibility to be studied elsewehere. %, namely one such pair for standard electromagnetism.

The feasibility of the separation of angular momentum into orbital and spin parts in a gauge-invariant manner, as well as its operational meaning, have long been subject to some level of discussion, particularly in a quantum context \cite{darwin-1932,vanEnk1994spinOrbital,Leader-2018,Bialynicki-Birula-2011,vanEnk1994commutationRules,Leader-2014}. As stated earlier in the paper, quantum aspects lie beyond the scope of this work and we do not dwell further on this matter, apart from noting that our analysis is done in the Coulomb-$\ell$ gauge (or equivalently for the transverse normal modes of the field \cite[Sec.~B$_\text{I}$]{cohen-tannoudji1997photonsAtoms}), the condition that has been found to be in best empirical agreement with observations for the standard electromagnetic field \cite{Leader-2018}.

As a complement, we include in Appendix~\ref{app:spin_components} a ``canonical'' derivation of the spin components  extended to the generalized multivectorial electromagnetic field. Ignoring the quantum aspects, we have used as a basis Sections~12 and~16 of Wentzel's treatise on quantum field theory \cite{wentzel1949quantumTheoryFields}, one of the first book treatments of the subject. Our analysis bypasses the canonical tensor that Wentzel makes use of, so the appropriate adaptations have been made. As expected, the final formulas obtained with this extended analysis coincide with~\eqref{eq:Sbell} and~\eqref{eq:Sbell_I}.

% The spatial angular momentum $\Lb$ can also be split into two parts, the orbital angular momentum and the intrinsic angular momentum, in turn associated with the spin of the particle or field. For the electromagnetic field, where particles have no rest frame \ldots

% Connection back to spin and QM / polarizations, number of dimensions

% Spin Jackson Problems 7.27, 7.28, 7.29; Section 10.3

\subsection{Spin and Orbital Angular Momentum in a Complex-valued Circular Polarization Basis}

\newcommand{\ebfLij}{\ebf_{-}^{\scriptscriptstyle{I}}}
\newcommand{\ebfRij}{\ebf_{+}^{\scriptscriptstyle{I}}}
\newcommand{\xifLij}{\xi_{-}^{\scriptscriptstyle{I}}}
\newcommand{\xifRij}{\xi_{+}^{\scriptscriptstyle{I}}}
\newcommand{\deltaLij}{\partial_{\xi_-^{\scriptscriptstyle{I}}}}
\newcommand{\deltaRij}{\partial_{\xi_+^{\scriptscriptstyle{I}}}}
\newcommand{\ebfLijcc}{\ebf_{-}^{\scriptscriptstyle{I}}}
\newcommand{\ebfRijcc}{\ebf_{+}^{\scriptscriptstyle{I}}}
\newcommand{\xifLijcc}{\xi_{-}^{\scriptscriptstyle{I}}}
\newcommand{\xifRijcc}{\xi_{+}^{\scriptscriptstyle{I}}}
\newcommand{\deltaLijcc}{\partial_{\xi_-^{\scriptscriptstyle{I}}}^{*}}
\newcommand{\deltaRijcc}{\partial_{\xi_+^{\scriptscriptstyle{I}}}^{*}}

From the definition of the $\odot$ product in~\eqref{eq:odot-ij}, the $I$-th component $S_I^\ell$ of the spin bivector $\Sbell$ in~\eqref{eq:Sbell} is given by
\begin{equation}\label{eq:S_I_ell}
	S_I^\ell = -j2\pi\sigma(\ell,\ell^c)\int_{\mathbf\Xi_\ell}\negphantom{\scriptscriptstyle \ell} \frac{\drm\xi_{\ell^c}}{2\chi_\ell}  \, 
	\Bigl(\bigl(\Delta_{ii}\ebf_i\lintprod \hat{\Ab}(\xibfplus)\bigr)^* \cdot \bigl(\hat{\Ab}(\xibfplus) \rintprod\ebf_j\Delta_{jj}\bigr) - \text{cc}  \Bigr), 
\end{equation}
where $I = (i,j)$.
The component $S_I^\ell$ adopts a particularly transparent form in the complex-valued circular-polarization basis. 
%The form of this new basis depends on the values of $\Delta_{ii}$ and $\Delta_{jj}$. If  $\Delta_{ii} = \Delta_{jj}$, and 
For any $\varphi$, let the right- and left-handed basis elements, respectively denoted by $\ebfRij$ and $\ebfLij$, be given by
\begin{subequations}\label{eq:jones_vectors}
\begin{gather}\label{eq:jones_vectors-R}
	\ebfRij = \cos\varphi\,\Delta_{ii}\ebf_i -  j \sin\varphi\,\Delta_{jj}\ebf_j, \\
	\ebfLij = -\sin\varphi\,\Delta_{ii}\ebf_i - j \cos\varphi\,\Delta_{jj}\ebf_j.\label{eq:jones_vectors-L}
\end{gather}
\end{subequations}
%These vectors satisfy the orthonormality conditions ${\ebfRij}^*\cdot\ebfRij = {\ebfLij}^*\cdot\ebfLij = \Delta_{ii} = \Delta_{jj}$ and ${\ebfRij}^*\cdot\ebfLij = 0$, as well as the relationships $j{\ebfRij}^*\wedge\ebfRij = -j{\ebfLij}^*\wedge\ebfLij = \sin(2\varphi)\,\ebf_I$. 
Note  that the symbol $j$ is used to represent both the imaginary unit and one of the components of $I$, a possible source of confusion in expressions as~\eqref{eq:jones_vectors} and others below.
These vectors satisfy the orthonormality relations $\smash{{\ebfRij}^*\cdot\ebfRij = \cos^2\varphi\Delta_{ii} + \sin^2\varphi\Delta_{jj}}$, $\smash{{\ebfLij}^*\cdot\ebfLij = \sin^2\varphi\Delta_{ii} + \cos^2\varphi\Delta_{jj}}$ and $\smash{{\ebfRij}^*\cdot\ebfLij = \sin\varphi\cos\varphi(\Delta_{jj}-\Delta_{ii})}$, as well as the relationships ${\ebfRij}\wedge\ebfRij = {\ebfLij}\wedge\ebfLij = 0$ and $j{\ebfRij}\wedge\ebfLij = \Delta_{II}\ebf_I$. 
The transformation in~\eqref{eq:jones_vectors} has determinant $\Delta_{ii}\Delta_{jj}$ and the inverse transformation is given by
%The inverse transformations corresponding to the definitions of $\ebfRij$ and $\ebfRij$ in~\eqref{eq:jones_vectors} are given by
\begin{subequations}\label{eq:jones_vectors_inv}
\begin{gather}\label{eq:jones_vectors_inv-i}
	\Delta_{ii}\ebf_i  = \cos\varphi\,\ebfRij-\sin\varphi\,\ebfLij, \\
	\Delta_{ii}\ebf_j  = j(\sin\varphi\,\ebfRij+\cos\varphi\,\ebfLij).\label{eq:jones_vectors_inv-j}
\end{gather}
\end{subequations}
The basis elements for $\varphi = \frac{\pi}{4}$ appears in the analysis of helicity and circular polarization \cite[Problem~7.27]{jackson}; for $\varphi = 0$, and apart from a factor $-j$, we recover the standard basis, i.\,e.~linear polarization.

When we substitute these expressions for $\ebf_i$ and $\ebf_j$ in~\eqref{eq:S_I_ell} we have to take into account that the complex-conjugate operation acting on the potential also affects the basis elements. For the standard space-time basis, this observation is irrelevant since the basis elements are real-valued. However, the polarization vectors are complex-valued and we need to use $\ebf_i^*$ rather than $\ebf_i$ in~\eqref{eq:jones_vectors_inv-i}. With this observation, the component $\smash{S_I^\ell}$ is given by
\begin{align}
	S_I^\ell &= -j2\pi\sigma(\ell,\ell^c)\int_{\mathbf\Xi_\ell}\negphantom{\scriptscriptstyle \ell} \frac{\drm\xi_{\ell^c}}{2\chi_\ell}  \, 
	\Bigl(j\bigl((\cos\varphi\,\ebfRij-\sin\varphi\,\ebfLij)^*\lintprod \hat{\Ab}^*(\xibfplus)\bigr) \cdot \bigl(\hat{\Ab}(\xibfplus) \rintprod(\sin\varphi\,\ebfRij+\cos\varphi\,\ebfLij)\bigr) - \text{cc}  \Bigr) \\
%	 &= 2\pi\sigma(\ell,\ell^c)\int_{\mathbf\Xi_\ell}\negphantom{\scriptscriptstyle \ell} \frac{\drm\xi_{\ell^c}}{2\chi_\ell}  \, 
%	\Bigl(j\bigl((\ebfRij-\ebfLij)\lintprod \hat{\Ab}^*(\xibfplus)\bigr) \cdot \bigl(\hat{\Ab}(\xibfplus) \rintprod(\ebfRij+\ebfLij)\bigr) - \text{cc}  \Bigr),
%	\\
%	&= -j2\pi\sigma(\ell,\ell^c)\int_{\mathbf\Xi_\ell}\negphantom{\scriptscriptstyle \ell} \frac{\drm\xi_{\ell^c}}{2\chi_\ell}  \, 
%	\Bigl(j\bigl((\cos\varphi\,\ebfRij-\sin\varphi\,\ebfLij)^*\lintprod \hat{\Ab}^*(\xibfplus)\bigr) \cdot \bigl(\hat{\Ab}(\xibfplus) \rintprod(\sin\varphi\,\ebfRij+\cos\varphi\,\ebfLij)\bigr) - \text{cc}  \Bigr) \\
%	&= -j2\pi\sigma(\ell,\ell^c)\int_{\mathbf\Xi_\ell}\negphantom{\scriptscriptstyle \ell} \frac{\drm\xi_{\ell^c}}{2\chi_\ell}  \, 
%	\Bigl(j\bigl((\cos\varphi\,\ebfRij-\sin\varphi\,\ebfLij)^*\lintprod \hat{\Ab}^*(\xibfplus)\bigr) \cdot \bigl(\hat{\Ab}(\xibfplus) \rintprod(\sin\varphi\,\ebfRij+\cos\varphi\,\ebfLij)\bigr) - \text{cc}  \Bigr) \\
	&=-j 2\pi\sigma(\ell,\ell^c)\int_{\mathbf\Xi_\ell}\negphantom{\scriptscriptstyle \ell} \frac{\drm\xi_{\ell^c}}{2\chi_\ell}  \, 
	\Bigl(j\sin(2\varphi)\,\bigl({\ebfRij}^*\lintprod \hat{\Ab}^*(\xibfplus)\bigr) \cdot \bigl(\hat{\Ab}(\xibfplus) \rintprod \ebfRij\bigr) - \bigl({\ebfLij}^*\lintprod \hat{\Ab}^*(\xibfplus)\bigr) \cdot \bigl(\hat{\Ab}(\xibfplus) \rintprod \ebfLij\bigr) + \notag \\
	 &\phantom{=-j 2\pi\sigma(\ell,\ell^c)\int_{\mathbf\Xi_\ell}\negphantom{\scriptscriptstyle \ell} \frac{\drm\xi_{\ell^c}}{2\chi_\ell}  \, 
	\Bigl(j} + j \cos(2\varphi)\,\bigl({\ebfRij}^*\lintprod \hat{\Ab}^*(\xibfplus)\bigr) \cdot \bigl(\hat{\Ab}(\xibfplus) \rintprod \ebfLij\bigr) + \text{cc} \Bigr),\label{eq:S_I_ell_jones_varphi}
\end{align}
where we have grouped common terms under the assumption that the fields $\hat{\Ab}^*(\xibfplus)$ and $\hat{\Ab}(\xibfplus)$ commute, as it corresponds to a classical theory. For the choice $\varphi = \pi/4$, the basis elements satisfy $\smash{{\ebfRij}^*\cdot\ebfRij = {\ebfLij}^*\cdot\ebfLij = \frac{1}{2}(\Delta_{ii} + \Delta_{jj})}$ and $\smash{{\ebfRij}^*\cdot\ebfLij = \frac{1}{2}(\Delta_{jj}-\Delta_{ii})}$, and 
 the components $\smash{S_I^\ell}$ adopt a particularly simple form,
\begin{align}\label{eq:S_I_ell_jones}
	S_I^\ell &= 2\pi\sigma(\ell,\ell^c)\int_{\mathbf\Xi_\ell}\negphantom{\scriptscriptstyle \ell} \frac{\drm\xi_{\ell^c}}{2\chi_\ell}  \, 
	\Bigl(\,\bigl({\ebfRij}^*\lintprod \hat{\Ab}^*(\xibfplus)\bigr) \cdot \bigl(\hat{\Ab}(\xibfplus) \rintprod \ebfRij\bigr) - \bigl({\ebfLij}^*\lintprod \hat{\Ab}^*(\xibfplus)\bigr) \cdot \bigl(\hat{\Ab}(\xibfplus) \rintprod \ebfLij\bigr) \Bigr).
\end{align}
 This formula extends a similar result for the standard electromagnetic field \cite[Problem~7.27]{jackson}, and expresses the spin as the sum of independent right- and left-handed components. For other values of  $\varphi$, the basis components are mixed.

The $I$-th component $L_I^\ell$, with $\ell\notin I$, of the orbital angular momentum bivector $\Lbell$ is given by~\eqref{eq:L_I_ell}.
%For the sake of compactness, we consider only the case $\Delta_{ii} = \Delta_{jj}$. A similar analysis holds for $\Delta_{ii} \neq \Delta_{jj}$, using the appropriate basis transformation.
For the basis change in~\eqref{eq:jones_vectors} with $\varphi = \pi/4$, the frequency vector components transform are expressed as a function of $\xifRij$ and $\xifLij$ in terms of the Hermitian inverse of the transformation matrix, i.\,e.~
\begin{subequations}\label{eq:jones_vectors_inv_comp}
\begin{gather}\label{eq:jones_vectors_inv_comp-i}
	\xi_i  = \frac{1}{\sqrt{2}}(\xifRij-\xifLij), \\
	\xi_j  = \frac{1}{\sqrt{2}}j(\xifRij+\xifLij),\label{eq:jones_vectors_inv_comp-j}
\end{gather}
\end{subequations}
and similarly for $\smash{\partial_{\xi_i}}$ and $\smash{\partial_{\xi_j}}$. Again, the symbol $j$ doubly represents a coordinate label in the left-hand side and the imaginary unit in the right-hand side of~\eqref{eq:jones_vectors_inv_comp-j}.
We therefore can express the orbital angular momentum component $L_I^\ell$ in~\eqref{eq:L_I_ell} in terms of the coefficients in the circular-polarization basis in~\eqref{eq:jones_vectors_inv_comp} as
%Apart from factor $\pi(-1)^{r}\sigma(\ell,\ell^c)\Delta_{ii}\Delta_{jj}$,
\begin{align}\label{eq:L_I_ell_jones}
L_I^\ell &= j\pi(-1)^{r}\sigma(\ell,\ell^c)\Delta_{ii}
	\int_{\mathbf\Xi_\ell}\negphantom{\scriptscriptstyle \ell} \frac{\drm\xi_{\ell^c}}{2\chi_\ell}  \, 
	\frac 12\Bigl(-j(\xifRij-\xifLij) \bigl((\deltaRij+\deltaLij)\hat\vp(\xibfplus)\bigr)^*\cdot \hat{\Ab}(\xibfplus) + \notag \\
	&\phantom{j\pi(-1)^{r}\sigma(\ell,\ell^c)\Delta_{ii}\Delta_{jj}
	\int_{\mathbf\Xi_\ell}\negphantom{\scriptscriptstyle \ell} \frac{\drm\xi_{\ell^c}}{2\chi_\ell}  \, 
	\frac 12\Bigl(-j} 
	- j(\xifRij+\xifLij) \bigl((\deltaRij-\deltaLij)\hat\vp(\xibfplus)\bigr)^*\cdot \hat{\Ab}(\xibfplus)  - \text{cc}  \Bigr) \\
&= 2\pi(-1)^{r}\sigma(\ell,\ell^c)\Delta_{ii}
	\int_{\mathbf\Xi_\ell}\negphantom{\scriptscriptstyle \ell} \frac{\drm\xi_{\ell^c}}{2\chi_\ell}  \, 
	\Re\Bigl(\xifRij\bigl(\deltaRij\hat\vp(\xibfplus)\bigr)^*\cdot \hat{\Ab}(\xibfplus) -  \xifLij \bigl(\deltaLij\hat\vp(\xibfplus)\bigr)^*\cdot \hat{\Ab}(\xibfplus) \Bigr), 
%\\
%&= j
%	\int_{\mathbf\Xi_\ell}\negphantom{\scriptscriptstyle \ell} \frac{\drm\xi_{\ell^c}}{2\chi_\ell}  \, 
%	\Bigl(j\xifRij\bigl(\deltaRijcc\hat\vp^*(\xibfplus)\bigr)\cdot \hat{\Ab}(\xibfplus) - j \xifLij \bigl(\deltaLijcc\hat\vp^*(\xibfplus)\bigr)\cdot \hat{\Ab}(\xibfplus)  - \text{cc}  \Bigr) \\
%&= 
%	\int_{\mathbf\Xi_\ell}\negphantom{\scriptscriptstyle \ell} \frac{\drm\xi_{\ell^c}}{2\chi_\ell}  \, 
%	\Bigl(\xifRij\bigl(\deltaLij\hat\vp^*(\xibfplus)\bigr)\cdot \hat{\Ab}(\xibfplus) - \xifLij \bigl(\deltaRij\hat\vp^*(\xibfplus)\bigr)\cdot \hat{\Ab}(\xibfplus)  + \text{cc}  \Bigr) \\
%&=
%	\int_{\mathbf\Xi_\ell}\negphantom{\scriptscriptstyle \ell} \frac{\drm\xi_{\ell^c}}{2\chi_\ell}  \, 
%	\Bigl(\xifRij \deltaLij\lvert\hat{\Ab}(\xibfplus)\rvert^2 - \xifLij \deltaRij\lvert\hat{\Ab}(\xibfplus)\rvert^2 \Bigr)
\end{align}
a formula reminiscent of that of the spin for the standard electromagnetic field \cite[Problem~7.27]{jackson}. As we have seen throughout the previous pages, a large number of standard results in the analysis of angular momentum for free electromagnetic fields naturally extend to arbitrary number of space-time dimensions and multivector field grade.
This brief discussion on the orbital angular momentum and the spin of the generalized electromagnetic field and their relationship to complex-valued circular polarizations, for generic values of $r$, $k$, and $n$, concludes the paper. 
The remainder is devoted to appendices with details or proofs of several results mentioned earlier in the paper.

\appendix

\renewcommand{\thesection}{\Alph{section}}

\section{Proof of the Stokes Theorem}
\label{app:stokes_theorem}

In this appendix, we prove the following statement: the flux of a tensor field $\Mcb$,  antisymmetric in the second and third components and with basis elements given by $\wbf_{i,I} = \ebf_{i} \otimes \ebf_{I}$, across the boundary $\partial\mathcal{V}^{m}$ of an $m$-dimensional hypersurface $\mathcal{V}^{m}$ is equal to the flux of the divergence of $\Mcb$ across $\mathcal{V}^{m}$ for any $m \leq k+n$, and in particular for $m = k+n$. This Stokes theorem thus gives
\begin{equation}\label{eq:flux-T-app}
	\int_{\mathcal{V}^{k+n}}\negphantom{\scriptscriptstyle +n}\drm^{k+n}\xb^{\hodgeinv}\lintprod(\deltabf \times \Mcb) = \int_{\partial\mathcal{V}^{k+n}}\negphantom{\scriptscriptstyle +n}\drm^{k+n-1}\xb^{\hodgeinv}\times \Mcb.
\end{equation}

We prove~\eqref{eq:flux-T-app} thanks to the generalized Stokes theorem for differential forms~\cite[pp.~80]{Cartan}, 
\begin{equation} \label{eq:stokes-forms}
 \int_{\mathcal{V}}\drm \omega= \int_{\partial \mathcal{V}} \omega	,
\end{equation}
where $\omega$ is a differential form and $\drm \omega$ is its exterior derivative, corresponding to the operator
\begin{equation}\label{eq:ext-der-form}
\drm = \sum_{j \in \mathcal{I}} \drm x_j \partial_j.
\end{equation}
The procedure we follow is almost identical to what was done in~\cite[Sec.~3.4--3.5]{colombaro2019introductionSpaceTimeExteriorCalculus}, and it starts by identifying $\omega$ with the integrand on the right-hand side of~\eqref{eq:flux-T-app}. Using~\eqref{eq:34} with $\Mcb = \Mcb_{\xcr}$, we have
\begin{align}
\omega &= \sum_{m\in\mathcal{I}} \drm x_{{m}^c}\Delta_{mm} \sigma(m,  {m}^c) \ebf_m \times \sum_{i,j,\ell\in\mathcal{I}} (x_i - \alpha_i) T_{\varepsilon(j,\ell)} \sigma(i,\ell) \wbf_{j, \varepsilon(i,\ell)} \\
&= \sum_{i,j,\ell\in\mathcal{I}} \drm x_{{j}^c} \sigma(j,{j}^c) (x_i - \alpha_i) T_{\varepsilon(j,\ell)} \sigma(i,\ell) \ebf_{\varepsilon(i,\ell)}, \label{eq:omegaform}
\end{align}
having applied $\ebf_m \times \wbf_{j, \varepsilon(i,\ell)} = \Delta_{mj} \ebf_{\varepsilon(i,\ell)}$.
We then let the exterior derivative in~\eqref{eq:ext-der-form} act on~\eqref{eq:omegaform} to obtain
\begin{align}
\drm \omega = \sum_{m\in\mathcal{I}} \sum_{i,j,\ell\in\mathcal{I}} \drm x_m \drm x_{{j}^c} \sigma(j,{j}^c)\sigma(i,\ell) \partial_m \bigl( (x_i - \alpha_i) T_{\varepsilon(j,\ell)} \bigr) \ebf_{\varepsilon(i,\ell)}. 
\end{align}
Since ${j}^c \in\mathcal{I}_{k+n-1}$, we can identify $m$ with $j$ and write $\drm x_{\varepsilon(j, {j}^c)} = \drm x_{j} \drm x_{{j}^c} \sigma(j,{j}^c)$ to obtain
\begin{align} \label{eq:domega1}
\drm \omega &= \sum_{i,j,\ell\in\mathcal{I}} \drm x_{\varepsilon(j, {j}^c)} \sigma(i,\ell) \partial_j \bigl( (x_i - \alpha_i) T_{\varepsilon(j,\ell)} \bigr) \ebf_{\varepsilon(i,\ell)} .
\end{align}

In parallel, we identify $\drm \omega$ in~\eqref{eq:stokes-forms} with the integrand of the left-hand side of~\eqref{eq:flux-T-app}, which can be expanded as
\begin{align}
\drm \omega &= \sum_{i,j,\ell\in\mathcal{I}} \drm x_{\varepsilon(j, {j}^c)} \sigma(i,\ell) \partial_j \bigl( (x_i - \alpha_i) T_{\varepsilon(j,\ell)} \bigr) \ebf_{\varepsilon(i,\ell)},
\end{align}
namely the same expression as~\eqref{eq:domega1}, therefore proving~\eqref{eq:flux-T-app}.

\section{Flux of the Angular-Momentum Tensor}
\label{app:flux-T}

\subsection{Computation of the Angular-Momentum Flux}
\label{app:computationOmegaell}

Writing $\xb - \xcr= (x_\ell - \alpha_\ell)\ebf_\ell + \xb_{\bar\ell} - \xcr_{\bar\ell}$, we may split the flux in~\eqref{eq:proofT5} as a weighted sum, namely
\begin{align}
\Omegaell &= \kappa_\ell (x_\ell - \alpha_\ell)  \Icb_\ell + \kappa_\ell  \Icb_{\bar\ell,+} - \kappa_\ell  \Icb_{\bar\ell,-}, \label{eq:proofT5a}
\end{align}
where the bivector-valued integrals $\Icb_{\ell}$, $\Icb_{\bar\ell,+}$, and $\Icb_{\bar\ell,-}$ are, respectively, given by
\begin{gather}
	\Icb_{\ell} = \int_{\mathbf R^{k+n-1}} \negphantom{\scriptscriptstyle -1} \drm x_{\ell^c} 
\iint_{\mathbf\Xi_\ell \times \mathbf\Xi_\ell}\negphantom{\scriptscriptstyle \ell}
\frac{\drm\xi_{\ell^c}}{2\chi_\ell} \frac{\drm\xi'_{\ell^c}}{2\chi'_\ell} \, 
e^{j2\pi(\xibf_{\bar\ell}+\xibf_{\bar\ell}')\cdot \xb_{\bar\ell}}
\ebf_\ell \times\Bigl( \ebf_\ell \boxwedge \bigl(\hat{\mf}^\ell(\xibf_{\bar\ell})\odot\hat{\mf}^\ell(\xibf'_{\bar\ell})+\hat{\mf}^\ell(\xibf_{\bar\ell})\owedge\hat{\mf}^\ell(\xibf'_{\bar\ell})\bigr)\Bigr), \label{eq:proofT5b} \\
	\Icb_{\bar\ell,+} = \int_{\mathbf R^{k+n-1}} \negphantom{\scriptscriptstyle -1} \drm x_{\ell^c} 
\iint_{\mathbf\Xi_\ell \times \mathbf\Xi_\ell}\negphantom{\scriptscriptstyle \ell}
\frac{\drm\xi_{\ell^c}}{2\chi_\ell} \frac{\drm\xi'_{\ell^c}}{2\chi'_\ell} \, 
e^{j2\pi(\xibf_{\bar\ell}+\xibf_{\bar\ell}')\cdot \xb_{\bar\ell}}
\ebf_\ell \times\Bigl(\xb_{\bar\ell} \boxwedge \bigl(\hat{\mf}^\ell(\xibf_{\bar\ell})\odot\hat{\mf}^\ell(\xibf'_{\bar\ell})+\hat{\mf}^\ell(\xibf_{\bar\ell})\owedge\hat{\mf}^\ell(\xibf'_{\bar\ell})\bigr)\Bigr), \label{eq:proofT5c} \\
	\Icb_{\bar\ell,-} = \int_{\mathbf R^{k+n-1}} \negphantom{\scriptscriptstyle -1} \drm x_{\ell^c} 
\iint_{\mathbf\Xi_\ell \times \mathbf\Xi_\ell}\negphantom{\scriptscriptstyle \ell}
\frac{\drm\xi_{\ell^c}}{2\chi_\ell} \frac{\drm\xi'_{\ell^c}}{2\chi'_\ell} \, 
e^{j2\pi(\xibf_{\bar\ell}+\xibf_{\bar\ell}')\cdot \xb_{\bar\ell}}
\ebf_\ell \times\Bigl( \xcr_{\bar\ell} \boxwedge \bigl(\hat{\mf}^\ell(\xibf_{\bar\ell})\odot\hat{\mf}^\ell(\xibf'_{\bar\ell})+\hat{\mf}^\ell(\xibf_{\bar\ell})\owedge\hat{\mf}^\ell(\xibf'_{\bar\ell})\bigr)\Bigr). \label{eq:proofT5d}
\end{gather}
We next evaluate these integrals, starting with $\Icb_\ell$. Interchanging the integration order of frequency and space-time, we evaluate the integral of $e^{j2\pi(\xibf_{\bar\ell}+\xibf'_{\bar\ell})\cdot \xb_{\bar\ell}}$ over space-time $\mathbf R^{k+n-1}$ as the $(k+n-1)$-multidimensional Dirac delta. After integration over $\xi'_{\ell^c}$ to remove this Dirac delta, we directly obtain
\begin{equation}
	\Icb_\ell = \label{eq:proofIell}
\int_{\mathbf\Xi_\ell}\negphantom{\scriptscriptstyle \ell}
\frac{\drm\xi_{\ell^c}}{4\chi_\ell^2}  \, 
\ebf_\ell \times\Bigl(\ebf_{\ell} \boxwedge \bigl(\hat{\mf}^\ell(\xibf_{\bar\ell})\odot\hat{\mf}^\ell(-\xibf_{\bar\ell})+\hat{\mf}^\ell(\xibf_{\bar\ell})\owedge\hat{\mf}^\ell(-\xibf_{\bar\ell})\bigr)\Bigr).
\end{equation}
It will prove convenient to define an integral $\Icb_{m}$ for $m\in\mathcal{I}$, replacing $\ebf_\ell$ inside the parentheses in~\eqref{eq:proofIell} by $\ebf_m$, 
\begin{equation}
	\Icb_{m} = \label{eq:proofIm}
\int_{\mathbf\Xi_\ell}\negphantom{\scriptscriptstyle \ell}
\frac{\drm\xi_{\ell^c}}{4\chi_\ell^2}  \, 
\ebf_\ell \times\Bigl(\ebf_m  \boxwedge \bigl( \hat{\mf}^\ell(\xibf_{\bar\ell}) \odot\hat{\mf}^\ell(-\xibf_{\bar\ell}) + \hat{\mf}^\ell(\xibf_{\bar\ell})\owedge\hat{\mf}^\ell(-\xibf_{\bar\ell})\bigr)\Bigr).
\end{equation}
The integral $\Icb_{\bar\ell,-}$ in~\eqref{eq:proofT5d} can now be evaluated in a similar way to $\Icb_\ell$ to obtain
\begin{equation}
	\Icb_{\bar\ell,-} = \label{eq:proofIellm}
	\sum_{t\in\mathcal{I}\setminus \ell} \alpha_t \Icb_{t},
\end{equation}
where $\Icb_{t}$ is given by~\eqref{eq:proofIm} setting $m = t$. 

As for the integral $\Icb_{\bar\ell,+}$ in~\eqref{eq:proofT5c}, we first rewrite the formula for $\Icb_{\bar\ell,+}$ by interchanging the integration order of frequency and space-time, using linearity and making some minor rearrangements and algebraic manipulations, as 
\begin{gather}
	\Icb_{\bar\ell,+} = \label{eq:proofIellp}
\iint_{\mathbf\Xi_\ell \times \mathbf\Xi_\ell}\negphantom{\scriptscriptstyle \ell}
\drm\xi_{\ell^c}\drm\xi'_{\ell^c} \, 
\ebf_\ell \times \Biggl(\int_{\mathbf R^{k+n-1}} \negphantom{\scriptscriptstyle -1} \drm x_{\ell^c} e^{j2\pi(\xibf_{\bar\ell}+\xibf_{\bar\ell}')\cdot \xb_{\bar\ell}}
\xb_{\bar\ell} \boxwedge  \biggl(\frac{\hat{\mf}^\ell(\xibf_{\bar\ell})\odot\hat{\mf}^\ell(\xibf'_{\bar\ell})+\hat{\mf}^\ell(\xibf_{\bar\ell})\owedge\hat{\mf}^\ell(\xibf'_{\bar\ell})}{4\chi_\ell\chi_\ell'}\biggr) \Biggr).
\end{gather}
Under the usual assumptions that the fields vanish sufficiently fast at infinity, 
the space-time integral in~\eqref{eq:proofIellp} can be evaluated by integration by parts in terms of a derivative of the Dirac delta as 
\begin{gather}
	\int_{\mathbf R^{k+n-1}} \negphantom{\scriptscriptstyle -1} \drm x_{\ell^c} e^{j2\pi(\xibf_{\bar\ell}+\xibf_{\bar\ell}')\cdot \xb_{\bar\ell}}
\xb_{\bar\ell} = \frac{1}{j2\pi}\deltabf_{\xibf_{\bar\ell}} \bigl(\delta(\xibf_{\bar\ell}+\xibf_{\bar\ell}') \bigr),
\end{gather}
where the vector-derivative operator $\deltabf_{\xibf_{\bar\ell}}$ is given by
\begin{equation}\label{eq:main_75}
	\deltabf_{\xibf_{\bar\ell}} = \sum_{t\in \mathcal I\setminus \ell} \Delta_{tt}\ebf_t \partial_{\xi_{t}}.
\end{equation}
Now, an extension of the proof in~\cite[p.~26]{Gelfand1966} to our multidimensional bivector-valued integrals in~\eqref{eq:proofIellp} shows that the derivative of the Dirac delta can be evaluated as
\begin{align}
	\Icb_{\bar\ell,+} &= 
\frac{1}{j2\pi}\iint_{\mathbf\Xi_\ell \times \mathbf\Xi_\ell}\negphantom{\scriptscriptstyle \ell}
\drm\xi_{\ell^c}\drm\xi'_{\ell^c} \, 
\ebf_\ell \times \Biggl(\deltabf_{\xibf_{\bar\ell}} \bigl(\delta(\xibf_{\bar\ell}+\xibf_{\bar\ell}') \bigr) \boxwedge  \biggl(\frac{\hat{\mf}^\ell(\xibf_{\bar\ell})\odot\hat{\mf}^\ell(\xibf'_{\bar\ell})+\hat{\mf}^\ell(\xibf_{\bar\ell})\owedge\hat{\mf}^\ell(\xibf'_{\bar\ell})}{4\chi_\ell\chi_\ell'}\biggr) \Biggr) \\
	&= \label{eq:proofIellp-2}
-\frac{1}{j2\pi}\int_{\mathbf\Xi_\ell}\negphantom{\scriptscriptstyle \ell}
\drm\xi_{\ell^c} \, 
\ebf_\ell \times \Biggl(\deltabf_{\xibf_{\bar\ell}}  \boxwedge  \biggl(\frac{\hat{\mf}^\ell(\xibf_{\bar\ell})\odot\hat{\mf}^\ell(\xibf'_{\bar\ell})+\hat{\mf}^\ell(\xibf_{\bar\ell})\owedge\hat{\mf}^\ell(\xibf'_{\bar\ell})}{4\chi_\ell\chi_\ell'}\biggr) \Biggr)\Bigg|_{\xibf_{\bar\ell}' = -\xibf_{\bar\ell}}.
\end{align}
Using the definition in~\eqref{eq:main_75}, we 
can express~\eqref{eq:proofIellp-2} as
\begin{align}
	\Icb_{\bar\ell,+} &= \label{eq:proofIellp-3}
-\frac{1}{j2\pi}\sum_{t\in\mathcal{I}\setminus\ell}  \Delta_{tt} \Icb_{t,+},
\end{align}
where the bivector-valued integrals $\Icb_{t,+}$ are given by
\begin{align} \label{eq:proofItp}
	\Icb_{t,+} = \int_{\mathbf\Xi_\ell}\negphantom{\scriptscriptstyle \ell}
\frac{\drm\xi_{\ell^c}}{2\chi_\ell}  \, 
\ebf_\ell \times\Biggl(\ebf_t  \boxwedge \Biggl(\partial_{\xi_{t}}\biggl(\frac{\hat{\mf}^\ell(\xibf_{\bar\ell})}{2\chi_\ell}\biggr)\odot\hat{\mf}^\ell(-\xibf_{\bar\ell}) + \partial_{\xi_{t}}\biggl(\frac{\hat{\mf}^\ell(\xibf_{\bar\ell})}{2\chi_\ell}\biggr)\owedge\hat{\mf}^\ell(-\xibf_{\bar\ell})\Biggr)\Biggr).
\end{align}
Finally, we evaluate the derivative of $\hat{\mf}^\ell(\xibf_{\bar\ell})/(2\chi_\ell)$ as
\begin{align}
	\partial_{\xi_{t}}\biggl(\frac{\hat{\mf}^\ell(\xibf_{\bar\ell})}{2\chi_\ell}\biggr) = \frac{\partial_{\xi_{t}}\hat{\mf}^\ell(\xibf_{\bar\ell})}{2\chi_\ell} + \Delta_{\ell\ell}\Delta_{tt}\xi_{t}\frac{\hat{\mf}^\ell(\xibf_{\bar\ell})}{2\chi_\ell^3}.
\end{align}
Using this expression, we may therefore rewrite~\eqref{eq:proofItp} as
\begin{align}
	\Icb_{t,+} = \label{eq:proofIellp-4}
	\Icb_{t,1} + \Delta_{\ell\ell}\Delta_{tt}\Icb_{t,0},
\end{align}
where $\Icb_{t,1}$ and $\Icb_{t,0}$ are, respectively, given by
\begin{gather}
	\Icb_{t,1} = \label{eq:proofIt1}
\int_{\mathbf\Xi_\ell}\negphantom{\scriptscriptstyle \ell}
\frac{\drm\xi_{\ell^c}}{4\chi_\ell^2}  \, 
\ebf_\ell \times\Bigl(\ebf_t  \boxwedge \Bigl(\bigl(\partial_{\xi_{t}}\hat{\mf}^\ell(\xibf_{\bar\ell})\bigr)\odot\hat{\mf}^\ell(-\xibf_{\bar\ell}) + \bigl(\partial_{\xi_{t}}\hat{\mf}^\ell(\xibf_{\bar\ell})\bigr)\owedge\hat{\mf}^\ell(-\xibf_{\bar\ell})\Bigr)\Bigr) \\
	\Icb_{t,0} = \label{eq:proofIt0}
\int_{\mathbf\Xi_\ell}\negphantom{\scriptscriptstyle \ell}
\frac{\drm\xi_{\ell^c}}{4\chi_\ell^4}  \, 
\xi_t\,\ebf_\ell \times\Bigl(\ebf_{t} \boxwedge \bigl(\hat{\mf}^\ell(\xibf_{\bar\ell})\odot\hat{\mf}^\ell(-\xibf_{\bar\ell})+\hat{\mf}^\ell(\xibf_{\bar\ell})\owedge\hat{\mf}^\ell(-\xibf_{\bar\ell})\bigr)\Bigr).
\end{gather}

Substituting~\eqref{eq:proofIell},~\eqref{eq:proofIellp-3},~\eqref{eq:proofIellp-4}, and~\eqref{eq:proofIellm} back into~\eqref{eq:proofT5a}, we obtain 
\begin{equation}\label{eq:phi_ell_I}
\Omegaell = 
\kappa_\ell x_\ell \Icb_\ell - \frac{\kappa_\ell}{j2\pi}\sum_{t\in\mathcal{I}\setminus\ell} (\Delta_{tt}\Icb_{t,1} + \Delta_{\ell\ell}\Icb_{t,0}) - \kappa_\ell  \sum_{m\in\mathcal{I}} \alpha_m \Icb_{m},
\end{equation}
where $\Icb_m$, for $m \in \mathcal{I}$ is given in~\eqref{eq:proofIm}, and $\Icb_{t,1}$ and $\Icb_{t,0}$ are, respectively, given by~\eqref{eq:proofIt1} and~\eqref{eq:proofIt0}.

The three bivector-valued integrands in~\eqref{eq:proofIm},~\eqref{eq:proofIt1} and~\eqref{eq:proofIt0} are of the form $\smash{\ebf_\ell \times \bigl(\ebf_{m} \boxwedge \mathbf{B}\bigr)}$, for some index $m$ and some symmetric rank-2 tensor $\mathbf{B}$. As for some indices $\smash{I = (i_1,i_2) \in \mathcal{J}_2}$ it holds that $\smash{\ebf_\ell \times \bigl(\ebf_{m} \boxwedge \ubf_I\bigr) = 0}$, only some components of the tensor $\mathbf{B}$ contribute to the integral. To determine which components of $\mathbf{B}$ contribute to the integral, we compute the double product $\smash{\ebf_\ell \times \bigl(\ebf_{m} \boxwedge \mathbf{B}\bigr)}$ with the definition of the product $\boxwedge$ in~\eqref{eq:boxwedge}, 
\begin{align}
	\ebf_\ell \times \bigl(\ebf_{m} \boxwedge \mathbf{B}\bigr) &= \sum_{I\in\mathcal{J}_2} B_I\,\ebf_\ell \times\bigl(\ebf_{m} \boxwedge \ubf_I\bigr) \\
	&= \sum_{I\in\mathcal{J}_2}\sum_{I^\pi \in I!} B_I\sigma(m,i_2^\pi)\,\ebf_\ell \times\bigl(\ebf_{i_1^\pi}\otimes\ebf_{\varepsilon(m,i_2^\pi)}\bigr) \\
	&= \Delta_{\ell\ell}\negphantomhalf{i}\sum_{j\in\mathcal{I}\setminus m}\negphantomhalf{i} B_{\varepsilon(\ell,j)}\sigma(m,j)\,\ebf_{\varepsilon(m,j)}, \label{eq:double_product}
\end{align}
where we have used that $I$ and its permutation $I^\pi$ must be such that $i_1^\pi = \ell$, i.\,e.~that $I$ must be of the form $\varepsilon(\ell,j)$ for some $j\in\mathcal{I}$. This observation fixes also the permutation $I^\pi = (\ell,j)$. Besides, we can remove $j = m$ from the summation as $\sigma(m,m) = 0$. For each $m$, we need thus consider only the components $B_{\varepsilon(\ell,j)}$, where $j \neq m$.

\textit{Computation of $\Icb_m$.}
In~\eqref{eq:proofIell}, the tensor $\mathbf{B}$ mentioned in the previous paragraph is given by
\begin{equation}
\mathbf{B} = \hat{\mf}^\ell(\xibf_{\bar\ell})\odot\hat{\mf}^\ell(-\xibf_{\bar\ell})+\hat{\mf}^\ell(\xibf_{\bar\ell})\owedge\hat{\mf}^\ell(-\xibf_{\bar\ell}).
\label{eq:B_for_Iell}
\end{equation}
In Section~\ref{app:B_ellj_ell} we evaluate its components $B_{\varepsilon(\ell,j)}$ needed in~\eqref{eq:double_product}. Substituting~\eqref{eq:proof-B-ell-j-final} in~\eqref{eq:double_product} gives
\begin{equation}\label{eq:B-ell-j-final-double_product}
	\ebf_\ell \times \bigl(\ebf_{m} \boxwedge \mathbf{B}\bigr) = 2\beta_r\Delta_{\ell\ell}\chi_{\ell}\negphantomhalf{\ell}\sum_{j\in\mathcal{I}\setminus m} \sigma(m,j)\,\ebf_{\varepsilon(m,j)}\sum_{\sigma \in \mathcal{S}}
\sigma \,\xisigmaj \lvert\hat{\Ab}(\xibfsigma)\rvert^2,
\end{equation}
where $\beta_r$ is given by
\begin{equation}
	\beta_r = 4\pi^2(-1)^{r-1},
\end{equation}
and with some abuse of notation, $\sigma$ denotes in this equation both the signature of a permutation and a sign.
Substituting~\eqref{eq:B-ell-j-final-double_product} back in~\eqref{eq:proofIell} gives
\begin{gather}
	\Icb_m = \label{eq:proofIell-2}
\beta_r\Delta_{\ell\ell}\negphantomhalf{\ell}\sum_{j\in\mathcal{I}\setminus m}\negphantomhalf{\ell}\sigma(m,j)\,\ebf_{\varepsilon(m,j)}\int_{\mathbf\Xi_\ell}\negphantom{\scriptscriptstyle \ell}
\frac{\drm\xi_{\ell^c}}{2\chi_\ell}  \, 
\bigl(\xiplusj \lvert\hat{\Ab}(\xibfplus)\rvert^2 - \ximinusj \lvert\hat{\Ab}(\xibfminus)\rvert^2\bigr).
\end{gather}

Assume that $j \neq \ell$, so that $\xisigmaj = \xi_j$, regardless of the value of $\sigma$. Then, 
splitting the integral in two, and making a change of variables $\zetabf_{\bar\ell} = -\xibf_{\bar\ell}$ in the integral with $\xibfminus$ yields
\begin{align}
-\int_{\mathbf\Xi_\ell}\negphantom{\scriptscriptstyle \ell}\frac{\drm\xi_{\ell^c}}{2\chi_\ell}  \, \xi_j \lvert\hat{\Ab}(\xibfminus)\rvert^2 
&= 
-\int_{\mathbf\Xi_\ell}\negphantom{\scriptscriptstyle \ell}\frac{\drm\zeta_{\ell^c}}{2\chi_\ell}  \, (-\zeta_{j}) \lvert\hat{\Ab}(-\zetabf_{\bar\ell} - \chi_\ell\ebf_\ell)\rvert^2 \\
&= 
\int_{\mathbf\Xi_\ell}\negphantom{\scriptscriptstyle \ell}\frac{\drm\zeta_{\ell^c}}{2\chi_\ell}  \, \zeta_{j} \lvert\hat{\Ab}(\zetabfplus)\rvert^2,\label{eq:proof-Iell-2-3}
\end{align}
since $-\zetabf_{\bar\ell} - \chi_\ell\ebf_\ell = - \zetabfplus$, and $\lvert\hat{\Ab}(\zetabfplus)\rvert^2 = \hat{\Ab}(\zetabfplus)\hat{\Ab}^*(\zetabfplus) = \hat{\Ab}^*(-\zetabfplus)\hat{\Ab}(-\zetabfplus) = \lvert\hat{\Ab}(-\zetabfplus)\rvert^2$ thanks to the hermiticity of $\hat{\Ab}(\zetabfplus)$. The second summand in the integral in~\eqref{eq:proofIell-2} coincides with the first. 

If $j = \ell$, then $\xisigmaj = \sigma\chi_\ell$, and the integral in~\eqref{eq:proofIell-2} is given by
\begin{gather}\label{eq:proofIell-2b}
\int_{\mathbf\Xi_\ell}\negphantom{\scriptscriptstyle \ell}
\frac{\drm\xi_{\ell^c}}{2\chi_\ell}  \, 
\bigl(\chi_\ell \lvert\hat{\Ab}(\xibfplus)\rvert^2 + \chi_\ell \lvert\hat{\Ab}(\xibfminus)\rvert^2\bigr).
\end{gather}
Splitting the integral in two, and making a change of variables $\zetabf_{\bar\ell} = -\xibf_{\bar\ell}$ in the integral with $\xibfminus$ shows that the second integral in~\eqref{eq:proofIell-2b} coincides with the first one, as it happened in~\eqref{eq:proof-Iell-2-3}.

Substituting~\eqref{eq:proof-Iell-2-3} and~\eqref{eq:proofIell-2b} back in~\eqref{eq:proofIell-2} gives the final expression for $\Icb_m$, namely
\begin{align}
	\Icb_m &= \label{eq:proofIell-3}
2\beta_r\Delta_{\ell\ell}\negphantomhalf{\ell}\sum_{j\in\mathcal{I}\setminus m}\negphantomhalf{\ell}\sigma(m,j)\,\ebf_{\varepsilon(m,j)}\int_{\mathbf\Xi_\ell}\negphantom{\scriptscriptstyle \ell}
\frac{\drm\xi_{\ell^c}}{2\chi_\ell}  \, 
\xiplusj \lvert\hat{\Ab}(\xibfplus)\rvert^2 \\
	&= \label{eq:proofIell-3b}
2\beta_r\Delta_{\ell\ell}\ebf_m\wedge\int_{\mathbf\Xi_\ell}\negphantom{\scriptscriptstyle \ell}
\frac{\drm\xi_{\ell^c}}{2\chi_\ell}  \, 
\xibfplus \lvert\hat{\Ab}(\xibfplus)\rvert^2,
\end{align}
where we have used that $\ebf_{\varepsilon(m,j)} = \sigma(m,j)\ebf_m\wedge\ebf_j$, that $\ebf_m \wedge \ebf_m = 0$, and the decomposition $\xibfplus = \xibf_{\bar\ell} + \chi_\ell\ebf_\ell$.

With the definition $\kappa_\ell = -\frac 12\Delta_{\ell\ell}\sigma (\ell , \ell^c)$,
the bivector-valued integral $\Icb_m$ can be expressed in terms of the energy-momentum flux $\smash{\Piell}$ in~\eqref{eq:Pi_ell} across the $(k+n)$-dimensional half space-time $\mathcal{V}_\ell^{k+n}$ of  fixed $x_\ell$, for $\ell \in \{0,\dotsc,k+n-1\}$, in~\eqref{eq:integr-region} as 
\begin{equation}
	\Icb_m = \frac{1}{\kappa_\ell} \ebf_m\wedge \Piell.
\end{equation}

\textit{Computation of $\Icb_{t,0}$.}
In~\eqref{eq:proofIt0}, $m =t$ while the tensor $\mathbf{B}$ is again given by~\eqref{eq:B_for_Iell}. Substituting the components $B_{\varepsilon(\ell,j)}$ in~\eqref{eq:proof-B-ell-j-final}
into~\eqref{eq:double_product} with $m = t$, and then back in~\eqref{eq:proofIt0} yields an analogous equation to~\eqref{eq:proofIell-2}, namely
%\begin{equation}
%	\ebf_\ell \times \bigl(\ebf_{t} \boxwedge \mathbf{B}\bigr) = 8\pi^2(-1)^{r-1}\Delta_{\ell\ell}\chi_{\ell}\negphantomhalf{t}\sum_{j\in\mathcal{I}\setminus t}\negphantomhalf{t} \sum_{\sigma \in \mathcal{S}}
%\sigma \,\xisigmaj \lvert\hat{\Ab}(\xibfsigma)\rvert^2\sigma(t,j)\,\ebf_{\varepsilon(t,j)}.
%\end{equation}
\begin{gather}
	\Icb_{t,0} = \label{eq:proofIt0-2}
\beta_r\Delta_{\ell\ell}\negphantomhalf{t}\sum_{j\in\mathcal{I}\setminus t}\negphantomhalf{t}\sigma(t,j)\,\ebf_{\varepsilon(t,j)}
\int_{\mathbf\Xi_\ell}\negphantom{\scriptscriptstyle \ell}
\frac{\drm\xi_{\ell^c}}{2\chi_\ell^3}  \, 
\xi_t\, \bigl(\xiplusj \lvert\hat{\Ab}(\xibfplus)\rvert^2 - \ximinusj \lvert\hat{\Ab}(\xibfminus)\rvert^2\bigr).
\end{gather}
It proves convenient to split the integral in two and separate the cases $j \neq \ell$ and $j = \ell$. In the first case, i.\,e.~$j \neq \ell$, noting first that $\smash{\ximinusj = \xiplusj = \xi_j}$, making a change of variables  $\zetabf_{\bar\ell} = -\xibf_{\bar\ell}$ in the integral with $\xibfminus$ gives
\begin{align}
\int_{\mathbf\Xi_\ell}\negphantom{\scriptscriptstyle \ell}
\frac{\drm\xi_{\ell^c}}{2\chi_\ell^3}  \, 
\xi_t\, \xi_{j} \lvert\hat{\Ab}(\xibfminus)\rvert^2 &= 
\int_{\mathbf\Xi_\ell}\negphantom{\scriptscriptstyle \ell}
\frac{\drm\zeta_{\ell^c}}{2\chi_\ell^3}  \, 
(-\zeta_t)\, (-\zeta_{j}) \lvert\hat{\Ab}(\zetabfplus)\rvert^2\\
&= 
\int_{\mathbf\Xi_\ell}\negphantom{\scriptscriptstyle \ell}
\frac{\drm\zeta_{\ell^c}}{2\chi_\ell^3}  \, 
\zeta_t\, \zeta_{j} \lvert\hat{\Ab}(\zetabfplus)\rvert^2, 
\end{align}
which cancels out with the integral $\xibfplus$ and the integral in~\eqref{eq:proofIt0-2} vanishes for~$j \neq \ell$. 
If~$j = \ell$, $\smash{\ximinusl = -\xiplusl = -\chi_\ell}$, unaffected by the change of variables  $\zetabf_{\bar\ell} = -\xibf_{\bar\ell}$. The integral with $\xibfminus$ gives thus
\begin{gather}
\int_{\mathbf\Xi_\ell}\negphantom{\scriptscriptstyle \ell}
\frac{\drm\xi_{\ell^c}}{2\chi_\ell^3}  \, 
\xi_t(-\chi_\ell)\, \lvert\hat{\Ab}(\xibfminus)\rvert^2 = -
\int_{\mathbf\Xi_\ell}\negphantom{\scriptscriptstyle \ell}
\frac{\drm\zeta_{\ell^c}}{2\chi_\ell^2}  \, 
(-\zeta_t)\, \lvert\hat{\Ab}(\zetabfplus)\rvert^2, 
\end{gather}
and the total integral in~\eqref{eq:proofIt0-2} vanishes too. Therefore, 
%The only non-zero summand in~\eqref{eq:proofIt0-2} is thus $j = \ell$, and $\Icb_{t,0}$ is given by
\begin{gather}
	\Icb_{t,0} = 0.\label{eq:proofIt0-3}
%8\pi^2(-1)^{r-1}\Delta_{\ell\ell}\sigma(t,\ell)\,\ebf_{\varepsilon(t,\ell)}
%\int_{\mathbf\Xi_\ell}\negphantom{\scriptscriptstyle \ell}
%\frac{\drm\xi_{\ell^c}}{2\chi_\ell}  \, 
%\frac{\xi_t}{\chi_\ell}\, \lvert\hat{\Ab}(\xibfplus)\rvert^2.
\end{gather}

\textit{Computation of $\Icb_{t,1}$.}
In~\eqref{eq:proofIt1}, the index $m$ is again $m =t$, while the tensor $\mathbf{B}$ is now given by
\begin{equation}
\mathbf{B} = \bigl(\partial_{\xi_{t}}\hat{\mf}^\ell(\xibf_{\bar\ell})\bigr)\odot\hat{\mf}^\ell(-\xibf_{\bar\ell}) + \bigl(\partial_{\xi_{t}}\hat{\mf}^\ell(\xibf_{\bar\ell})\bigr)\owedge\hat{\mf}^\ell(-\xibf_{\bar\ell}).
%\hat{\mf}^\ell(\xibf_{\bar\ell})\odot\hat{\mf}^\ell(-\xibf_{\bar\ell})+\hat{\mf}^\ell(\xibf_{\bar\ell})\owedge\hat{\mf}^\ell(-\xibf_{\bar\ell}).
\label{eq:B_for_It}
\end{equation}
Substituting the double product~\eqref{eq:double_product} in~\eqref{eq:proofIt1} gives
\begin{align}\label{app:B_ellj_t-2}
	\Icb_{t,1} = 
\Delta_{\ell\ell}\sigma(t,\ell)\,\ebf_{\varepsilon(t,\ell)}
\int_{\mathbf\Xi_\ell}\negphantom{\scriptscriptstyle \ell}\frac{\drm\xi_{\ell^c}}{4\chi_\ell^2}  \, 
B_{\ell\ell} + \Delta_{\ell\ell}\sum_{j\in\mathcal{I}\setminus {\ell, t}}\negphantomhalf{i}\sigma(t,j)\,\ebf_{\varepsilon(t,j)}
\int_{\mathbf\Xi_\ell}\negphantom{\scriptscriptstyle \ell}\frac{\drm\xi_{\ell^c}}{4\chi_\ell^2}  \, 
B_{\varepsilon(\ell,j)}.
\end{align}

In Section~\ref{app:B_ellj_t} we evaluate the components $B_{\varepsilon(\ell,j)}$ needed in~\eqref{eq:double_product}, namely $\ell = j$ and $\ell\neq j$. Substituting the expression of $B_{\ell\ell}$ in~\eqref{eq:proof-Blll-It} into the first integral in~\eqref{app:B_ellj_t-2}, and expanding the sum over $\sigma$ gives
\begin{align}\label{app:B_ellj_t-3}
	\int_{\mathbf\Xi_\ell}\negphantom{\scriptscriptstyle \ell} \frac{\drm\xi_{\ell^c}}{4\chi_\ell^2}  \, {B}_{\ell\ell} 
	= \beta_r\int_{\mathbf\Xi_\ell}\negphantom{\scriptscriptstyle \ell} \frac{\drm\xi_{\ell^c}}{4\chi_\ell^2}  \, 
	\Bigl( -&\underbrace{\Delta_{\ell\ell}\Delta_{tt} \xi_{t} \big\lvert\hat{\Ab}(\xibfplus)\big\rvert^2}_\text{1a} - \underbrace{\Delta_{\ell\ell}\Delta_{tt} \xi_{t} \big\lvert\hat{\Ab}(\xibfminus)\big\rvert^2}_\text{1b} + 2\underbrace{\chi_{\ell}^2 \bigl(\partial_{\xi_{t}}\hat\vp (\xibfplus)\bigr)\cdot \hat{\Ab}^*(\xibfplus)}_\text{2a} + 
	\notag \\ 
	& + 2 \underbrace{\chi_{\ell}^2 \bigl(\partial_{\xi_{t}}\hat\vp (\xibfminus)\bigr)\cdot \hat{\Ab}^*(\xibfminus)}_\text{2b} - \underbrace{ e^{j4\pi\Delta_{\ell\ell}\chi_\ell x_\ell}\Delta_{\ell\ell}\Delta_{tt} \xi_t\bigl( \hat{\Ab}(\xibfplus) \cdot \hat{\Ab}^*(\xibfminus)\bigr)}_\text{3a} - 
	\notag \\ 
	& -\underbrace{e^{-j4\pi\Delta_{\ell\ell}\chi_\ell x_\ell}\Delta_{\ell\ell}\Delta_{tt} \xi_t\bigl( \hat{\Ab}(\xibfminus) \cdot \hat{\Ab}^*(\xibfplus)\bigr)}_\text{3b}
	\Bigr).
\end{align}
We split the integrand in~\eqref{app:B_ellj_t-3} into three terms, respectively, indexed by 1, 2, and 3, each with consecutive pairs of summands labelled by a and b. In the integrand with label 1b, the change of variables  $\zetabf_{\bar\ell} = -\xibf_{\bar\ell}$ has opposite sign to the contribution from 1a, so the first integral is zero. Then, each of the integrands with labels 3a and 3b is an odd function of the integration variable $\xibf_{\bar\ell}$, as can be verified with the change of variables  $\zetabf_{\bar\ell} = -\xibf_{\bar\ell}$. Indeed, $\smash{\hat\vp (\xibfplus)}$ transforms into $\smash{\hat\vp^* (\xibfminus)}$ and (resp.~$\smash{\hat\vp (\xibfminus)}$) into~$\smash{\hat\vp^* (\xibfplus)}$) and therefore the third integral is zero too. It only remains the second integral, which can be expressed with the usual change of variables $\zetabf_{\bar\ell} = -\xibf_{\bar\ell}$ in 2b as 
\begin{align}\label{app:B_ellj_t-4}
	\int_{\mathbf\Xi_\ell}\negphantom{\scriptscriptstyle \ell} \frac{\drm\xi_{\ell^c}}{4\chi_\ell^2}  \, {B}_{\ell\ell} 
	= 4\pi^2(-1)^{r-1}\int_{\mathbf\Xi_\ell}\negphantom{\scriptscriptstyle \ell} \frac{\drm\xi_{\ell^c}}{2\chi_\ell}  \, 
	\chi_{\ell} \Bigl( \bigl(\partial_{\xi_{t}}\hat\vp (\xibfplus)\bigr)\cdot \hat{\Ab}^*(\xibfplus) - \text{cc}
	\Bigr),
\end{align}
where $\text{cc}$ denotes the complex conjugate.

Proceeding in a similar manner, substituting the expression for ${B}_{\varepsilon(\ell,j)}$ in~\eqref{eq:proof-Bllj-It} into the second integral in~\eqref{app:B_ellj_t-2}, and splitting the integrand into three terms, respectively, indexed by 1, 2, and 3, each with consecutive pairs of summands labelled by a and b, gives
\begin{align}\label{app:B_ellj_t-6}
	\int_{\mathbf\Xi_\ell}\negphantom{\scriptscriptstyle \ell} \frac{\drm\xi_{\ell^c}}{4\chi_\ell^2}  \, {B}_{\varepsilon(\ell,j)}
	= 4\pi^2\int_{\mathbf\Xi_\ell}\negphantom{\scriptscriptstyle \ell} \frac{\drm\xi_{\ell^c}}{4\chi_\ell}\Bigl(
	\Delta&_{tt}\Bigl(\underbrace{\bigl(\hat{\Ab}^*(\xibfplus) \odot \hat{\Ab}(\xibfplus) \bigr)\bigl|_{tj} - \bigl(\hat{\Ab}^*(\xibfplus) \odot \hat{\Ab}(\xibfplus) \bigr)\bigl|_{jt}}_{1a}\Bigr) - \notag \\
	&\qquad- \Delta_{tt}\Bigl(\underbrace{\bigl(\hat{\Ab}^*(\xibfminus) \odot \hat{\Ab}(\xibfminus) \bigr)\bigl|_{tj} - \bigl(\hat{\Ab}^*(\xibfminus) \odot \hat{\Ab}(\xibfminus) \bigr)\bigl|_{jt}}_{1b}\Bigr) - \notag \\
	&
	- 2(-1)^{r}\underbrace{\xi_j \bigl(\partial_{\xi_{t}}\hat\vp (\xibfplus)\bigr)\cdot \hat{\Ab}^*(\xibfplus)}_{2a}  + 2(-1)^{r}\underbrace{\xi_j \bigl(\partial_{\xi_{t}}\hat\vp (\xibfminus)\bigr)\cdot \hat{\Ab}^*(\xibfminus)}_{2b} - \notag \\
	%&\hphantom{= 4\pi^2\sum_{\sigma \in \mathcal{S}} \Bigl(\Delta_{tt}}
	&-\underbrace{e^{j4\pi\Delta_{\ell\ell}\chi_\ell x_\ell}\Delta_{tt}\Bigl(\bigl(\hat{\Ab}^*(\xibfminus) \odot \hat{\Ab}(\xibfplus) \bigr)\bigl|_{jt} + \bigl(\hat{\Ab}^*(\xibfminus) \odot \hat{\Ab}(\xibfplus) \bigr)\bigl|_{tj}\Bigr)}_{3a} + \notag \\
	 &+\underbrace{e^{-j4\pi\Delta_{\ell\ell}\chi_\ell x_\ell}\Delta_{tt}\Bigl(\bigl(\hat{\Ab}^*(\xibfplus) \odot \hat{\Ab}(\xibfminus) \bigr)\bigl|_{jt} + \bigl(\hat{\Ab}^*(\xibfplus) \odot \hat{\Ab}(\xibfminus) \bigr)\bigl|_{tj}
	 \Bigr)}_{3b}.
\end{align}
As before, we consider separately the integrands. If the quantities $\smash{\hat{\Ab}(\xibfplus)}$ and $\smash{\hat{\Ab}^*(\xibfplus)}$ commute, the change of variables  $\zetabf_{\bar\ell} = -\xibf_{\bar\ell}$ in the integrand 1b gives the integrand 1a with an opposite sign; this sign cancels the minus sign in front. Besides, for commuting quantities, the second summand of 1a (and of 1b) is the complex conjugate of the first summand.
The same change of variables  $\zetabf_{\bar\ell} = -\xibf_{\bar\ell}$  shows that the integrand 2b (resp.~3b) is the complex conjugate of 2a (resp.~3a). Similarly, the same change of variables and commutativity assumption applied in the second integrand of 3b shows that the second summand coincides with the first one. We therefore have
\begin{align}\label{app:B_ellj_t-7}
	\int_{\mathbf\Xi_\ell}\negphantom{\scriptscriptstyle \ell} \frac{\drm\xi_{\ell^c}}{4\chi_\ell^2}  \, {B}_{\varepsilon(\ell,j)}
	= 4\pi^2\int_{\mathbf\Xi_\ell}\negphantom{\scriptscriptstyle \ell} \frac{\drm\xi_{\ell^c}}{2\chi_\ell}\Bigl(
	\Delta&_{tt}\Bigl(\bigl(\hat{\Ab}^*(\xibfplus) \odot \hat{\Ab}(\xibfplus) \bigr)\bigl|_{tj} - \text{cc}  \Bigr) - (-1)^{r}\xi_j \Bigl(\bigl(\partial_{\xi_{t}}\hat\vp (\xibfplus)\bigr)\cdot \hat{\Ab}^*(\xibfplus) - \text{cc} \Bigr) + \notag \\
	%&\hphantom{= 4\pi^2\sum_{\sigma \in \mathcal{S}} \Bigl(\Delta_{tt}}
	& +\Delta_{tt}\Bigl(e^{-j4\pi\Delta_{\ell\ell}\chi_\ell x_\ell}\bigl(\hat{\Ab}^*(\xibfplus) \odot \hat{\Ab}(\xibfminus) \bigr)\bigl|_{tj}- \text{cc} \Bigr).
\end{align}
Note that the rank-2 tensors with components $\smash{\bigl(\hat{\Ab}^*(\xibfsigmao) \odot \hat{\Ab}(\xibfsigmat) \bigr)\bigl|_{tj} - \text{ cc}}$ are actually antisymmetric in the indices $t$ and $j$ and can thus be seen as a bivector component with element basis $\ebf_{tj}$.

Combining~\eqref{app:B_ellj_t-4} and~\eqref{app:B_ellj_t-7} back into~\eqref{app:B_ellj_t-2} gives
\begin{align}%\label{app:B_ellj_t-8}
	\Icb_{t,1} = 4\pi^2&\Delta_{\ell\ell}\Delta_{tt}\sum_{j\in\mathcal{I}\setminus {\ell, t}}\negphantomhalf{i}\sigma(t,j)\,\ebf_{\varepsilon(t,j)}
	  \int_{\mathbf\Xi_\ell}\negphantom{\scriptscriptstyle \ell} \frac{\drm\xi_{\ell^c}}{2\chi_\ell}  \, 
	\Bigl(\bigl(\hat{\Ab}^*(\xibfplus) \odot \hat{\Ab}(\xibfplus) \bigr)\bigl|_{tj} - \text{cc}  \Bigr) +
	\notag \\ 
	&+ \beta_r\Delta_{\ell\ell}\sigma(t,\ell)\,\ebf_{\varepsilon(t,\ell)}
	\int_{\mathbf\Xi_\ell}\negphantom{\scriptscriptstyle \ell} \frac{\drm\xi_{\ell^c}}{2\chi_\ell}  \, 
	\chi_{\ell} \Bigl( \bigl(\partial_{\xi_{t}}\hat\vp (\xibfplus)\bigr)\cdot \hat{\Ab}^*(\xibfplus) - \text{cc} \Bigr) + \notag\\   
	&+ \beta_r\Delta_{\ell\ell}\sum_{j\in\mathcal{I}\setminus {\ell, t}}\negphantomhalf{i}\sigma(t,j)\,\ebf_{\varepsilon(t,j)}
	 \int_{\mathbf\Xi_\ell}\negphantom{\scriptscriptstyle \ell} \frac{\drm\xi_{\ell^c}}{2\chi_\ell}  \, 
	\xi_{j} \Bigl( \bigl(\partial_{\xi_{t}}\hat\vp (\xibfplus)\bigr)\cdot \hat{\Ab}^*(\xibfplus)  - \text{cc}  \Bigr)  +
	\notag \\ 
	&+ 4\pi^2\Delta_{\ell\ell}\Delta_{tt}\sum_{j\in\mathcal{I}\setminus {\ell, t}}\negphantomhalf{i}\sigma(t,j)\,\ebf_{\varepsilon(t,j)}
	  \int_{\mathbf\Xi_\ell}\negphantom{\scriptscriptstyle \ell} \frac{\drm\xi_{\ell^c}}{2\chi_\ell}  \, 
	\Bigl(e^{-j4\pi\Delta_{\ell\ell}\chi_\ell x_\ell}\bigl(\hat{\Ab}^*(\xibfplus) \odot \hat{\Ab}(\xibfminus) \bigr)\bigl|_{tj} - \text{cc}  \Bigr).
	\label{app:B_ellj_t-9}
\end{align}
Let us denote by $\smash{\Icb_{t,1}^{1}}$ the first summand in~\eqref{app:B_ellj_t-9}; the second and third terms in~\eqref{app:B_ellj_t-9} can be grouped into a single summation, denoted by $\smash{\Icb_{t,1}^{2}}$, over $j \in\mathcal{I}\setminus {t}$. We denote by $\smash{\Icb_{t,1}^{3}}$ the remaining summand in~\eqref{app:B_ellj_t-9}. 
As $\ebf_{\varepsilon(t,j)} = -\sigma(t,j)\ebf_j\wedge\ebf_t$ and $\ebf_t\wedge\ebf_t = 0$, the contribution of $\Icb_{t,1}^1$ to the flux is given by 
\begin{align}\label{app:B_ellj_t-13}
	 - \frac{\kappa_\ell}{j2\pi} \sum_{t\in\mathcal{I}\setminus\ell}  \Delta_{tt}\Icb_{t,1}^1 &= j 2\pi\kappa_\ell\Delta_{\ell\ell} \sum_{t\in\mathcal{I}\setminus\ell}  \sum_{j\in\mathcal{I}\setminus {\ell, t}}\negphantomhalf{i}\sigma(t,j)\,\ebf_{\varepsilon(t,j)}
	  \int_{\mathbf\Xi_\ell}\negphantom{\scriptscriptstyle \ell} \frac{\drm\xi_{\ell^c}}{2\chi_\ell}  \, 
	\Bigl(\bigl(\hat{\Ab}^*(\xibfplus) \odot \hat{\Ab}(\xibfplus) \bigr)\bigl|_{tj} - \text{cc}  \Bigr) \\
	&= j 2\pi\kappa_\ell\Delta_{\ell\ell} \sum_{t\in\mathcal{I}}  \sum_{j\in\mathcal{I}\setminus t}\negphantomhalf{i}\sigma(t,j)\,\ebf_{\varepsilon(t,j)}
	  \int_{\mathbf\Xi_\ell}\negphantom{\scriptscriptstyle \ell} \frac{\drm\xi_{\ell^c}}{2\chi_\ell}  \, 
	\Bigl(\bigl(\hat{\Ab}^*(\xibfplus) \odot \hat{\Ab}(\xibfplus) \bigr)\bigl|_{tj} - \text{cc}  \Bigr) \label{app:B_ellj_t-14} \\
	&= j 4\pi\kappa_\ell\Delta_{\ell\ell} \sum_{t,j\in\mathcal{I}:t < j}\negphantomhalf{i}\ebf_{\varepsilon(t,j)}
	  \int_{\mathbf\Xi_\ell}\negphantom{\scriptscriptstyle \ell} \frac{\drm\xi_{\ell^c}}{2\chi_\ell}  \, 
	\Bigl(\bigl(\hat{\Ab}^*(\xibfplus) \odot \hat{\Ab}(\xibfplus) \bigr)\bigl|_{tj} - \text{cc}  \Bigr),\label{app:B_ellj_t-14b}
\end{align}
where we have extended the summations  to $t = \ell$ and $j = \ell$ since these added terms are zero in the Coulomb-$\ell$ gauge in~\eqref{app:B_ellj_t-14} and noted that every ordered list of non-repeated index pairs appears twice in the summation in~\eqref{app:B_ellj_t-14b}. Interpreting $\hat{\Ab}^*(\xibfplus) \odot \hat{\Ab}(\xibfplus)  - \text{cc}$ as a bivector, we obtain
\begin{align}
	 - \frac{\kappa_\ell}{j2\pi} \sum_{t\in\mathcal{I}\setminus\ell}  \Delta_{tt}\Icb_{t,1}^1 
	&= -j 2\pi\sigma(\ell,\ell^c)
	\int_{\mathbf\Xi_\ell}\negphantom{\scriptscriptstyle \ell} \frac{\drm\xi_{\ell^c}}{2\chi_\ell}  \, 
	\Bigl(\hat{\Ab}^*(\xibfplus) \odot \hat{\Ab}(\xibfplus)  - \text{cc}  \Bigr).\label{app:B_ellj_t-14c}
\end{align}

Proceeding in a similar manner, 
we can rewrite $\smash{\Icb_{t,1}^{2}}$ as 
\begin{align}\label{app:B_ellj_t-10}
	 \Icb_{t,1}^{2} &= \beta_r\Delta_{\ell\ell}
	 \int_{\mathbf\Xi_\ell}\negphantom{\scriptscriptstyle \ell} \frac{\drm\xi_{\ell^c}}{2\chi_\ell}  \, 
	 \xibfplus\wedge  \Bigl( \ebf_t\bigl(\partial_{\xi_{t}}\hat\vp^*(\xibfplus)\bigr)\cdot \hat{\Ab}(\xibfplus)  - \text{cc} \Bigr) \\
	 &= \beta_r\Delta_{\ell\ell}
	 \int_{\mathbf\Xi_\ell}\negphantom{\scriptscriptstyle \ell} \frac{\drm\xi_{\ell^c}}{2\chi_\ell}  \, 
	 \xibfplus\wedge \Bigl( \bigl(\ebf_t\partial_{\xi_{t}}\otimes\hat\vp^*(\xibfplus)\bigr)\times \hat{\Ab}(\xibfplus)  - \text{cc}  \Bigr),\label{app:B_ellj_t-11}
\end{align}
where we have also used that $\ebf_t\bigl(\partial_{\xi_{t}}\hat\vp^*(\xibfplus)\bigr)\cdot \hat{\Ab}(\xibfplus) = \bigl(\ebf_t\partial_{\xi_{t}}\otimes\hat\vp^*(\xibfplus)\bigr)\times \hat{\Ab}(\xibfplus)$, as can be seen by direct computation.
Getting back to~\eqref{eq:phi_ell_I}, and summing over $t\in\mathcal{I}\setminus\ell$, this first term of~\eqref{app:B_ellj_t-9} contributes to the flux as
\begin{equation}\label{app:B_ellj_t-12}
	- \frac{\kappa_\ell}{j2\pi} \sum_{t\in\mathcal{I}\setminus\ell}  \Delta_{tt}\Icb_{t,1}^2 = j\pi(-1)^{r}\sigma(\ell,\ell^c)
	\int_{\mathbf\Xi_\ell}\negphantom{\scriptscriptstyle \ell} \frac{\drm\xi_{\ell^c}}{2\chi_\ell}  \, 
	\xibfplus\wedge \Bigl( \bigl(\deltabf_{\xibf_{\bar\ell}}\otimes\hat\vp^*(\xibfplus)\bigr)\times \hat{\Ab}(\xibfplus)  - \text{cc}  \Bigr). 
\end{equation}
It is straightforward to verify that this expression is indeed a bivector, regardless of the value of $r$.

Analogously, the contribution of $\Icb_{t,1}^3$ is given by 
\begin{align}\label{app:B_ellj_t-15}
	 - \frac{\kappa_\ell}{j2\pi} \sum_{t\in\mathcal{I}\setminus\ell}  \Delta_{tt}\Icb_{t,1}^3 &= 
	 -j\pi\sigma(\ell,\ell^c)\sum_{t\in\mathcal{I}\setminus\ell} \sum_{j\in\mathcal{I}\setminus {\ell, t}}\negphantomhalf{i}\sigma(t,j)\,\ebf_{\varepsilon(t,j)}
	  \int_{\mathbf\Xi_\ell}\negphantom{\scriptscriptstyle \ell} \frac{\drm\xi_{\ell^c}}{2\chi_\ell}  \, 
	\Bigl(e^{-j4\pi\Delta_{\ell\ell}\chi_\ell x_\ell}\bigl(\hat{\Ab}^*(\xibfplus) \odot \hat{\Ab}(\xibfminus) \bigr)\bigl|_{tj} - \text{cc}  \Bigr).
\end{align}
Under the change of variable $\zetabf_{\bar\ell} = -\xibf_{\bar\ell}$ and the commutativity assumption, this equation becomes
\begin{align}\label{app:B_ellj_t-15b}
	 - \frac{\kappa_\ell}{j2\pi} \sum_{t\in\mathcal{I}\setminus\ell}  \Delta_{tt}\Icb_{t,1}^3 &= 
	 -j\pi\sigma(\ell,\ell^c)\sum_{t\in\mathcal{I}\setminus\ell} \sum_{j\in\mathcal{I}\setminus {\ell, t}}\negphantomhalf{i}\sigma(t,j)\,\ebf_{\varepsilon(t,j)}
	  \int_{\mathbf\Xi_\ell}\negphantom{\scriptscriptstyle \ell} \frac{\drm\xi_{\ell^c}}{2\chi_\ell}  \, 
	\Bigl(e^{-j4\pi\Delta_{\ell\ell}\chi_\ell x_\ell}\bigl(\hat{\Ab}^*(\xibfplus) \odot \hat{\Ab}^*(\xibfminus) \bigr)\bigl|_{jt} - \text{cc}  \Bigr).
\end{align}
Renaming now $t$ as $j'$ and $j$ as $t'$, this equation coincides with~\eqref{app:B_ellj_t-15}, except that $\sigma(t,j)$ picks a minus sign.  The integrand is thus an odd function of $\xibf_{\bar\ell}$ and the integral in~\eqref{app:B_ellj_t-15} is zero, and therefore
\begin{align}\label{app:B_ellj_t-16}
	 - \frac{\kappa_\ell}{j2\pi} \sum_{t\in\mathcal{I}\setminus\ell}  \Delta_{tt}\Icb_{t,1}^3 &= 0.
\end{align}

Combining $\Icb_m$ in~\eqref{eq:proofIell-3b} with~\eqref{eq:proofIt0-3},~\eqref{app:B_ellj_t-12},~\eqref{app:B_ellj_t-14}, and~\eqref{app:B_ellj_t-15} into~\eqref{eq:phi_ell_I} gives 
\begin{align}\label{app:B_ellj_t-17}
\Omegaell =  (x_\ell\ebf_{\ell}-\xcr)\wedge\Piell + j\pi&(-1)^{r}\sigma(\ell,\ell^c)
	\int_{\mathbf\Xi_\ell}\negphantom{\scriptscriptstyle \ell} \frac{\drm\xi_{\ell^c}}{2\chi_\ell}  \, 
	\xibfplus\wedge \Bigl( \bigl(\deltabf_{\xibf_{\bar\ell}}\otimes\hat\vp^*(\xibfplus)\bigr)\times \hat{\Ab}(\xibfplus)  - \text{cc}  \Bigr) - \notag \\
& -j 2\pi\sigma(\ell,\ell^c)
	\int_{\mathbf\Xi_\ell}\negphantom{\scriptscriptstyle \ell} \frac{\drm\xi_{\ell^c}}{2\chi_\ell}  \, 
	\Bigl(\hat{\Ab}^*(\xibfplus) \odot \hat{\Ab}(\xibfplus)  - \text{cc}  \Bigr),
\end{align}
where $\smash{\Piell}$ is the energy-momentum flux across the region in~\eqref{eq:Pi_ell}.

\subsection{Evaluation of the Tensor Components $B_\varepsilon(\ell,j)$ for $\Icb_m$ and $\Icb_{t,0}$ (Lorenz Gauge)}
\label{app:B_ellj_ell}

We start by listing some useful identities relating interior and wedge products \cite[Sec.~2.2]{colombaro2020generalizedMaxwellEquations}. 
Given two vectors $\ub$ and $\vb$ and a $r$-vector $\wb$, the following expressions hold
\begin{gather} 
%\label{eq:equiv-double-interior}
%\ub \lintprod (\wb \rintprod \vb) = (\ub \lintprod \wb) \rintprod \vb , \\
\ub \lintprod (\vb \lintprod \wb) = -\vb \lintprod (\ub \lintprod \wb), \label{eq:equiv-double-interior-2}
\end{gather}
and
\begin{equation} \label{eq:equiv-lint-wedge}
\ub \lintprod (\vb \wedge \wb) = (-1)^r (\ub\cdot\vb) \wb + \vb \wedge (\ub \lintprod \wb) .
\end{equation}
In addition, given two vectors $\vb$ and $\vb'$ and two $r$-vectors $\wb$ and $\wb'$, then it holds that
\begin{gather} 
%\label{eq:equiv-wedge-int}
%\left( \vb \wedge \wb \right) \cdot \left( \wb' \wedge \vb' \right) = (-1)^r \left( \vb \cdot \vb' \right) \left( \wb \cdot \wb' \right)
%+ \left( \vb' \lintprod \wb \right) \cdot \left( \wb' \rintprod \vb \right), \\
\left( \vb \wedge \wb \right) \cdot \left( \vb' \wedge \wb' \right) + \left( \vb' \lintprod \wb \right) \cdot \left( \vb \lintprod \wb' \right) = \left( \vb \cdot \vb' \right) \left( \wb \cdot \wb' \right). \label{eq:equiv-wedge-int-2}
\end{gather}

Also, for a vector $\ub$, an $(r-1)$-vector $\vb$ and an $r$-vector $\wb$, it holds that
\begin{gather} 
%\label{eq:proof-wedge-dot-}
%(\ub \wedge \vb) \cdot \wb = \vb \cdot (\wb \rintprod \ub), \\
(\ub \wedge \vb) \cdot \wb = (-1)^{r-1} (\ub \lintprod \wb) \cdot \vb. \label{eq:proof-wedge-dot-2}
\end{gather}

In the tensor definition in~\eqref{eq:B_for_Iell} we need both $\hat{\mf}^\ell(\xibf_{\bar\ell})$, given in~\eqref{eq:proofT4c}, and $\hat{\mf}^\ell(-\xibf_{\bar\ell})$. The latter is evaluated noting that the real-valuedness of the field implies that $\hat{\mf} (-\xibf) =\hat{\mf}^* (\xibf)$ as follows,
\begin{align}
\hat{\mf}^\ell(-\xibf_{\bar\ell}) &= e^{j2\pi\Delta_{\ell\ell}\chi_\ell x_\ell} \, \hat{\mf} (-\xibfplus) + e^{-j2\pi\Delta_{\ell\ell}\chi_\ell x_\ell} \, \hat{\mf} (-\xibfminus) \\
&= e^{j2\pi\Delta_{\ell\ell}\chi_\ell x_\ell} \, \hat{\mf}^*(\xibfminus) + e^{-j2\pi\Delta_{\ell\ell}\chi_\ell x_\ell} \, \hat{\mf}^*(\xibfplus) \\
&= \sum_{\sigma\in\mathcal{S}}e^{-j2\pi\Delta_{\ell\ell}\sigma\chi_\ell x_\ell} \, \hat{\mf}^*(\xibfsigma). \label{eq:proofT4e}
\end{align}
Substituting the definition of $\hat{\mf}^\ell$ given in~\eqref{eq:proofT4c} together with~\eqref{eq:proofT4e} in the tensor definition~\eqref{eq:B_for_Iell}, we obtain
\begin{equation}
\mathbf{B} = \sum_{\sigma_1,\sigma_2 \in \mathcal{S}}
e^{j2\pi\Delta_{\ell\ell}(\sigma_1-\sigma_2)\chi_\ell x_\ell}\Bigl( 
\hat{\mf} (\xibfsigmao)\odot \hat{\mf}^* (\xibfsigmat) + \hat{\mf} (\xibfsigmao)\owedge \hat{\mf}^* (\xibfsigmat)\Bigr).
\label{eq:proof-B-sigma1-sigma2}
\end{equation}

Let us define $\mathbf{B}^{\sigma_1\sigma_2}$ as $\smash{\mathbf{B}^{\sigma_1\sigma_2} =  
\hat{\mf} (\xibfsigmao)\odot \hat{\mf}^* (\xibfsigmat) + \hat{\mf} (\xibfsigmao)\owedge \hat{\mf}^* (\xibfsigmat)}$; we need to evaluate the components $\smash{{B}_{\varepsilon(\ell,j)}^{\sigma_1\sigma_2}}$. Using the definition of the products $\odot$ and $\owedge$ in~\eqref{eq:odot-ij} and~\eqref{eq:owedge-ij}, the $(i,j)$-th component $\smash{{B}_{ij}^{\sigma_1\sigma_2}}$ is given by
\begin{align}
	{B}_{ij}^{\sigma_1\sigma_2} %&= \left( \hat{\mf}(\xibfsigmao) \odot \hat{\mf}^* (\xibfsigmat) + \hat{\mf}(\xibfsigmao) \owedge \hat{\mf}^* (\xibfsigmat) \right) _{ij} \label{eq:proof-Is1} \\ 
	&=  \Delta_{ii}\Delta_{jj} \Big( (\ebf_{i}\lintprod \hat{\mf}(\xibfsigmao))\cdot (\hat{\mf}^*(\xibfsigmat)\rintprod \ebf_{j}) + (\ebf_{i}\wedge\hat{\mf}(\xibfsigmao))\cdot(\hat{\mf}^*(\xibfsigmat) \wedge \ebf_{j}) \Big) \\
	&= \Delta_{ii}\Delta_{jj} \Big( (\ebf_{i}\lintprod \hat{\mf}(\xibfsigmao))\cdot (\hat{\mf}^*(\xibfsigmat)\rintprod \ebf_{j}) + 
(-1)^r\Delta_{ij}   \hat{\mf}(\xibfsigmao) \cdot  \hat{\mf}^*(\xibfsigmat) 
+(\ebf_{j}\lintprod\hat{\mf}(\xibfsigmao))\cdot(\hat{\mf}^*(\xibfsigmat) \rintprod \ebf_{i}) \Big) \label{eq:proof-Is2} \\
	&= (-1)^{r-1}\Delta_{ii}\Delta_{jj} \Bigl( \underbrace{(\ebf_{i}\lintprod \hat{\mf}(\xibfsigmao))\cdot (\ebf_{j} \lintprod \hat{\mf}^*(\xibfsigmat))}_{= \alpha_{ij}^{\sigma_1\sigma_2}} 
	+ \underbrace{(\ebf_{j}\lintprod\hat{\mf}(\xibfsigmao)) \cdot (\ebf_{i} \lintprod \hat{\mf}^*(\xibfsigmat) )}_{= \alpha_{ji}^{\sigma_1\sigma_2}} 
	-\Delta_{ij}   \hat{\mf}(\xibfsigmao) \cdot  \hat{\mf}^*(\xibfsigmat) \Bigr), \label{eq:proof-Is3}
\end{align}
where we have applied~\eqref{eq:equiv-wedge-int-2} in~\eqref{eq:proof-Is2} and the identity $\vb \lintprod \wb = (-1)^{\gr(\vb)(\gr(\wb)+\gr(\vb))} \wb \rintprod \vb$ \cite[Eq.~(21)]{colombaro2020generalizedMaxwellEquations} in~\eqref{eq:proof-Is2} and~\eqref{eq:proof-Is3}, and have defined the quantities $\alpha_{ij}^{\sigma_1\sigma_2}$ for ease of presentation. 

Substituting the potential in the Fourier domain, $\hat{\mf}(\xibf) = j2\pi \xibf\wedge \hat{\vp}(\xibf)$,  and subsequently using~\eqref{eq:equiv-wedge-int-2} yields
\begin{align}
	\hat{\mf}(\xibfsigmao) \cdot  \hat{\mf}^*(\xibfsigmat) &= 4\pi^2\bigl(\xibfsigmao \wedge \hat{\Ab}(\xibfsigmao)\bigr) \cdot \bigl(\xibfsigmat \wedge \hat{\Ab}^*(\xibfsigmat)\bigr) \\
	&= 4\pi^2\Bigl(
	(\xibfsigmao\cdot\xibfsigmat)\bigl( \hat{\Ab}(\xibfsigmao) \cdot \hat{\Ab}^*(\xibfsigmat)\bigr) -
	\bigl(\xibfsigmat \lintprod \hat{\Ab}(\xibfsigmao)\bigr) \cdot \bigl(\xibfsigmao \lintprod \hat{\Ab}^*(\xibfsigmat)\bigr) 
	\Bigr)
	. \label{eq:I_ij_4}
\end{align}
We note that the wave-equation condition $\xibfsigma\cdot\xibfsigma = 0$, together with the identity $\xibfsigmao =  \xibfsigmat + (\sigma_1-\sigma_2) \chi_\ell\ebf_\ell$, with $\sigma_1,\sigma_2\in\{+1,-1\}$, implies that 
\begin{equation} \label{eq:xibf+and-}
	\xibfsigmao\cdot\xibfsigmat = (\sigma_1\sigma_2-1)\Delta_{\ell\ell} \chi^2_\ell.
\end{equation}
Also, the Lorenz-gauge condition in the Fourier domain, namely
%\begin{equation}
$\xibfsigma \lintprod \hat{\Ab}(\xibfsigma)= 0 = \xibfsigma \lintprod \hat{\Ab}^*(\xibfsigma)$,
%\end{equation}
implies that 
\begin{gather}\label{eq:xibf+and--gauge-1}
\xibfsigmao \lintprod \hat{\Ab}(\xibfsigmat) = (\sigma_1-\sigma_2)\chi_\ell\ebf_\ell \lintprod \hat{\Ab}(\xibfsigmat), \\
\xibfsigmao \lintprod \hat{\Ab}^*(\xibfsigmat) = (\sigma_1-\sigma_2)\chi_\ell\ebf_\ell \lintprod \hat{\Ab}^*(\xibfsigmat).\label{eq:xibf+and--gauge-2}
\end{gather}
Substituting~\eqref{eq:xibf+and-}--\eqref{eq:xibf+and--gauge-2} back into~\eqref{eq:I_ij_4} yields
\begin{align}
	\hat{\mf}(\xibfsigmao) \cdot  \hat{\mf}^*(\xibfsigmat) &= 4\pi^2\Bigl(
	(\sigma_1\sigma_2-1)\Delta_{\ell\ell} \chi^2_\ell\bigl( \hat{\Ab}(\xibfsigmao) \cdot \hat{\Ab}^*(\xibfsigmat)\bigr) +
	(\sigma_1-\sigma_2)^2\chi_\ell^2\bigl(\ebf_\ell \lintprod \hat{\Ab}(\xibfsigmao)\bigr) \cdot \bigl(\ebf_\ell \lintprod \hat{\Ab}^*(\xibfsigmat)\bigr) 
	\Bigr)
	.\label{eq:I_ij_4b}
\end{align}

Turning back our attention to $\smash{\alpha_{ij}^{\sigma_1\sigma_2}}$, substitution of the potential in the Fourier domain, $\smash{\hat{\mf}(\xibf) = j2\pi \xibf\wedge \hat{\vp}(\xibf)}$,  followed by the use of~\eqref{eq:equiv-lint-wedge}, gives
\begin{align}
	\ebf_{i}\lintprod \hat{\mf}(\xibfsigmao) %&= j2\pi\ebf_{i}\lintprod \bigl(\xibfsigmao \wedge \hat{\Ab}(\xibfsigmao)\bigr) \\
	&= j2\pi\bigl((-1)^{r-1}\bigl(\ebf_i \cdot \xibfsigmao) \hat{\Ab}(\xibfsigmao) + \xibfsigmao \wedge\bigl(\ebf_i \lintprod\hat{\Ab}(\xibfsigmao) \bigr)\bigr)  \\
	&= -j2\pi\bigl((-1)^{r}\Delta_{ii}\xisigmaoi \hat{\Ab}(\xibfsigmao) - \xibfsigmao \wedge \bigl(\ebf_i \lintprod\hat{\Ab}(\xibfsigmao) \bigr)\bigr). \label{eq:163a}
\end{align}
We therefore have for $\smash{\alpha_{ij}^{\sigma_1\sigma_2}}$ (and similarly for $\smash{\alpha_{ji}^{\sigma_1\sigma_2}}$), apart from a factor $4\pi^2$
\begin{align}
	\alpha_{ij}^{\sigma_1\sigma_2} &\propto \Bigl((-1)^{r}\Delta_{ii}\xisigmaoi \hat{\Ab}(\xibfsigmao) - \bigl(\xibfsigmao \wedge \bigl(\ebf_i \lintprod\hat{\Ab}(\xibfsigmao) \bigr)\bigr)\Bigr)\cdot\Bigl((-1)^{r}\Delta_{jj}\xisigmatj \hat{\Ab}^*(\xibfsigmat) - \bigl(\xibfsigmat \wedge \bigl(\ebf_j \lintprod\hat{\Ab}^*(\xibfsigmat) \bigr)\bigr)\Bigr) \\
	&= \Delta_{ii}\Delta_{jj}\xisigmaoi \xisigmatj \hat{\Ab}(\xibfsigmao)\cdot\hat{\Ab}^*(\xibfsigmat) - (-1)^{r}\Delta_{ii}\xisigmaoi \hat{\Ab}(\xibfsigmao)\cdot\bigl(\xibfsigmat \wedge \bigl(\ebf_j \lintprod\hat{\Ab}^*(\xibfsigmat) \bigr)\bigr)\Bigr) \notag \\
	&\hphantom{= \Delta_{ii}} - (-1)^{r}\Delta_{jj}\xisigmatj \bigl(\xibfsigmao \wedge \bigl(\ebf_i \lintprod\hat{\Ab}(\xibfsigmao) \bigr)\cdot\hat{\Ab}^*(\xibfsigmat)\bigr) + \bigl(\xibfsigmao \wedge \bigl(\ebf_i \lintprod\hat{\Ab}(\xibfsigmao) \bigr)\bigr)\cdot \bigl(\xibfsigmat \wedge \bigl(\ebf_j \lintprod\hat{\Ab}^*(\xibfsigmat) \bigr)\bigr)
	.
\end{align}	
Using~\eqref{eq:proof-wedge-dot-2} in the second and third summands and~\eqref{eq:equiv-wedge-int-2} in the fourth summand, we have apart from a factor $4\pi^2$
\begin{align}
	\alpha_{ij}^{\sigma_1\sigma_2} &\propto \Delta_{ii}\Delta_{jj}\xisigmaoi \xisigmatj \hat{\Ab}(\xibfsigmao)\cdot\hat{\Ab}^*(\xibfsigmat) - \notag \\
	&\hphantom{\Delta_{ii}} - \Delta_{ii}\xisigmaoi \bigl(\xibfsigmat \lintprod \hat{\Ab}(\xibfsigmao)\bigr) \cdot\bigl(\ebf_j \lintprod\hat{\Ab}^*(\xibfsigmat) \bigr) - \Delta_{jj}  \xisigmatj\bigl( \xibfsigmao \lintprod \hat{\Ab}^*(\xibfsigmat)\bigr) \cdot \bigl(\ebf_i \lintprod\hat{\Ab}(\xibfsigmao) \bigr) \notag \\
	&\hphantom{\Delta_{ii}}  + (\xibfsigmao \cdot \xibfsigmat)  \bigl(\ebf_i \lintprod\hat{\Ab}(\xibfsigmao) \bigr)\cdot \bigl(\ebf_j \lintprod\hat{\Ab}^*(\xibfsigmat) \bigr)
	- \bigl(\xibfsigmat \lintprod \bigl(\ebf_i \lintprod\hat{\Ab}(\xibfsigmao) \bigr)\bigr)\cdot \bigl(\xibfsigmao \lintprod \bigl(\ebf_j \lintprod\hat{\Ab}^*(\xibfsigmat) \bigr)\bigr) \\
	&= \Delta_{ii}\Delta_{jj}\xisigmaoi \xisigmatj \hat{\Ab}(\xibfsigmao)\cdot\hat{\Ab}^*(\xibfsigmat) - \notag \\
	&\hphantom{\Delta_{ii}} - \Delta_{ii}\xisigmaoi \bigl(\xibfsigmat \lintprod \hat{\Ab}(\xibfsigmao)\bigr) \cdot\bigl(\ebf_j \lintprod\hat{\Ab}^*(\xibfsigmat) \bigr) -\Delta_{jj}  \xisigmatj\bigl( \xibfsigmao \lintprod \hat{\Ab}^*(\xibfsigmat)\bigr) \cdot \bigl(\ebf_i \lintprod\hat{\Ab}(\xibfsigmao) \bigr) \notag \\
	&\hphantom{\Delta_{ii}}  + (\xibfsigmao \cdot \xibfsigmat)  \bigl(\ebf_i \lintprod\hat{\Ab}(\xibfsigmao) \bigr)\cdot \bigl(\ebf_j \lintprod\hat{\Ab}^*(\xibfsigmat) \bigr)
	- \bigl(\ebf_i \lintprod \bigl(\xibfsigmat \lintprod\hat{\Ab}(\xibfsigmao) \bigr)\bigr)\cdot \bigl(\ebf_j \lintprod \bigl(\xibfsigmao \lintprod\hat{\Ab}^*(\xibfsigmat) \bigr)\bigr),
	\label{eq:alpha_ij_3}
\end{align}	
where we have used~\eqref{eq:equiv-double-interior-2} to swap the product order between $\xibfsigmat$ and $\ebf_i$ and between $\xibfsigmao$ and $\ebf_j$.
Finally, substituting the identities~\eqref{eq:xibf+and-}--\eqref{eq:xibf+and--gauge-2} into~\eqref{eq:alpha_ij_3} gives
\begin{align}
	\alpha_{ij}^{\sigma_1\sigma_2} &\propto \Delta_{ii}\Delta_{jj}\xisigmaoi \xisigmatj \hat{\Ab}(\xibfsigmao)\cdot\hat{\Ab}^*(\xibfsigmat) - \Delta_{ii}\xisigmaoi(\sigma_2-\sigma_1)\chi_\ell \bigl(\ebf_\ell \lintprod \hat{\Ab}(\xibfsigmao)\bigr) \cdot\bigl(\ebf_j \lintprod\hat{\Ab}^*(\xibfsigmat) \bigr) - \notag \\ 
	&\hphantom{\propto \Delta_{ii}}
	- \Delta_{jj}  \xisigmatj(\sigma_1-\sigma_2)\chi_\ell\bigl( \ebf_\ell \lintprod \hat{\Ab}^*(\xibfsigmat)\bigr) \cdot \bigl(\ebf_i \lintprod\hat{\Ab}(\xibfsigmao) \bigr) + (\sigma_1\sigma_2-1)\Delta_{\ell\ell} \chi^2_\ell \bigl(\ebf_i \lintprod\hat{\Ab}(\xibfsigmao) \bigr)\cdot \bigl(\ebf_j \lintprod\hat{\Ab}^*(\xibfsigmat) \bigr) + \notag \\
	&\hphantom{\propto \Delta_{ii}}  
	+ (\sigma_1-\sigma_2)^2\chi_\ell^2\bigl(\ebf_i \lintprod \bigl(\ebf_\ell \lintprod\hat{\Ab}(\xibfsigmao) \bigr)\bigr)\cdot \bigl(\ebf_j \lintprod \bigl(\ebf_\ell \lintprod\hat{\Ab}^*(\xibfsigmat) \bigr)\bigr).
	\label{eq:alpha_ij_3b}
\end{align}	

We continue our evaluation of $\smash{B^{\sigma_1\sigma_2}_{ij}}$ by considering separately the cases $\sigma_1 = \sigma_2$ and $\sigma_1 \neq \sigma_2$. First, for $\sigma_1 = \sigma_2 = \sigma$, and 
combining~\eqref{eq:alpha_ij_3b}, both for $\smash{\alpha_{ij}^{\sigma_1\sigma_2}}$ and $\smash{\alpha_{ji}^{\sigma_1\sigma_2}}$, and~\eqref{eq:I_ij_4b}, we express ${B}_{ij}^{\sigma\sigma}$ in~\eqref{eq:proof-Is3} as
\begin{align}
	{B}_{ij}^{\sigma\sigma} &= 4\pi^2(-1)^{r-1}\Delta_{ii}\Delta_{jj} 2\Delta_{ii}\Delta_{jj}\xisigmai \xisigmaj \hat{\Ab}(\xibfsigma)\cdot\hat{\Ab}^*(\xibfsigma) \\
	&= 8\pi^2(-1)^{r-1}\xisigmai \xisigmaj \lvert\hat{\Ab}(\xibfsigma)\rvert^2. \label{eq:proof-Bij-sigma-sigma}
\end{align}
For ${\sigma}_1 = \sigma$, and $\sigma_2 = -\sigma = \bar\sigma$, combining~\eqref{eq:alpha_ij_3b} and~\eqref{eq:I_ij_4b} apart from a factor $4\pi^2(-1)^{r-1}\Delta_{ii}\Delta_{jj}$, we express ${B}_{ij}^{\sigma\bar\sigma}$ in~\eqref{eq:proof-Is3} as 
\begin{align}
	{B}_{ij}^{\sigma\bar\sigma} \propto  
	\Delta_{ii}&\Delta_{jj}\xisigmai \xisigmabj \hat{\Ab}(\xibfsigma)\cdot\hat{\Ab}^*(\xibfsigmab) + 2 \Delta_{ii}\xisigmai\sigma\chi_\ell \bigl(\ebf_\ell \lintprod \hat{\Ab}(\xibfsigma)\bigr) \cdot\bigl(\ebf_j \lintprod\hat{\Ab}^*(\xibfsigmab) \bigr) - \notag \\ 
	& - 2\Delta_{jj}  \xisigmabj\sigma\chi_\ell\bigl( \ebf_\ell \lintprod \hat{\Ab}^*(\xibfsigmab)\bigr) \cdot \bigl(\ebf_i \lintprod\hat{\Ab}(\xibfsigma) \bigr) - 2\Delta_{\ell\ell} \chi^2_\ell \bigl(\ebf_i \lintprod\hat{\Ab}(\xibfsigma) \bigr)\cdot \bigl(\ebf_j \lintprod\hat{\Ab}^*(\xibfsigmab) \bigr) + \notag \\
	& + 4\chi_\ell^2\bigl(\ebf_i \lintprod \bigl(\ebf_\ell \lintprod\hat{\Ab}(\xibfsigma) \bigr)\bigr)\cdot \bigl(\ebf_j \lintprod \bigl(\ebf_\ell \lintprod\hat{\Ab}^*(\xibfsigmab) \bigr)\bigr) + \notag \\
	& +\Delta_{ii}\Delta_{jj}\xisigmaj \xi_{\bar\sigma,i} \hat{\Ab}(\xibfsigma)\cdot\hat{\Ab}^*(\xibfsigmab) + 2 \Delta_{jj}\xisigmaj\sigma\chi_\ell \bigl(\ebf_\ell \lintprod \hat{\Ab}(\xibfsigma)\bigr) \cdot\bigl(\ebf_i \lintprod\hat{\Ab}^*(\xibfsigmab) \bigr) - \notag \\ 
	& - 2\Delta_{ii}  \xi_{\bar\sigma,i}\sigma\chi_\ell\bigl( \ebf_\ell \lintprod \hat{\Ab}^*(\xibfsigmab)\bigr) \cdot \bigl(\ebf_j \lintprod\hat{\Ab}(\xibfsigma) \bigr) - 2\Delta_{\ell\ell} \chi^2_\ell \bigl(\ebf_j \lintprod\hat{\Ab}(\xibfsigma) \bigr)\cdot \bigl(\ebf_i \lintprod\hat{\Ab}^*(\xibfsigmab) \bigr) + \notag \\
	& + 4\chi_\ell^2\bigl(\ebf_j \lintprod \bigl(\ebf_\ell \lintprod\hat{\Ab}(\xibfsigma) \bigr)\bigr)\cdot \bigl(\ebf_i \lintprod \bigl(\ebf_\ell \lintprod\hat{\Ab}^*(\xibfsigmab) \bigr)\bigr) + \notag \\
	& - \Delta_{ij}\Bigl( -2\Delta_{\ell\ell} \chi^2_\ell\bigl( \hat{\Ab}(\xibfsigma) \cdot \hat{\Ab}^*(\xibfsigmab)\bigr) +
	4\chi_\ell^2\bigl(\ebf_\ell \lintprod \hat{\Ab}(\xibfsigma)\bigr) \cdot \bigl(\ebf_\ell \lintprod \hat{\Ab}^*(\xibfsigmab)\bigr) \Bigr). \label{eq:alpha_ij_5}
\end{align}
We now set $i = \ell$, and note that $\ebf_\ell \lintprod \bigl(\ebf_\ell \lintprod\hat{\Ab}(\xibfsigma) \bigr) = 0$, $\xisigmal = \sigma\chi_\ell$, and $\xisigmabl = -\sigma\chi_\ell$, to simplify~\eqref{eq:alpha_ij_5} as 
\begin{align}
	{B}_{ij}^{\sigma\bar\sigma} \propto  
	\Delta_{\ell\ell}&\Delta_{jj}\sigma\chi_\ell \xisigmabj \hat{\Ab}(\xibfsigma)\cdot\hat{\Ab}^*(\xibfsigmab) + 2 \Delta_{\ell\ell}\chi_\ell^2\bigl(\ebf_\ell \lintprod \hat{\Ab}(\xibfsigma)\bigr) \cdot\bigl(\ebf_j \lintprod\hat{\Ab}^*(\xibfsigmab) \bigr) - \notag \\ 
	& - 2\Delta_{jj}  \xisigmabj\sigma\chi_\ell\bigl( \ebf_\ell \lintprod \hat{\Ab}^*(\xibfsigmab)\bigr) \cdot \bigl(\ebf_\ell \lintprod\hat{\Ab}(\xibfsigma) \bigr) - 2\Delta_{\ell\ell} \chi^2_\ell \bigl(\ebf_\ell \lintprod\hat{\Ab}(\xibfsigma) \bigr)\cdot \bigl(\ebf_j \lintprod\hat{\Ab}^*(\xibfsigmab) \bigr) + \notag \\
	& -\Delta_{\ell\ell}\Delta_{jj}\xisigmaj \sigma\chi_\ell \hat{\Ab}(\xibfsigma)\cdot\hat{\Ab}^*(\xibfsigmab) + 2 \Delta_{jj}\xisigmaj\sigma\chi_\ell \bigl(\ebf_\ell \lintprod \hat{\Ab}(\xibfsigma)\bigr) \cdot\bigl(\ebf_\ell \lintprod\hat{\Ab}^*(\xibfsigmab) \bigr) - \notag \\ 
	& + 2\Delta_{\ell\ell} \chi_\ell^2\bigl( \ebf_\ell \lintprod \hat{\Ab}^*(\xibfsigmab)\bigr) \cdot \bigl(\ebf_j \lintprod\hat{\Ab}(\xibfsigma) \bigr) - 2\Delta_{\ell\ell} \chi^2_\ell \bigl(\ebf_j \lintprod\hat{\Ab}(\xibfsigma) \bigr)\cdot \bigl(\ebf_\ell \lintprod\hat{\Ab}^*(\xibfsigmab) \bigr) + \notag \\
	& - \Delta_{\ell j}\Bigl( -2\Delta_{\ell\ell} \chi^2_\ell\bigl( \hat{\Ab}(\xibfsigma) \cdot \hat{\Ab}^*(\xibfsigmab)\bigr) +
	4\chi_\ell^2\bigl(\ebf_\ell \lintprod \hat{\Ab}(\xibfsigma)\bigr) \cdot \bigl(\ebf_\ell \lintprod \hat{\Ab}^*(\xibfsigmab)\bigr)\Bigr).\label{eq:alpha_ij_6}
\end{align}
Now, cancelling several common terms, Eq.~\eqref{eq:alpha_ij_6} becomes
\begin{align}
	{B}_{ij}^{\sigma\bar\sigma} \propto  
	\Delta_{\ell\ell}&\Delta_{jj}\sigma\chi_\ell (\xisigmabj - \xisigmaj) \hat{\Ab}(\xibfsigma)\cdot\hat{\Ab}^*(\xibfsigmab)  - 2\Delta_{jj}  (\xisigmabj - \xisigmaj) \sigma\chi_\ell\bigl( \ebf_\ell \lintprod \hat{\Ab}^*(\xibfsigmab)\bigr) \cdot \bigl(\ebf_\ell \lintprod\hat{\Ab}(\xibfsigma) \bigr) - \notag \\
	& - \Delta_{\ell j}\Bigl( -2\Delta_{\ell\ell} \chi^2_\ell\bigl( \hat{\Ab}(\xibfsigma) \cdot \hat{\Ab}^*(\xibfsigmab)\bigr) +
	4\chi_\ell^2\bigl(\ebf_\ell \lintprod \hat{\Ab}(\xibfsigma)\bigr) \cdot \bigl(\ebf_\ell \lintprod \hat{\Ab}^*(\xibfsigmab)\bigr)\Bigr).\label{eq:alpha_ij_7}
\end{align}
At this point, we need to distinguish two separate possibilities: $j \neq \ell$ and $j = \ell$. If $j \neq \ell$, it holds that $\smash{\xisigmaj = \xisigmabj}$ and $\Delta_{\ell j} = 0$, and therefore~\eqref{eq:alpha_ij_7} vanishes. When $j = \ell$, then $\smash{\xisigmaj = -\xisigmabj = \sigma\chi_\ell}$, and $\Delta_{\ell j} = \Delta_{\ell\ell}$ and~\eqref{eq:alpha_ij_7} vanishes too. We conclude that $\smash{{B}_{\varepsilon(\ell,j)}^{\sigma\bar\sigma} = 0}$ if $\sigma \neq\bar\sigma$, regardless of the value of $j$. Moreover, the same steps~\eqref{eq:alpha_ij_6}--\eqref{eq:alpha_ij_7} similarly prove that ${B}_{\varepsilon(\ell,i)}^{\sigma\bar\sigma} = 0$ if $\sigma \neq\bar\sigma$ for any $i$.

Since the only nonzero contribution is given by ${B}_{ij}^{\sigma\sigma}$ in \eqref{eq:proof-Bij-sigma-sigma}, substituting this latter equation in~\eqref{eq:proof-B-sigma1-sigma2} yields
\begin{equation}
{B}_{\varepsilon(\ell,j)} = 8\pi^2(-1)^{r-1}\chi_{\ell}\sum_{\sigma \in \mathcal{S}}
\sigma \,\xisigmaj \lvert\hat{\Ab}(\xibfsigma)\rvert^2.
\label{eq:proof-B-ell-j-final}
\end{equation}

\subsection{Evaluation of the Tensor Components $B_\varepsilon(\ell,j)$ for $\Icb_{t,1}$ (Coulomb-$\ell$ Gauge)}
\label{app:B_ellj_t}

As this section follows similar steps to those in the previous one, the presentation is streamlined somewhat.

Substituting the expressions for $\hat{\mf}^\ell(\xibfsigma)$ and $\hat{\mf}^\ell(-\xibfsigma)$, respectively, given in~\eqref{eq:proofT4c} and~\eqref{eq:proofT4e} in the tensor definition~\eqref{eq:B_for_It}, we obtain
\begin{equation}
\mathbf{B} = \sum_{\sigma_1,\sigma_2 \in \mathcal{S}}
e^{j2\pi\Delta_{\ell\ell}(\sigma_1-\sigma_2)\chi_\ell x_\ell}
\mathbf{B}^{\sigma_1\sigma_2},
\label{eq:proof-B-sigma1-sigma2-It}
\end{equation}
where the rank-2 symmetric tensor $\mathbf{B}^{\sigma_1\sigma_2}$ is defined as 
\begin{equation}
	\mathbf{B}^{\sigma_1\sigma_2} =  
\bigl(\partial_{\xi_{t}}\hat{\mf} (\xibfsigmao)\bigr)\odot \hat{\mf}^* (\xibfsigmat) + \bigl(\partial_{\xi_{t}}\hat{\mf} (\xibfsigmao)\bigr)\owedge \hat{\mf}^* (\xibfsigmat).
\end{equation}
Following the same steps as in~\eqref{eq:proof-Is2}--\eqref{eq:proof-Is3}, the $(i,j)$-th component $\smash{{B}_{ij}^{\sigma_1\sigma_2}}$ is given by
\begin{align}
	{B}_{ij}^{\sigma_1\sigma_2} %&= \left( \bigl(\partial_{\xi_{t}}\hat{\mf} (\xibfsigmao)\bigr) \odot \hat{\mf}^* (\xibfsigmat) + \bigl(\partial_{\xi_{t}}\hat{\mf} (\xibfsigmao)\bigr) \owedge \hat{\mf}^* (\xibfsigmat) \right) _{ij} \\ 
%	&=  \Delta_{ii}\Delta_{jj} \Big( (\ebf_{i}\lintprod \hat{\mf}(\xibfsigmao))\cdot (\hat{\mf}^*(\xibfsigmat)\rintprod \ebf_{j}) + (\ebf_{i}\wedge\hat{\mf}(\xibfsigmao))\cdot(\hat{\mf}^*(\xibfsigmat) \wedge \ebf_{j}) \Big) \\
%	&= \Delta_{ii}\Delta_{jj} \Big( (\ebf_{i}\lintprod \hat{\mf}(\xibfsigmao))\cdot (\hat{\mf}^*(\xibfsigmat)\rintprod \ebf_{j}) +  (-1)^r\Delta_{ij}   \hat{\mf}(\xibfsigmao) \cdot  \hat{\mf}^*(\xibfsigmat)  + (\ebf_{j}\lintprod\hat{\mf}(\xibfsigmao))\cdot(\hat{\mf}^*(\xibfsigmat) \rintprod \ebf_{i}) \Big) \label{eq:proof-Is2-It} \\
	&= (-1)^{r-1}\Delta_{ii}\Delta_{jj} \Bigl(  \alpha_{ij}^{\sigma_1\sigma_2} + \alpha_{ji}^{\sigma_1\sigma_2} - \Delta_{ij}   \bigl(\partial_{\xi_{t}}\hat{\mf} (\xibfsigmao)\bigr) \cdot  \hat{\mf}^*(\xibfsigmat) \Bigr), \label{eq:proof-Is3-It}
\end{align}
where $\smash{\alpha_{ij}^{\sigma_1\sigma_2}}$, and similarly $\smash{\alpha_{ji}^{\sigma_1\sigma_2}}$, is given by
\begin{equation}
	\alpha_{ij}^{\sigma_1\sigma_2} = \Bigl(\ebf_{i}\lintprod \bigl(\partial_{\xi_{t}}\hat{\mf} (\xibfsigmao)\bigr)\Bigr)\cdot \Bigl(\ebf_{j} \lintprod \hat{\mf}^*(\xibfsigmat)\Bigr).
\end{equation}

Substituting the potential in the Fourier domain, we obtain
\begin{equation}
	\partial_{\xi_{t}}\hat{\mf} (\xibfsigmao) = j2\pi\ebf_t\wedge\hat\vp (\xibfsigmao) +  j2\pi\xibfsigmao\wedge\bigl(\partial_{\xi_{t}}\hat\vp (\xibfsigmao)\bigr).
\end{equation}
Using this identity together with~\eqref{eq:equiv-wedge-int-2} allows us to write the last term in~\eqref{eq:proof-Is3-It}, apart from a factor $4\pi^2$, as
\begin{align}
	\bigl(\partial_{\xi_{t}}\hat{\mf} (\xibfsigmao)\bigr) \cdot  \hat{\mf}^*(\xibfsigmat) &\propto 
	\bigl(\ebf_{t} \wedge \hat{\Ab}(\xibfsigmao)\bigr) \cdot \bigl(\xibfsigmat \wedge \hat{\Ab}^*(\xibfsigmat)\bigr)
	+
	\bigl(\xibfsigmao \wedge \bigl(\partial_{\xi_{t}}\hat\vp (\xibfsigmao)\bigr)\bigr) \cdot \bigl(\xibfsigmat \wedge \hat{\Ab}^*(\xibfsigmat)\bigr)
	\\
	&= 
	(\ebf_t \cdot\xibfsigmat)\bigl( \hat{\Ab}(\xibfsigmao) \cdot \hat{\Ab}^*(\xibfsigmat)\bigr) -
	\bigl(\xibfsigmat \lintprod \hat{\Ab}(\xibfsigmao)\bigr) \cdot \bigl(\ebf_t  \lintprod \hat{\Ab}^*(\xibfsigmat)\bigr) + \notag \\
	&\phantom{= (\ebf_t}
	+ (\xibfsigmao\cdot\xibfsigmat)\bigl( \bigl(\partial_{\xi_{t}}\hat\vp (\xibfsigmao)\bigr) \cdot \hat{\Ab}^*(\xibfsigmat)\bigr) -
	\bigl(\xibfsigmat \lintprod \bigl(\partial_{\xi_{t}}\hat\vp (\xibfsigmao)\bigr)\bigr) \cdot \bigl(\xibfsigmao \lintprod \hat{\Ab}^*(\xibfsigmat)\bigr) 
	. \label{eq:I_ij_4-It}
\end{align}

In the Coulomb-$\ell$-gauge, for which $\smash{\ebf_\ell\lintprod\hat\vp = 0}$, the conditions~\eqref{eq:xibf+and--gauge-1}  and~\eqref{eq:xibf+and--gauge-2} become
\begin{gather}\label{eq:xibf+and--coul-gauge-1}
\xibfsigmao \lintprod \hat{\Ab}(\xibfsigmat) = 0, \\
\xibfsigmao \lintprod \hat{\Ab}^*(\xibfsigmat) = 0.\label{eq:xibf+and--coul-gauge-2}
\end{gather}
Substituting the wave equation condition~\eqref{eq:xibf+and-} together with \eqref{eq:xibf+and--coul-gauge-1}--\eqref{eq:xibf+and--coul-gauge-2}  into~\eqref{eq:I_ij_4-It} gives, apart from a factor $4\pi^2$,
\begin{align}
	\bigl(\partial_{\xi_{t}}\hat{\mf} (\xibfsigmao)\bigr)  \cdot  \hat{\mf}^*(\xibfsigmat) 
	&\propto 
	\Delta_{tt} \xisigmatt\bigl( \hat{\Ab}(\xibfsigmao) \cdot \hat{\Ab}^*(\xibfsigmat)\bigr) 
	+ (\sigma_1\sigma_2-1)\Delta_{\ell\ell} \chi^2_\ell\bigl( \bigl(\partial_{\xi_{t}}\hat\vp (\xibfsigmao)\bigr) \cdot \hat{\Ab}^*(\xibfsigmat)\bigr). 
	\label{eq:FF_ij_4b-It}
\end{align}

With a similar substitution of the potential in the Fourier domain followed by~\eqref{eq:equiv-lint-wedge}, we write 
\begin{align}
	\ebf_{i}\lintprod \bigl(\partial_{\xi_{t}}\hat{\mf} (\xibfsigmao)\bigr) &= j2\pi\Bigl(\ebf_{i}\lintprod \bigl(\ebf_t \wedge \hat{\Ab}(\xibfsigmao)\bigr) + \ebf_{i}\lintprod \bigl(\xibfsigmao \wedge \bigl(\partial_{\xi_{t}}\hat\vp (\xibfsigmao)\bigr)\bigr)\Bigr) \\
	&= j2\pi\Bigl((-1)^{r-1}(\ebf_i \cdot \ebf_t) \hat{\Ab}(\xibfsigmao) + \ebf_t \wedge\bigl(\ebf_i \lintprod\hat{\Ab}(\xibfsigmao) \bigr) + \notag \\
	&\phantom{= j2\pi\Bigl(} + (-1)^{r-1}\bigl(\ebf_i \cdot \xibfsigmao) \bigl(\partial_{\xi_{t}}\hat\vp (\xibfsigmao)\bigr) + \xibfsigmao \wedge \bigl(\ebf_i \lintprod \bigl(\partial_{\xi_{t}}\hat\vp (\xibfsigmao)\bigr) \bigr)\Bigr)  \\
	&=- j2\pi\Bigl(
	(-1)^{r}\Delta_{it} \hat{\Ab}(\xibfsigmao) - \ebf_t \wedge \bigl(\ebf_i \lintprod \hat{\Ab}(\xibfsigmao) \bigr) + \notag \\
	&\phantom{= -j2\pi\Bigl(} +
	(-1)^{r}\Delta_{ii}\xisigmaoi \bigl(\partial_{\xi_{t}}\hat\vp (\xibfsigmao)\bigr) - \xibfsigmao \wedge \bigl(\ebf_i \lintprod \bigl(\partial_{\xi_{t}}\hat\vp (\xibfsigmao)\bigr) \bigr)
	\Bigr).\label{eq:163b}
\end{align}
Combining~\eqref{eq:163b} with~\eqref{eq:163a}, we therefore have for $\alpha_{ij}^{\sigma_1\sigma_2}$ (and similarly for $\alpha_{ji}^{\sigma_1\sigma_2}$), apart from a factor $4\pi^2$, 
\begin{align}
	\alpha_{ij}^{\sigma_1\sigma_2} &\propto \Bigl((-1)^{r}\Delta_{it} \hat{\Ab}(\xibfsigmao) - \ebf_t \wedge \bigl(\ebf_i \lintprod \hat{\Ab}(\xibfsigmao) \bigr) + 
	(-1)^{r}\Delta_{ii}\xisigmaoi \bigl(\partial_{\xi_{t}}\hat\vp (\xibfsigmao)\bigr) - \xibfsigmao \wedge \bigl(\ebf_i \lintprod \bigl(\partial_{\xi_{t}}\hat\vp (\xibfsigmao)\bigr) \bigr)\Bigr) \cdot \notag \\
	&\hphantom{=\Delta_{it}}
	\cdot \Bigl((-1)^{r}\Delta_{jj}\xisigmatj \hat{\Ab}^*(\xibfsigmat) - \bigl(\xibfsigmat \wedge \bigl(\ebf_j \lintprod\hat{\Ab}^*(\xibfsigmat) \bigr)\bigr)\Bigr) \\
	&= \Delta_{it} \Delta_{jj}\xisigmatj \hat{\Ab}(\xibfsigmao)\cdot\hat{\Ab}^*(\xibfsigmat) - (-1)^{r}\Delta_{it} \hat{\Ab}(\xibfsigmao)\cdot\bigl(\xibfsigmat \wedge \bigl(\ebf_j \lintprod\hat{\Ab}^*(\xibfsigmat) \bigr)\bigr) -
	\notag \\
	&\hphantom{=\Delta_{it}} - (-1)^{r}\Delta_{jj}\xisigmatj \bigl(\ebf_t \wedge \bigl(\ebf_i \lintprod \hat{\Ab}(\xibfsigmao) \bigr)\bigr)\cdot\hat{\Ab}^*(\xibfsigmat) + \bigl(\ebf_t \wedge \bigl(\ebf_i \lintprod \hat{\Ab}(\xibfsigmao) \bigr)\bigr)\cdot\bigl(\xibfsigmat \wedge \bigl(\ebf_j \lintprod\hat{\Ab}^*(\xibfsigmat) \bigr)\bigr) + \notag\\
	&\hphantom{=\Delta_{it}} + \Delta_{ii}\Delta_{jj}\xisigmaoi\xisigmatj \bigl(\partial_{\xi_{t}}\hat\vp (\xibfsigmao)\bigr)\cdot \hat{\Ab}^*(\xibfsigmat) - (-1)^{r}\Delta_{ii}\xisigmaoi \bigl(\partial_{\xi_{t}}\hat\vp (\xibfsigmao)\bigr)\cdot\bigl(\xibfsigmat \wedge \bigl(\ebf_j \lintprod\hat{\Ab}^*(\xibfsigmat) \bigr)\bigr) - \notag\\
	&\hphantom{=\Delta_{it}} - (-1)^{r}\Delta_{jj}\xisigmatj \bigl(\xibfsigmao \wedge \bigl(\ebf_i \lintprod \bigl(\partial_{\xi_{t}}\hat\vp (\xibfsigmao)\bigr) \bigr)\bigr)\cdot\hat{\Ab}^*(\xibfsigmat)  + \notag\\
	&\hphantom{=\Delta_{it}} + \bigl(\xibfsigmao \wedge \bigl(\ebf_i \lintprod \bigl(\partial_{\xi_{t}}\hat\vp (\xibfsigmao)\bigr) \bigr)\bigr)\cdot\bigl(\xibfsigmat \wedge \bigl(\ebf_j \lintprod\hat{\Ab}^*(\xibfsigmat) \bigr)\bigr).
\end{align}	

Using~\eqref{eq:proof-wedge-dot-2} in the second, third, sixth, and seventh summands and~\eqref{eq:equiv-wedge-int-2} in the fourth and eighth ones, together with~\eqref{eq:equiv-double-interior-2} to swap the order of the interior products between $\xibfsigmat$ and $\ebf_i$,  $\xibfsigmao$ and $\ebf_j$, and $\ebf_t$ and $\ebf_j$, we obtain
\begin{align}
	\alpha_{ij}^{\sigma_1\sigma_2} &\propto 
	\Delta_{it} \Delta_{jj}\xisigmatj \hat{\Ab}(\xibfsigmao)\cdot\hat{\Ab}^*(\xibfsigmat) - \Delta_{it} \bigl(\ebf_j \lintprod\hat{\Ab}^*(\xibfsigmat) \bigr)\cdot\bigl(\xibfsigmat \lintprod \hat{\Ab}(\xibfsigmao) \bigr) -
	\notag \\
	&\hphantom{=\Delta_{it}} - \Delta_{jj}\xisigmatj \bigl(\ebf_t \lintprod \hat{\Ab}^*(\xibfsigmat)\bigr)\cdot\bigl(\ebf_i \lintprod \hat{\Ab}(\xibfsigmao) \bigr) + (\ebf_t \cdot \xibfsigmat ) \bigl(\ebf_i \lintprod \hat{\Ab}(\xibfsigmao) \bigr)\cdot\bigl(\ebf_j \lintprod\hat{\Ab}^*(\xibfsigmat) \bigr) -
	\notag \\
	&\hphantom{=\Delta_{it}} -  \bigl(\ebf_j \lintprod \bigl(\ebf_t \lintprod\hat{\Ab}^*(\xibfsigmat) \bigr)\bigr)\cdot\bigl(\ebf_i \lintprod \bigl( \xibfsigmat \lintprod \hat{\Ab}(\xibfsigmao) \bigr)\bigr) + \Delta_{ii}\Delta_{jj}\xisigmaoi\xisigmatj \bigl(\partial_{\xi_{t}}\hat\vp (\xibfsigmao)\bigr)\cdot \hat{\Ab}^*(\xibfsigmat) - \notag \\
	&\hphantom{=\Delta_{it}}  - \Delta_{ii}\xisigmaoi \bigl(\ebf_j \lintprod\hat{\Ab}^*(\xibfsigmat) \bigr)\cdot\bigl(\xibfsigmat \lintprod \bigl(\partial_{\xi_{t}}\hat\vp (\xibfsigmao)\bigr) \bigr) - \Delta_{jj}\xisigmatj \bigl(\xibfsigmao \lintprod \hat{\Ab}^*(\xibfsigmat) \bigr)\bigr)\cdot\bigl(\ebf_i \lintprod \bigl(\partial_{\xi_{t}}\hat\vp (\xibfsigmao)\bigr) +
	\notag \\
	&\hphantom{=\Delta_{it}} + \bigl(\xibfsigmao\cdot \xibfsigmat\bigr) \bigl( \ebf_i \lintprod \bigl(\partial_{\xi_{t}}\hat\vp (\xibfsigmao)\bigr) \bigr)\cdot\bigl( \ebf_j \lintprod\hat{\Ab}^*(\xibfsigmat) \bigr) - 
	\notag \\
	&\hphantom{=\Delta_{it}} 
	-\bigl( \ebf_j \lintprod \bigl(\xibfsigmao\lintprod\hat{\Ab}^*(\xibfsigmat) \bigr) \bigr)\bigr)\cdot\bigl(\ebf_i  \lintprod \bigl( \xibfsigmat \lintprod \bigl(\partial_{\xi_{t}}\hat\vp (\xibfsigmao)\bigr) \bigr). \label{eq:alpha_ij_3-It}
\end{align}

By taking the derivative of~\eqref{eq:xibf+and--coul-gauge-1} with respect to $\xi_t$, with $t \neq \ell$, we have
\begin{equation}\label{eq:xibf+and--gauge-1-der}
	\xibfsigmao \lintprod \bigl( \partial_{\xi_t}\hat{\Ab}(\xibfsigmat)\bigr) = - \ebf_{t} \lintprod \hat{\Ab}(\xibfsigmat).
\end{equation}
Substituting~\eqref{eq:xibf+and-},~\eqref{eq:xibf+and--gauge-1-der}, and~\eqref{eq:xibf+and--coul-gauge-1}--\eqref{eq:xibf+and--coul-gauge-2} back into~\eqref{eq:alpha_ij_3-It}, this equation simplifies to
\begin{align}
	\alpha_{ij}^{\sigma_1\sigma_2} &\propto 
	\Delta_{it} \Delta_{jj}\xisigmatj \hat{\Ab}(\xibfsigmao)\cdot\hat{\Ab}^*(\xibfsigmat) + (\sigma_1\sigma_2-1)\Delta_{\ell\ell} \chi^2_\ell \bigl( \ebf_i \lintprod \bigl(\partial_{\xi_{t}}\hat\vp (\xibfsigmao)\bigr) \bigr)\cdot\bigl( \ebf_j \lintprod\hat{\Ab}^*(\xibfsigmat) \bigr) + 
	\notag \\
	&\hphantom{=\Delta_{it}} + \Delta_{ii}\xisigmaoi \bigl(\ebf_j \lintprod\hat{\Ab}^*(\xibfsigmat) \bigr)\cdot\bigl(\ebf_{t} \lintprod \hat{\Ab}(\xibfsigmao) \bigr) - 
	\Delta_{jj}\xisigmatj \bigl(\ebf_t \lintprod \hat{\Ab}^*(\xibfsigmat)\bigr)\cdot\bigl(\ebf_i \lintprod \hat{\Ab}(\xibfsigmao) \bigr) +
		\notag \\
	&\hphantom{=\Delta_{it}} + \Delta_{tt} \xi_{t}  \bigl(\ebf_i \lintprod \hat{\Ab}(\xibfsigmao) \bigr)\cdot\bigl(\ebf_j \lintprod\hat{\Ab}^*(\xibfsigmat) \bigr) + \Delta_{ii}\Delta_{jj}\xisigmaoi\xisigmatj \bigl(\partial_{\xi_{t}}\hat\vp (\xibfsigmao)\bigr)\cdot \hat{\Ab}^*(\xibfsigmat). 	
	\label{eq:alpha_ij_3b-It}
\end{align}

As in the analysis of the tensor components for $I_\ell$ and $\Icb_{t,0}$ in~\ref{app:B_ellj_ell}, 
we continue our evaluation of $\smash{B^{\sigma_1\sigma_2}_{ij}}$ by considering separately the cases $\sigma_1 = \sigma_2$ and $\sigma_1 \neq \sigma_2$. First, for $\sigma_1 = \sigma_2 = \sigma$, 
combining~\eqref{eq:alpha_ij_3b-It} for $\smash{\alpha_{ij}^{\sigma_1\sigma_2}}$ and $\smash{\alpha_{ji}^{\sigma_1\sigma_2}}$ with~\eqref{eq:FF_ij_4b-It} and~\eqref{eq:xibf+and--gauge-1-der}, we  write the tensor component ${B}_{ij}^{\sigma\sigma}$ in~\eqref{eq:proof-Is3-It} apart from a factor $4\pi^2(-1)^{r-1}\Delta_{ii}\Delta_{jj}$ as
\begin{align}
	{B}_{ij}^{\sigma\sigma} &\propto  
	\Delta_{it} \Delta_{jj}\xisigmaj \hat{\Ab}(\xibfsigma)\cdot\hat{\Ab}^*(\xibfsigma) + \Delta_{jt} \Delta_{ii}\xisigmai \hat{\Ab}(\xibfsigma)\cdot\hat{\Ab}^*(\xibfsigma) + 
	\notag \\
	&\hphantom{=\Delta_{it}} + \Delta_{ii}\xisigmai \bigl(\ebf_j \lintprod\hat{\Ab}^*(\xibfsigma) \bigr)\cdot\bigl(\ebf_{t} \lintprod \hat{\Ab}(\xibfsigma) \bigr) - 
	\Delta_{jj}\xisigmaj \bigl(\ebf_t \lintprod \hat{\Ab}^*(\xibfsigma)\bigr)\cdot\bigl(\ebf_i \lintprod \hat{\Ab}(\xibfsigma) \bigr) +
	\notag \\
	&\hphantom{=\Delta_{it}} + \Delta_{jj}\xisigmaj \bigl(\ebf_i \lintprod\hat{\Ab}^*(\xibfsigma) \bigr)\cdot\bigl(\ebf_{t} \lintprod \hat{\Ab}(\xibfsigma) \bigr) - 
	\Delta_{ii}\xisigmai \bigl(\ebf_t \lintprod \hat{\Ab}^*(\xibfsigma)\bigr)\cdot\bigl(\ebf_j \lintprod \hat{\Ab}(\xibfsigma) \bigr) -
		\notag \\
	&\hphantom{=\Delta_{it}} + \Delta_{tt} \xi_{t}  \bigl(\ebf_i \lintprod \hat{\Ab}(\xibfsigma) \bigr)\cdot\bigl(\ebf_j \lintprod\hat{\Ab}^*(\xibfsigma) \bigr) + \Delta_{tt} \xi_{t}  \bigl(\ebf_j \lintprod \hat{\Ab}(\xibfsigma) \bigr)\cdot\bigl(\ebf_i \lintprod\hat{\Ab}^*(\xibfsigma) \bigr) + 	\notag \\
	&\hphantom{=\Delta_{it}} + 2\Delta_{ii}\Delta_{jj}\xisigmai\xisigmaj \bigl(\partial_{\xi_{t}}\hat\vp (\xibfsigma)\bigr)\cdot \hat{\Ab}^*(\xibfsigma) - \Delta_{ij}\Delta_{tt} \xi_{t}\bigl( \hat{\Ab}(\xibfsigma) \cdot \hat{\Ab}^*(\xibfsigma)\bigr).
	\label{eq:proof-Bijss-It-1}
\end{align}

Evaluating~\eqref{eq:proof-Bijss-It-1} for $i = j = \ell$, and noting that $t \neq \ell$, gives
\begin{equation}
	{B}_{\ell\ell}^{\sigma\sigma} = 4\pi^2(-1)^{r} \Bigl( \Delta_{\ell\ell}\Delta_{tt} \xi_{t} \big\lvert\hat{\Ab}(\xibfsigma)\big\rvert^2 - 2\chi_{\ell}^2 \bigl(\partial_{\xi_{t}}\hat\vp (\xibfsigma)\bigr)\cdot \hat{\Ab}^*(\xibfsigma) \Bigr) . \label{eq:proof-Bllss-It}
\end{equation}
Similarly, evaluating~\eqref{eq:proof-Bijss-It-1} for $i = \ell$ and $j \neq \ell$, and noting that $t \neq \ell$ and $j \neq t$, gives
\begin{align}
	{B}_{\varepsilon(\ell,j)}^{\sigma\sigma} &= 4\pi^2(-1)^{r-1}\Delta_{jj} \Bigl(
	\sigma\chi_\ell \bigl(\ebf_j \lintprod\hat{\Ab}^*(\xibfsigma) \bigr)\cdot\bigl(\ebf_{t} \lintprod \hat{\Ab}(\xibfsigma) \bigr) - 
	\sigma\chi_\ell \bigl(\ebf_t \lintprod \hat{\Ab}^*(\xibfsigma)\bigr)\cdot\bigl(\ebf_j \lintprod \hat{\Ab}(\xibfsigma) \bigr) +
		\notag \\
	&\hphantom{= 4\pi^2(-1)^{r-1}\Delta_{jj} \Bigl(\sigma\chi_\ell} + 2\Delta_{jj}\sigma\chi_\ell\xi_j \bigl(\partial_{\xi_{t}}\hat\vp (\xibfsigma)\bigr)\cdot \hat{\Ab}^*(\xibfsigma)\Bigr) \\
	&= 4\pi^2 \Bigl(
	\Delta_{tt}\sigma\chi_\ell \Bigl(\bigl(\hat{\Ab}^*(\xibfsigma) \odot \hat{\Ab}(\xibfsigma) \bigr)\bigl|_{tj} - \bigl(\hat{\Ab}^*(\xibfsigma) \odot \hat{\Ab}(\xibfsigma) \bigr)\bigl|_{jt}\Bigr) - 2(-1)^{r}\sigma\chi_\ell\xi_j \bigl(\partial_{\xi_{t}}\hat\vp (\xibfsigma)\bigr)\cdot \hat{\Ab}^*(\xibfsigma)\Bigr),
	\label{eq:proof-Bljss-It}
\end{align}
where we have used the definition of the $\odot$ product in~\eqref{eq:odot-ij} and the identity $\ebf_i \lintprod \vp = (-1)^{r} \vp \rintprod \ebf_i$ \cite[Eq.~(21)]{colombaro2020generalizedMaxwellEquations}.

For $-\sigma_2 = \sigma_1 = \sigma$, combining~\eqref{eq:alpha_ij_3b-It} for $\smash{\alpha_{ij}^{\sigma\bar\sigma}}$ and $\smash{\alpha_{ji}^{\sigma\bar\sigma}}$ with~\eqref{eq:FF_ij_4b-It}, we can write the tensor component ${B}_{ij}^{\sigma\bar\sigma}$ in~\eqref{eq:proof-Is3-It} apart from a factor $4\pi^2(-1)^{r-1}\Delta_{ii}\Delta_{jj}$ as
\begin{align}
	B_{ij}^{\sigma\bar\sigma} &\propto \Delta_{it} \Delta_{jj}\xisigmabj \hat{\Ab}(\xibfsigma)\cdot\hat{\Ab}^*(\xibfsigmab) - 2\Delta_{\ell\ell} \chi^2_\ell \bigl( \ebf_i \lintprod \bigl(\partial_{\xi_{t}}\hat\vp (\xibfsigma)\bigr) \bigr)\cdot\bigl( \ebf_j \lintprod\hat{\Ab}^*(\xibfsigmab) \bigr) + 
	\notag \\
	&\hphantom{=\Delta_{it}} + \Delta_{ii}\xisigmai \bigl(\ebf_j \lintprod\hat{\Ab}^*(\xibfsigmab) \bigr)\cdot\bigl(\ebf_{t} \lintprod \hat{\Ab}(\xibfsigma) \bigr) - 
	\Delta_{jj}\xisigmabj \bigl(\ebf_t \lintprod \hat{\Ab}^*(\xibfsigmab)\bigr)\cdot\bigl(\ebf_i \lintprod \hat{\Ab}(\xibfsigma) \bigr) -
		\notag \\
	&\hphantom{=\Delta_{it}} + \Delta_{tt} \xi_{t}  \bigl(\ebf_i \lintprod \hat{\Ab}(\xibfsigma) \bigr)\cdot\bigl(\ebf_j \lintprod\hat{\Ab}^*(\xibfsigmab) \bigr) + \Delta_{ii}\Delta_{jj}\xisigmai\xisigmabj \bigl(\partial_{\xi_{t}}\hat\vp (\xibfsigma)\bigr)\cdot \hat{\Ab}^*(\xibfsigmab) + 
	\notag \\
	&\hphantom{=\Delta_{it}} +
	\Delta_{jt} \Delta_{ii}\xisigmabi \hat{\Ab}(\xibfsigma)\cdot\hat{\Ab}^*(\xibfsigmab) - 2\Delta_{\ell\ell} \chi^2_\ell \bigl( \ebf_j \lintprod \bigl(\partial_{\xi_{t}}\hat\vp (\xibfsigma)\bigr) \bigr)\cdot\bigl( \ebf_i \lintprod\hat{\Ab}^*(\xibfsigmab) \bigr) + 
	\notag \\
	&\hphantom{=\Delta_{it}} + \Delta_{jj}\xisigmaj \bigl(\ebf_i \lintprod\hat{\Ab}^*(\xibfsigmab) \bigr)\cdot\bigl(\ebf_{t} \lintprod \hat{\Ab}(\xibfsigma) \bigr) - 
	\Delta_{ii}\xisigmabi \bigl(\ebf_t \lintprod \hat{\Ab}^*(\xibfsigmab)\bigr)\cdot\bigl(\ebf_j \lintprod \hat{\Ab}(\xibfsigma) \bigr) -
		\notag \\
	&\hphantom{=\Delta_{it}} + \Delta_{tt} \xi_{t}  \bigl(\ebf_j \lintprod \hat{\Ab}(\xibfsigma) \bigr)\cdot\bigl(\ebf_i \lintprod\hat{\Ab}^*(\xibfsigmab) \bigr) + \Delta_{ii}\Delta_{jj}\xisigmaj\xisigmabi \bigl(\partial_{\xi_{t}}\hat\vp (\xibfsigma)\bigr)\cdot \hat{\Ab}^*(\xibfsigmab) -
		\notag \\
	&\hphantom{=\Delta_{it}}
	- \Delta_{ij}\Delta_{tt} \xi_t\bigl( \hat{\Ab}(\xibfsigma) \cdot \hat{\Ab}^*(\xibfsigmab)\bigr) 
	+ 2\Delta_{ij} \Delta_{\ell\ell} \chi^2_\ell\bigl( \bigl(\partial_{\xi_{t}}\hat\vp (\xibfsigma)\bigr) \cdot \hat{\Ab}^*(\xibfsigmab)\bigr).
	\label{eq:proof-Bljssb-It}
\end{align}
For $i =\ell$, using that $t \neq \ell$, the Coulomb-$\ell$-gauge condition $\ebf_\ell\lintprod\hat{\Ab}$ and its consequence $\ebf_\ell \lintprod (\partial_{\xi_{t}}\hat\vp) = 0$, yield
\begin{align}
	B_{\varepsilon(\ell,j)}^{\sigma\bar\sigma} &\propto  \Delta_{\ell\ell}\sigma\chi_\ell \bigl(\ebf_j \lintprod\hat{\Ab}^*(\xibfsigmab) \bigr)\cdot\bigl(\ebf_{t} \lintprod \hat{\Ab}(\xibfsigma) \bigr)  + 
	\Delta_{\ell\ell}\sigma\chi_\ell \bigl(\ebf_t \lintprod \hat{\Ab}^*(\xibfsigmab)\bigr)\cdot\bigl(\ebf_j \lintprod \hat{\Ab}(\xibfsigma) \bigr) +
		\notag \\
	&\hphantom{=\Delta_{it}} + \Delta_{\ell\ell}\Delta_{jj}\sigma\chi_\ell\xisigmabj \bigl(\partial_{\xi_{t}}\hat\vp (\xibfsigma)\bigr)\cdot \hat{\Ab}^*(\xibfsigmab) - \Delta_{\ell\ell}\Delta_{jj}\xisigmaj\sigma\chi_\ell \bigl(\partial_{\xi_{t}}\hat\vp (\xibfsigma)\bigr)\cdot \hat{\Ab}^*(\xibfsigmab) - 
	\notag \\
	&\hphantom{=\Delta_{it}} -
	\Delta_{jt} \Delta_{\ell\ell}\sigma\chi_\ell \hat{\Ab}(\xibfsigma)\cdot\hat{\Ab}^*(\xibfsigmab) - \Delta_{\ell j}\Delta_{tt} \xi_t\bigl( \hat{\Ab}(\xibfsigma) \cdot \hat{\Ab}^*(\xibfsigmab)\bigr) 
	+ 2\Delta_{\ell j} \Delta_{\ell\ell} \chi^2_\ell\bigl( \bigl(\partial_{\xi_{t}}\hat\vp (\xibfsigma)\bigr) \cdot \hat{\Ab}^*(\xibfsigmab)\bigr).
	\label{eq:proof-Bijssc-It-1}
\end{align}
If we also consider $j = \ell$, using again the Coulomb-$\ell$-gauge condition and that $t \neq \ell$ in~\eqref{eq:proof-Bijssc-It-1} gives
\begin{align}
	B_{\ell\ell}^{\sigma\bar\sigma} &= 4\pi^2(-1)^{r}\Delta_{\ell\ell}\Delta_{jj}\Bigl(2\chi_\ell^2 \bigl(\partial_{\xi_{t}}\hat\vp (\xibfsigma)\bigr)\cdot \hat{\Ab}^*(\xibfsigmab) + \Delta_{\ell\ell}\Delta_{tt} \xi_t\bigl( \hat{\Ab}(\xibfsigma) \cdot \hat{\Ab}^*(\xibfsigmab)\bigr) 
	- 2 \chi^2_\ell\bigl( \bigl(\partial_{\xi_{t}}\hat\vp (\xibfsigma)\bigr) \cdot \hat{\Ab}^*(\xibfsigmab)\bigr)\Bigr) \\
	&= 4\pi^2(-1)^{r}\Delta_{\ell\ell}\Delta_{tt} \xi_t\bigl( \hat{\Ab}(\xibfsigma) \cdot \hat{\Ab}^*(\xibfsigmab)\bigr).
	\label{eq:proof-Blllssc-It}
\end{align}
For $j \neq \ell$, noting that $t \neq \ell$ and $j \neq t$, we evaluate~\eqref{eq:proof-Bijssc-It-1} as
\begin{align}
	B_{\varepsilon(\ell,j)}^{\sigma\bar\sigma} &= 4\pi^2(-1)^{r-1}\Delta_{\ell\ell}\Delta_{jj}\Bigl(\Delta_{\ell\ell}\sigma\chi_\ell \bigl(\ebf_j \lintprod\hat{\Ab}^*(\xibfsigmab) \bigr)\cdot\bigl(\ebf_{t} \lintprod \hat{\Ab}(\xibfsigma) \bigr)  + 
	\Delta_{\ell\ell}\sigma\chi_\ell \bigl(\ebf_t \lintprod \hat{\Ab}^*(\xibfsigmab)\bigr)\cdot\bigl(\ebf_j \lintprod \hat{\Ab}(\xibfsigma) \bigr) +
		\notag \\
	&\hphantom{=\Delta_{it}} + \Delta_{\ell\ell}\Delta_{jj}\sigma\chi_\ell\xi_j \bigl(\partial_{\xi_{t}}\hat\vp (\xibfsigma)\bigr)\cdot \hat{\Ab}^*(\xibfsigmab) - \Delta_{\ell\ell}\Delta_{jj}\xi_j\sigma\chi_\ell \bigl(\partial_{\xi_{t}}\hat\vp (\xibfsigma)\bigr)\cdot \hat{\Ab}^*(\xibfsigmab)\Bigr) \\
	&= 4\pi^2(-1)^{r-1}\Delta_{jj}\Bigl(\sigma\chi_\ell \bigl(\ebf_j \lintprod\hat{\Ab}^*(\xibfsigmab) \bigr)\cdot\bigl(\ebf_{t} \lintprod \hat{\Ab}(\xibfsigma) \bigr)  + 
	\sigma\chi_\ell \bigl(\ebf_t \lintprod \hat{\Ab}^*(\xibfsigmab)\bigr)\cdot\bigl(\ebf_j \lintprod \hat{\Ab}(\xibfsigma) \bigr)\Bigr) \\
	&= -4\pi^2\Delta_{tt}\sigma\chi_\ell \Bigl(\bigl(\hat{\Ab}^*(\xibfsigmab) \odot \hat{\Ab}(\xibfsigma) \bigr)\bigl|_{jt} + \bigl(\hat{\Ab}^*(\xibfsigmab) \odot \hat{\Ab}(\xibfsigma) \bigr)\bigl|_{tj}\Bigr), 
	\label{eq:proof-Blljssc-It}
\end{align}
where we have used the definition of the $\odot$ product in~\eqref{eq:odot-ij} and the identity $\ebf_i \lintprod \vp = (-1)^{r} \vp \rintprod \ebf_i$ \cite[Eq.~(21)]{colombaro2020generalizedMaxwellEquations}. Note that the product $\odot$ could be replaced by $\owedge$ in~\eqref{eq:Sbell} with an overall change of sign, since the off-diagonal transposed components of both products coincide \cite[Eq.~(22)]{colombaro2020generalizedMaxwellEquations}.

Finally, using~\eqref{eq:proof-B-sigma1-sigma2} and, respectively, combining~\eqref{eq:proof-Bllss-It} and~\eqref{eq:proof-Blllssc-It}, and~\eqref{eq:proof-Bljss-It} and~\eqref{eq:proof-Blljssc-It}, we obtain
\begin{gather}
	B_{\ell\ell} = 4\pi^2(-1)^{r}\sum_{\sigma \in \mathcal{S}}  \Bigl( \Delta_{\ell\ell}\Delta_{tt} \xi_{t} \big\lvert\hat{\Ab}(\xibfsigma)\big\rvert^2 - 2\chi_{\ell}^2 \bigl(\partial_{\xi_{t}}\hat\vp (\xibfsigma)\bigr)\cdot \hat{\Ab}^*(\xibfsigma) + e^{j4\pi\Delta_{\ell\ell}\sigma\chi_\ell x_\ell}\Delta_{\ell\ell}\Delta_{tt} \xi_t\bigl( \hat{\Ab}(\xibfsigma) \cdot \hat{\Ab}^*(\xibfsigmab)\bigr) \Bigr),
	\label{eq:proof-Blll-It}
\end{gather}
and
\begin{gather}
	B_{\varepsilon(\ell,j)} = 4\pi^2\sum_{\sigma \in \mathcal{S}} \Bigl(
	\Delta_{tt}\sigma\chi_\ell \Bigl(\bigl(\hat{\Ab}^*(\xibfsigma) \odot \hat{\Ab}(\xibfsigma) \bigr)\bigl|_{tj} - \bigl(\hat{\Ab}^*(\xibfsigma) \odot \hat{\Ab}(\xibfsigma) \bigr)\bigl|_{jt}\Bigr) - 2(-1)^{r}\sigma\chi_\ell\xi_j \bigl(\partial_{\xi_{t}}\hat\vp (\xibfsigma)\bigr)\cdot \hat{\Ab}^*(\xibfsigma) \Bigr) \notag \\
	%&\hphantom{= 4\pi^2\sum_{\sigma \in \mathcal{S}} \Bigl(\Delta_{tt}}
	-e^{j4\pi\Delta_{\ell\ell}\sigma\chi_\ell x_\ell}\Delta_{tt}\sigma\chi_\ell\Bigl(\bigl(\hat{\Ab}^*(\xibfsigmab) \odot \hat{\Ab}(\xibfsigma) \bigr)\bigl|_{jt} + \bigl(\hat{\Ab}^*(\xibfsigmab) \odot \hat{\Ab}(\xibfsigma) \bigr)\bigl|_{tj} \Bigr)
	.
	\label{eq:proof-Bllj-It}
\end{gather}

\section{Spin Components: ``Canonical'' Analysis}
\label{app:spin_components}

\newcommand{\anm}{\Omegaell}
\newcommand{\anmi}{\Omega^\ell}
\newcommand{\spin}{\mathbf{S}}
\newcommand{\spini}{S}

For the standard electromagnetic field, the intrinsic angular momentum is defined only for the spatial components of the angular momentum bivector $\anm$. For the sake of simplicity, let $\xcr = 0$. For generic $\ell$, we study thus the components $\anmi_I$, with $I\in\mathcal{I}_2$, that do not include $\ell$, i.\,e.~$\ell\notin I$.  From \eqref{eq:flux_Til_alt}, and writing $I = (i,j)$, with $i,j\notin \ell$, we have to evaluate the following integral:
\begin{equation}\label{eq:51}
	\anmi_I = \sigma(\ell,\ell^c)\int_{\mathbf R^{k+n-1}}\negphantom{\scriptscriptstyle -1}\drm x_{\ell^c} \bigl(x_{i} T_{\varepsilon(\ell,j)} - x_{j}T_{\varepsilon(\ell,i)}\bigr).
\end{equation}
The following analysis is inspired by Sections~12 and~16 of Wentzel's book \cite{wentzel1949quantumTheoryFields}, which describe how to obtain the spin components from the canonical stress-energy-momentum tensor. Our analysis bypasses however the canonical tensor, and the appropriate adaptations have been made.
Using the expression of the non-diagonal components of the stress-energy-momentum tensor in \eqref{eq:emset_ij} in the integrand in \eqref{eq:51} gives
\begin{equation}\label{eq:52}
	-
	x_{i} \negphantom{\scriptscriptstyle \bar L}\sum_{\bar L\in{\mathcal{I}_{r-1}}:\ell,j\notin \bar L}\negphantom{\scriptscriptstyle \bar L}\Delta_{\bar L\bar L}\sigma(\bar L,\ell)\sigma(j,\bar L)\mfi_{\varepsilon(\ell,\bar L)}\mfi_{\varepsilon(j,\bar L)} 
	+ x_{j} \negphantom{\scriptscriptstyle \bar L}\sum_{\bar L\in{\mathcal{I}_{r-1}}:\ell,i\notin \bar L}\negphantom{\scriptscriptstyle \bar L}\Delta_{\bar L\bar L}\sigma(\bar L,\ell)\sigma(i,\bar L)\mfi_{\varepsilon(\ell,\bar L)}\mfi_{\varepsilon(i,\bar L)}.
\end{equation}
Let us split the summations over $\bar L$ into the cases where $i$ (resp.~$j$) belongs to $\bar L$ and those where it does not, and focus of the former. When $i$ (resp.~$j$) belongs to $\bar L$, we may define $L$ as a set in $\mathcal{I}_{r-2}$ such that $\ell,i,j\notin L$, so that the original set $\bar L$ is now given by $L \cup i$ (resp.~$L \cup j$). We rewrite \eqref{eq:52} accordingly as 
\begin{align}\label{eq:53a}
	-&\negphantom{\scriptscriptstyle L \in}\sum_{L\in{\mathcal{I}_{r-2}}:\ell,i,j\notin L}\negphantom{\notin\scriptscriptstyle L} x_{i} \Delta_{ii}\Delta_{LL}\sigma\bigl(\varepsilon(i,L),\ell\bigr)\sigma\bigl(j,\varepsilon(i,L)\bigr))\mfi_{\varepsilon(\ell,i,L)}\mfi_{\varepsilon(i,j,L)} 
	+ \negphantom{\scriptscriptstyle L\in}\sum_{L\in{\mathcal{I}_{r-2}}:\ell,i,j\notin L}\negphantom{\notin\scriptscriptstyle L} x_{j} \Delta_{jj}\Delta_{LL}\sigma\bigl(\varepsilon(j,L),\ell\bigr)\sigma\bigl(i,\varepsilon(j,L)\bigr)\mfi_{\varepsilon(\ell,j,L)}\mfi_{\varepsilon(i,j,L)} \\
	&= -\negphantom{\scriptscriptstyle L\in}\sum_{L\in{\mathcal{I}_{r-2}}:\ell,i,j\notin L}\negphantom{\notin\scriptscriptstyle L} \Delta_{LL}\Bigl(
	\Delta_{ii} \sigma\bigl(\varepsilon(i,L),\ell\bigr)\sigma\bigl(j,\varepsilon(L,i)\bigr) x_{i} \mfi_{\varepsilon(\ell,i,L)}\mfi_{\varepsilon(i,j,L)} 
	- \Delta_{jj}\sigma\bigl(\varepsilon(L,j),\ell\bigr)\sigma\bigl(i,\varepsilon(L,j)\bigr) x_{j} \mfi_{\varepsilon(\ell,j,L)}\mfi_{\varepsilon(i,j,L)}\Bigr).\label{eq:53}
\end{align}
From the definition of the field from the potential, we have
\begin{gather}
	\mfi_{\varepsilon(i,j,L)} = \Delta_{ii}\sigma\bigl(i,\varepsilon(j,L)\bigr)\partial_i \vpi_{\varepsilon(j,L)} + \dotsc \\
	\mfi_{\varepsilon(i,j,L)} = \Delta_{jj}\sigma\bigl(j,\varepsilon(i,L)\bigr)\partial_j \vpi_{\varepsilon(i,L)} + \dotsc. 
\end{gather}
Ignoring the terms in the dots, that do not contribute to the spin, and, respectively, substituting these two expressions in the two appearances of $\mfi_{\varepsilon(i,j,L)}$ in~\eqref{eq:53} gives the following expression for each summand
\begin{align}
	-\Delta_{LL}\sigma\bigl(j,\varepsilon(L,i)\bigr)\sigma\bigl(i,\varepsilon(j,L)\bigr)\Bigl(
	\sigma\bigl(\varepsilon(i,L),\ell\bigr) x_{i} \mfi_{\varepsilon(\ell,i,L)}\partial_i \vpi_{\varepsilon(j,L)}
	- \sigma\bigl(\varepsilon(L,j),\ell\bigr) x_{j} \mfi_{\varepsilon(\ell,j,L)}\partial_j \vpi_{\varepsilon(i,L)}\Bigr).\label{eq:59}
\end{align}
We continue with the following manipulations,
\begin{align}
	x_{i} \mfi_{\varepsilon(\ell,i,L)}\partial_i \vpi_{\varepsilon(j,L)} &= \partial_i \bigl(x_{i} \vpi_{\varepsilon(j,L)}\mfi_{\varepsilon(\ell,i,L)}\bigr) - \vpi_{\varepsilon(j,L)}\mfi_{\varepsilon(\ell,i,L)} - x_{i} \vpi_{\varepsilon(j,L)} \partial_i \mfi_{\varepsilon(\ell,i,L)}. \label{eq:60}
\end{align}
A similar expression holds for the second summand in \eqref{eq:59}. If we now neglect the third summand, as unrelated to the spin, and argue that the first is zero after integration in \eqref{eq:60}, substituting these back in \eqref{eq:59} yields
\begin{align}
	\Delta_{LL}\sigma\bigl(j,\varepsilon(L,i)\bigr)\sigma\bigl(i,\varepsilon(j,L)\bigr)\Bigl(
	\sigma\bigl(\varepsilon(i,L),\ell\bigr)\vpi_{\varepsilon(j,L)}\mfi_{\varepsilon(\ell,i,L)}
	- \sigma\bigl(\varepsilon(L,j),\ell\bigr)\vpi_{\varepsilon(i,L)}\mfi_{\varepsilon(\ell,j,L)}\Bigr).\label{eq:60b}
\end{align}
 
In the Coulomb-$\ell$-gauge, we also have that
\begin{gather}
	\mfi_{\varepsilon(\ell,i,L)} = \Delta_{\ell\ell}\sigma\bigl(\ell,\varepsilon(i,L)\bigr)\partial_\ell \vpi_{\varepsilon(i,L)}, \\
	\mfi_{\varepsilon(\ell,j,L)} = \Delta_{\ell\ell}\sigma\bigl(\ell,\varepsilon(j,L)\bigr)\partial_\ell \vpi_{\varepsilon(j,L)}, 
\end{gather}
and substituting these expressions in \eqref{eq:60b}, and using that $\sigma\bigl(\varepsilon(i,L),\ell\bigr)\sigma\bigl(\ell,\varepsilon(i,L)\bigr) = (-1)^{r-1}$, gives
\begin{align}
	\Delta_{\ell\ell}\Delta_{LL}&\sigma\bigl(j,\varepsilon(L,i)\bigr)\sigma\bigl(i,\varepsilon(j,L)\bigr)\Bigl(
	\sigma\bigl(\varepsilon(i,L),\ell\bigr)\sigma\bigl(\ell,\varepsilon(i,L)\bigr)\vpi_{\varepsilon(j,L)}\partial_\ell \vpi_{\varepsilon(i,L)}
	- \sigma\bigl(\varepsilon(L,j),\ell\bigr)\sigma\bigl(\ell,\varepsilon(j,L)\bigr)\vpi_{\varepsilon(i,L)}\partial_\ell \vpi_{\varepsilon(j,L)}\Bigr) = \\
	&= (-1)^{r-1}\Delta_{\ell\ell}\Delta_{LL}\sigma\bigl(j,\varepsilon(L,i)\bigr)\sigma\bigl(i,\varepsilon(j,L)\bigr)\bigl(
	\vpi_{\varepsilon(j,L)}\partial_\ell \vpi_{\varepsilon(i,L)}
	- \vpi_{\varepsilon(i,L)}\partial_\ell \vpi_{\varepsilon(j,L)}\bigr) \label{eq:60c} \\
	&= \Delta_{\ell\ell}\Delta_{LL}\sigma(L,i)\sigma(j,L)\bigl(
	\vpi_{\varepsilon(j,L)}\partial_\ell \vpi_{\varepsilon(i,L)}
	- \vpi_{\varepsilon(i,L)}\partial_\ell \vpi_{\varepsilon(j,L)}\bigr), \label{eq:60d}
\end{align}
where we have used that $-(-1)^{r}\sigma\bigl(j,\varepsilon(i,L)\bigr)\sigma\bigl(i,\varepsilon(j,L)\bigr) = \sigma(L,i)\sigma(j,L)$. Indeed, using~\cite[Appendix~A]{martinez2021exteriorAlgebraicStressEnergyMomentumTensor}, we have $\sigma(L,i)\sigma(j,L) = \sigma\bigl(i,\varepsilon(j,L)\bigr)\sigma\bigl(\varepsilon(i,L),j\bigr)$ and it also holds that $(-1)^{r-1}\sigma\bigl(j,\varepsilon(i,L)\bigr) = \sigma\bigl(\varepsilon(i,L),j\bigr)$.

Getting back to the integral in \eqref{eq:51}, the $\ell$-spin component $\spini_{ij}^\ell$ is then given by
\begin{equation}\label{eq:64}
	\spini_{ij}^\ell = -\Delta_{\ell\ell}\sigma(\ell,\ell^c)\negphantom{\scriptscriptstyle L}\sum_{L\in{\mathcal{I}_{r-2}}:\ell,i,j\notin L}\negphantom{\scriptscriptstyle L}\Delta_{LL} \sigma(L,i)\sigma(j,L)\int_{\mathbf R^{k+n-1}}\negphantom{\scriptscriptstyle -1}\drm x_{\ell^c} \Bigl(\vpi_{\varepsilon(i,L)}\partial_\ell \vpi_{\varepsilon(j,L)} - \vpi_{\varepsilon(j,L)}\partial_\ell\vpi_{\varepsilon(i,L)}\Bigr).
\end{equation}
For classical electromagnetism with $r = 2$, $k = 1$, $n = 3$, and $\ell = 0$, we have $L =\O$, $\sigma(i,j)\sigma(j,i) = -1$, and the spin components are given by the standard formula (e.\,g.~\cite[Eq.~(4.83)]{maggiore2005modernIntroductionQFT})
\begin{equation}\label{eq:64a}
	\int_{\mathbf R^{3}}\drm x_{123} \bigl(\vpi_{i}\partial_0 \vpi_{j} - \vpi_{j}\partial_0\vpi_{i}\bigr).
\end{equation}

Continuing with the integral in \eqref{eq:64}, we express the potentials in terms of their normal-mode decomposition:
\begin{gather}
	\vp (\xb) = \int_{\mathbf\Xi_\ell}\negphantom{\scriptscriptstyle \ell}
		\frac{\drm\xi_{\ell^c}}{2\chi_\ell} 
		e^{j2\pi\xibf_{\bar\ell}\cdot \xb_{\bar\ell}}
		\bigl(e^{j2\pi\Delta_{\ell\ell}\chi_\ell x_\ell} \, \hat{\vp} (\xibfplus) + e^{-j2\pi\Delta_{\ell\ell}\chi_\ell x_\ell} \, \hat{\vp} (\xibfminus)\bigr) \\
	\Delta_{\ell\ell}\partial_\ell\vp (\xb) = \int_{\mathbf\Xi_\ell}\negphantom{\scriptscriptstyle \ell} j2\pi 
		\frac{\drm\xi_{\ell^c}}{2\chi_\ell} 
		e^{j2\pi\xibf_{\bar\ell}\cdot \xb_{\bar\ell}}
		\bigl(\chi_\ell e^{j2\pi\Delta_{\ell\ell}\chi_\ell x_\ell} \, \hat{\vp} (\xibfplus) - \chi_\ell  e^{-j2\pi\Delta_{\ell\ell}\chi_\ell x_\ell} \, \hat{\vp} (\xibfminus)\bigr),
\end{gather}
where $\xibf_{\bar\ell} = \xibf -\xibf_\ell\ebf_\ell$ and similarly $\xb_{\bar\ell} = \xb -\xb_\ell\ebf_\ell$, $\chi_\ell = +\sqrt{-\Delta_{\ell\ell}  \xibf_{\bar\ell}\cdot\xibf_{\bar\ell}}$ and $\xibfpm = \xibf_{\bar{\ell}} \pm \chi_\ell\ebf_\ell$. Writing
\begin{equation}
	\hat{\vp}^{\ell,\pm}(\xibf_{\bar\ell}) = e^{j2\pi\Delta_{\ell\ell}\chi_\ell x_\ell} \, \hat{\vp} (\xibfplus) \pm e^{-j2\pi\Delta_{\ell\ell}\chi_\ell x_\ell} \, \hat{\vp} (\xibfminus),
\end{equation}
we may thus evaluate the integral by using a multidimensional Dirac function as
\begin{align}
	\Delta_{\ell\ell}\int_{\mathbf R^{k+n-1}}\negphantom{\scriptscriptstyle L}\drm x_{\ell^c} \vpi_{\varepsilon(i,L)}\partial_\ell \vpi_{\varepsilon(j,L)} &= 
	j2\pi\int_{\mathbf\Xi_\ell\times\mathbf\Xi_\ell}\negphantom{\scriptscriptstyle \ell} 
\frac{\drm \xi_{\ell^c} \drm \xi'_{\ell^c}}{4\chi_\ell\chi'_\ell} \chi_\ell' \int_{\mathbf R^{k+n-1}}\negphantom{\scriptscriptstyle L}\drm x_{\ell^c} e^{j2\pi(\xibf_{\bar\ell}+\xibf_{\bar\ell}')\cdot \xb_{\bar\ell}} \hat{\vpi}^{\ell,+}_{\varepsilon(i,L)}(\xibf_{\bar\ell})\hat{\vpi}^{\ell,-}_{\varepsilon(j,L)}(\xibf_{\bar\ell}'), \\
	&= j\pi\int_{\mathbf\Xi_\ell\times\mathbf\Xi_\ell}\negphantom{\scriptscriptstyle \ell} 
\frac{\drm \xi_{\ell^c} \drm \xi'_{\ell^c}}{2\chi_\ell} \delta(\xibf_{\bar\ell}+\xibf_{\bar\ell}') \hat{\vpi}^{\ell,+}_{\varepsilon(i,L)}(\xibf_{\bar\ell})\hat{\vpi}^{\ell,-}_{\varepsilon(j,L)}(\xibf_{\bar\ell}') \\
	&= j\pi\int_{\mathbf\Xi_\ell}\negphantom{\scriptscriptstyle \ell}
\frac{\drm \xi_{\ell^c} }{2\chi_\ell} \,\hat{\vpi}^{\ell,+}_{\varepsilon(i,L)}(\xibf_{\bar\ell})\hat{\vpi}^{\ell,-}_{\varepsilon(j,L)}(-\xibf_{\bar\ell}).\label{eq:69}
\end{align}
Expanding  the Fourier components of the potential in \eqref{eq:69} yields
\begin{gather}
	\hat{\vpi}^{\ell,+}_{\varepsilon(i,L)}(\xibf_{\bar\ell}) = e^{j2\pi\Delta_{\ell\ell}\chi_\ell x_\ell} \, \hat{\vpi}_{\varepsilon(i,L)} (\xibfplus) + e^{-j2\pi\Delta_{\ell\ell}\chi_\ell x_\ell} \, \hat{\vpi}_{\varepsilon(i,L)} (\xibfminus),\label{eq:73}
\end{gather}
and similarly
\begin{align}
	\hat{\vpi}^{\ell,-}_{\varepsilon(j,L)}(-\xibf_{\bar\ell}) &= e^{j2\pi\Delta_{\ell\ell}\chi_\ell x_\ell} \, \hat{\vpi}_{\varepsilon(j,L)} (-\xibf_{\bar\ell}+\chi_\ell\ebf_\ell) - e^{-j2\pi\Delta_{\ell\ell}\chi_\ell x_\ell} \, \hat{\vpi}_{\varepsilon(j,L)} (-\xibf_{\bar\ell}-\chi_\ell\ebf_\ell) \\
	&= e^{j2\pi\Delta_{\ell\ell}\chi_\ell x_\ell} \, \hat{\vpi}_{\varepsilon(j,L)} (-\xibfminus) - e^{-j2\pi\Delta_{\ell\ell}\chi_\ell x_\ell} \, \hat{\vpi}_{\varepsilon(j,L)} (-\xibfplus).\label{eq:74}
\end{align}
Taking the product of~\eqref{eq:73} and \eqref{eq:74} yields
\begin{align}
	e^{j4\pi\Delta_{\ell\ell}\chi_\ell x_\ell} \, &\hat{\vpi}_{\varepsilon(i,L)} (\xibfplus)\, \hat{\vpi}_{\varepsilon(j,L)} (-\xibfminus) - \hat{\vpi}_{\varepsilon(i,L)} (\xibfplus)\, \hat{\vpi}_{\varepsilon(j,L)} (-\xibfplus) + \notag \\
	&+ \hat{\vpi}_{\varepsilon(i,L)} (\xibfminus)\, \hat{\vpi}_{\varepsilon(j,L)} (-\xibfminus) - e^{-j4\pi\Delta_{\ell\ell}\chi_\ell x_\ell} \, \hat{\vpi}_{\varepsilon(i,L)} (\xibfminus) \, \hat{\vpi}_{\varepsilon(j,L)} (-\xibfplus).\label{eq:75}
\end{align}

Proceeding analogously with the second summand in~\eqref{eq:69}, $\vpi_{\varepsilon(j,L)}\partial_\ell\vpi_{\varepsilon(i,L)}$, gives
\begin{align}
	e^{j4\pi\Delta_{\ell\ell}\chi_\ell x_\ell} \, &\hat{\vpi}_{\varepsilon(j,L)} (\xibfplus)\, \hat{\vpi}_{\varepsilon(i,L)} (-\xibfminus) - \hat{\vpi}_{\varepsilon(j,L)} (\xibfplus)\, \hat{\vpi}_{\varepsilon(i,L)} (-\xibfplus) + \notag \\
	&+ \hat{\vpi}_{\varepsilon(j,L)} (\xibfminus)\, \hat{\vpi}_{\varepsilon(i,L)} (-\xibfminus) - e^{-j4\pi\Delta_{\ell\ell}\chi_\ell x_\ell} \, \hat{\vpi}_{\varepsilon(j,L)} (\xibfminus) \, \hat{\vpi}_{\varepsilon(i,L)} (-\xibfplus).\label{eq:76}
\end{align}
With the change of variable $\xibf_{\bar\ell} \to -\xibf_{\bar\ell}$, and noting that $-\xibfpm = -(\xibf_{\bar\ell}\pm\chi_\ell\ebf_\ell) = -\xibf_{\bar\ell}\mp\chi_\ell\ebf_\ell \to +\xibfmp$, we thus have
\begin{align}
	e^{j4\pi\Delta_{\ell\ell}\chi_\ell x_\ell} \, &\hat{\vpi}_{\varepsilon(j,L)} (-\xibfminus)\, \hat{\vpi}_{\varepsilon(i,L)} (\xibfplus) - \hat{\vpi}_{\varepsilon(j,L)} (-\xibfminus)\, \hat{\vpi}_{\varepsilon(i,L)} (\xibfminus) + \notag \\
	&+ \hat{\vpi}_{\varepsilon(j,L)} (-\xibfplus)\, \hat{\vpi}_{\varepsilon(i,L)} (\xibfplus) - e^{-j4\pi\Delta_{\ell\ell}\chi_\ell x_\ell} \, \hat{\vpi}_{\varepsilon(j,L)} (-\xibfplus) \, \hat{\vpi}_{\varepsilon(i,L)} (\xibfminus).\label{eq:77}
\end{align}

Combining~\eqref{eq:75} and~\eqref{eq:77} with its corresponding $-1$ sign, cancelling common terms (assuming that the relevant quantities commute), and grouping common terms yields the following,
\begin{align}
	- 2\bigl(\hat{\vpi}_{\varepsilon(i,L)} (\xibfplus)\, \hat{\vpi}_{\varepsilon(j,L)} (-\xibfplus) - \hat{\vpi}_{\varepsilon(i,L)} (\xibfminus)\, \hat{\vpi}_{\varepsilon(j,L)} (-\xibfminus)\bigr) = - 2\bigl(\hat{\vpi}_{\varepsilon(i,L)} (\xibfplus)\, \hat{\vpi}_{\varepsilon(j,L)}^* (\xibfplus) - \hat{\vpi}_{\varepsilon(i,L)} (\xibfminus)\, \hat{\vpi}_{\varepsilon(j,L)}^* (\xibfminus)\bigr). 
\end{align}
With the change of variable $\xibf_{\bar\ell} \to -\xibf_{\bar\ell}$, and noting again that $\xibfpm \to -\xibfmp$, we thus have as final result
\begin{align}\label{eq:82}
	- 2\bigl(\hat{\vpi}_{\varepsilon(i,L)} (\xibfplus)\, \hat{\vpi}_{\varepsilon(j,L)}^* (\xibfplus) - \text{cc}\bigr). 
\end{align}
Putting this equation back into~\eqref{eq:69} and then into~\eqref{eq:64} gives
\begin{equation}\label{eq:83}
	\spini_{ij}^\ell = j2\pi\sigma(\ell,\ell^c)\negphantom{\scriptscriptstyle L}\sum_{L\in{\mathcal{I}_{r-2}}:\ell,i,j\notin L}\negphantom{\scriptscriptstyle L}\Delta_{LL} \sigma(L,i)\sigma(j,L)\int_{\mathbf\Xi_\ell}\negphantom{\scriptscriptstyle \ell}
\frac{\drm \xi_{\ell^c} }{2\chi_\ell} \,\bigl(\hat{\vpi}_{\varepsilon(i,L)} (\xibfplus)\, \hat{\vpi}_{\varepsilon(j,L)}^* (\xibfplus) - \text{cc}\bigr).
\end{equation}
The $(i,j)$-th spin component $\spini_{ij}^\ell$  coincides with the corresponding bivector component~\eqref{eq:Sbell_I}.
%
% BibTeX users please use
 \bibliographystyle{spphys}
 \bibliography{physics.bib, IC-bib-th.bib}
\end{document}

% end of file template.tex